\documentclass[twocolumn]{emulateapj}	

\usepackage{longtable}
\usepackage{amsmath}
\usepackage{subfigure}
\usepackage{tikz} 
\def\checkmark{\tikz\fill[scale=0.4](0,.35) -- (.25,0) -- (1,.7) -- (.25,.15) -- cycle;} 

\begin{document}

\title{WIYN Open Cluster Study. LXXV. Testing the Metallicity Dependence of Stellar Lithium
Depletion Using Hyades-Aged Clusters.  1. Hyades \& Praesepe}
\author{Jeffrey D. Cummings$^1$}
\affil{Center for Astrophysical Sciences, Johns Hopkins University}
\affil{3400 N Charles Street}
\affil{Baltimore, MD 21218, USA}
\affil{jcummi19$@$jhu.edu}
\author{Constantine P. Deliyannis$^1$}\footnote{Visiting Astronomer, Kitt Peak National Observatory, National Optical
Astronomy Observatory, which is operated by the Association of Universities
for Research in Astronomy (AURA) under cooperative agreement with the
National Science Foundation.}
\affil{Department of Astronomy, Indiana University}
\affil{727 E 3rd Street}
\affil{Bloomington, IN 47405-7105, USA}
\affil{cdeliyan$@$indiana.edu}
\author{Ryan M. Maderak$^1$}
\affil{Department of Physics and Astronomy, Benedictine College}
\affil{1020 N 2nd Street}
\affil{Atchison, KS 66002}
\affil{rmaderak$@$benedictine.edu}
\and
\author{Aaron Steinhauer}
\affil{Department of Physics \& Astronomy, SUNY Geneseo}
\affil{1 College Circle}
\affil{Geneseo, NY 14454}
\affil{steinhau$@$geneseo.edu}

\begin{abstract}
WIYN/Hydra spectroscopy (at R$\sim$15,000) of the moderately
metal-rich Praesepe
and Hyades open clusters was used to study their main sequence (MS) iron
([Fe/H]) and lithium (A(Li)) abundances.  Self-consistent [Fe/H] and Li
analyses of these clusters of consistent age, which we re-evaluate,
confirms they have consistent [Fe/H] and provides a foundation to
investigate the poorly understood G-dwarf and F-dwarf Li-depletions.
Neither phenomenon agrees with standard stellar evolution theory, but
possible explanations abound. We supplement our A(Li) with previously
published results placed on a uniform abundance scale. This creates the
largest self-consistently analyzed sample of A(Li) in both the Hyades (90)
and Praesepe (110). For
each star, high-precision UBVRI photometry was used to determine a ten
color-based T$_{\rm eff}$ and then to test for photometric peculiarities
indicated by a large $\sigma_{\rm Teff}$ ($>$ 75 K). The stars with large
$\sigma_{\rm Teff}$ were predominantly found to be binaries or stars with
peculiar (apparent) A(Li). \textit{When considering only proper-motion members
that have low $\sigma_{\rm Teff}$ and are also photometrically consistent with
the cluster MS fiducial, each cluster has a more tightly
defined Li morphology than previously observed and the two clusters' A(Li)
are indistinguishable.} This suggests that clusters of consistent age and
metallicity may have consistent Li-depletion trends across a broad range
of T$_{\rm eff}$; no additional major parameters are required, at least
for these two clusters. We propose that the combined Hyades and Praesepe
data offer more rigorous constraints than does either cluster alone, and we
discuss newly-revealed features of the combined Li-T$_{\rm eff}$ trend.
\end{abstract}

\section{Introduction}

The study of Li in open clusters provides invaluable information about physical processes occurring in the interiors of stars.  This is because, inside stars, at T $\geq$ 2.5 million K both stable Li
isotopes are destroyed by (p,$\alpha$) reactions (Burbidge et~al.\ 1957; 
Bodenheimer 1965)\footnote[2]{Li$^6$ is a factor of 12 less
abundant in meteorites than Li$^7$ (Anders \& Grevesse 1988), and is more
fragile than Li$^7$ to the extent that its (measurable) survival past the
pre-main sequence is debatable (Deliyannis et~al.\ 1990; Pinsonneault et~al.\ 1990), thus we restrict attention to Li$^7$ (hereafter, ``Li"), which is
destroyed through the Li$^7$(p,$\alpha$)He$^4$ reaction.}, and thus Li survives only in the outermost layers of
stars.  The surface Li abundance traces the actions of the surface
convection zone (SCZ) and other mixing processes that can bring
Li-depleted material to the surface.  Analysis of Li in open
clusters provides information for a broad range of stellar masses at ages
that can be precisely determined, giving both the mass dependence and
timing of Li-altering processes.  Observations of open cluster dwarfs that span a broad range of age and metallicity have shown that surface Li abundances are depleted in most, if not nearly all, stars (e.g., Deliyannis 2000; Jeffries 2000; Sestito \& Randich 2005).  ``Standard'' stellar evolution theory (no rotation, no mass loss, no magnetic fields, and no diffusion) predicts that surface Li depletion begins during the pre-main sequence (PMS) phase when the SCZ is very deep, allowing surface Li to travel deep into the interior where it can be destroyed (Proffitt \& Michaud 1989; Deliyannis et~al.\ 1990, hereafter DDK; Pinsonneault 1997, hereafter P97).  The degree of surface Li depletion during the PMS depends on stellar mass and metallicity (DDK; P97; Deliyannis \& Demarque 1991).  Regarding the dependence on mass: Population I PMS stars that will become A and F dwarfs ($\ge$ 1.3 M$_{\odot}$) remain in the PMS phase relatively briefly, and the base of the SCZ just barely achieves a high enough temperature to destroy Li, so only a negligible amount of surface Li depletion occurs before they reach the main sequence (MS).  Lower-mass stars remain in the PMS phase longer and have a higher temperature (and density) at the base of their deeper SCZs, so they deplete a greater amount of surface Li.  The SCZs are shallower during the MS, so no further Li destruction occurs, except for (possibly very late G), K, and M dwarfs, whose SCZs remain deep with hot bases. Standard theory also predicts that Li depletion has a strong dependence on metallicity, where the more-metal-rich stars deplete surface Li more rapidly.  This is because a higher metallicity causes a higher opacity, which in turn causes a deeper SCZ.

A broad range of open cluster observations, however, show that in most clusters standard theory predicts insufficient Li depletion at a majority of, if not all, observed stellar temperature ranges (see the review in P97).  Especially surprising are the strikingly severe Li depletions (up to at least $\sim$2 dex) observed in mid-F cluster dwarfs (the ``Li dip" or ``Li gap''), which are expected to have \textit{negligible} standard PMS Li depletion.  This Li gap was first discovered in the $\sim$650 Myr-old Hyades (Boesgaard \& Tripicco 1986), followed by its discovery (Hobbs \& Pilachowski 1986) in the older NGC 752 (1.45$\pm$0.1 Gyr; Anthony-Twarog et~al.\ 2009) and in the Hyades-aged Praesepe (Soderblom et~al.\ 1993a).  The absence or near-absence of a Li gap in the $\sim$100 Myr-old solar metallicity Pleiades showed that the Li gap forms during the MS (Boesgaard, Budge, \& Ramsay 1988), perhaps starting as early as an age of $\sim$150 Myr based on Li data in M35 (Steinhauer \& Deliyannis 2004). 

Just cooler than the Li gap and leading into late F/early G dwarfs, the observed cluster A(Li) rise and form a short plateau before declining again in later G and K dwarfs.  The second major departure from standard theory is the magnitude of observed Li depletion in G (and K) dwarfs, which is also closely related to the 50-year old ``Solar Li Problem''.  Standard solar models predict a depletion factor of $\sim$3 (e.g., P97), whereas the Sun's actual depletion is $\sim$150 (e.g., King et~al.\ 1997), assuming the meteoritic A(Li) = 3.31 $\pm$ 0.04\footnote[3]{We adopt the standard notion of A(Li) = 12 + log(N$_{Li}$/N$_H$).} (Anders \& Grevesse 1989) as the initial solar value.  Similarly, G and K dwarfs in virtually all observed clusters older than the Pleiades have depleted far more Li than standard theory predicts.  Once again, the young Pleiades (Soderblom et~al.\ 1993b) helped identify this extra Li depletion (compared to the standard model's factor of $\sim$3) as primarily a MS phenomenon because, a) 1 M$_{\odot}$ dwarfs in the Pleiades have indeed depleted their Li by only a factor of $\sim$3 and b) the mean Li-T$_{\rm eff}$ trend for G and K dwarfs in the young Pleiades shows general agreement with the standard models of P97\footnote[4]{The P97 models employ the same input physics as the standard models with no convective overshoot in Chaboyer et~al.\ (1995a; hereafter C95a), but with updated opacities (from Iglesias \& Rogers 1991 to Rogers \& Iglesias 1992; Pinsonneault 2016, priv. comm.).  We thus refer the reader to C95a for details about the input physics of the P97 models.  Furthermore, the P97 models show a broader metallicity range (-0.20$\leq$[Fe/H]$\leq$+0.15) than the C95a models, which display only [Fe/H]=0.0 and +0.10.}.  The standard model, however, remains incomplete even at the young age of the Pleiades and \textit{cannot} produce the intrinsic scatter of $\sim$1 dex observed around its mean A(Li) trend.  Therefore, there are both PMS (Li scatter) and MS (continued Li depletion) effects that must be explained by non-standard mechanisms.  In principle, the same mechanism could explain both effects, but more than one mechanism might be at work.

Just hotter than the Li gap and leading into early F/late A dwarfs, the observed cluster A(Li) rise to levels near $\sim$3 dex.  Standard theory predicts no Li depletion in these and hotter dwarfs.  Indeed, a number of such dwarfs exhibit A(Li) near 3.0-3.3 dex, but others have suffered moderate to severe Li depletion (Hyades: Burkhart \& Coupry 1989; NGC 3680: Pasquini, Randich, S. \& Pallavicini 2001; Anthony-Twarog et~al.\ 2009; NGC 752: Hobbs \& Pilachowski 1986; Pilachowski \& Hobbs 1988; Sestito et~al.\ 2004; IC 4651: Pasquini et~al.\ 2004; Balachandran et~al.\ 1991).  Also, one star in the Hyades-aged cluster NGC 6633 is super-Li-rich with A(Li)=4.3, suggesting the possible existence of a ``Li Peak'' in stars just hotter than the Li gap that is caused by diffusion (Deliyannis, Steinhauer, \& Jeffries 2002).  Diffusion may also explain the observed low C, high Fe, and even higher Ni.  However, other elemental abundances (e.g., Al, S, and Ca) are better explained by accretion of circumstellar material (Laws \& Gonzalez 2003); perhaps both diffusion and accretion are needed to explain this star (Ashwell et~al.\ 2005).

Using more up to date standard physics, Somers \& Pinsonneault (2014,
hereafter SP14) have conducted a thorough investigation of how
uncertainties in the standard physics affects Li depletion in the standard
model.  This provides a key reference to define these additional Li depletion patterns.  SP14 find that the very high temperature sensitivity of Li depletion leads to important uncertainties in predicted standard depletion, including its metallicity dependence.  The dominating uncertainties result from the adopted equation of state and the adopted solar composition.  To help overcome these, SP14 empirically calibrated their standard model to the Li depletion observed in the Pleiades from 5500 to 6100 K.  The SP14 models are slightly more depleted than the P97 models (neither set have overshoot) and are more comparable to the standard models from C95a that have an overshoot of 0.05 pressure scale heights below the convection zone.  Regardless, while important uncertainties remain in the standard model's input physics, the loose agreement between the Pleiades's mean Li trend and standard theory suggests that, at least for solar-metallicity stars, the standard theory provides a sound foundation for building our understanding PMS Li depletion.  The critical remaining challenges are to understand what non-standard physical mechanisms can explain MS Li depletion in both G and F dwarfs and the significant Li scatter at constant T$_{\rm eff}$ observed in most open clusters.

One of the first non-standard mechanisms proposed to explain the Hyades A(Li) trend (and also the solar Li) involved mass loss occurring during the MS (Hobbs et~al.\ 1989), but Swenson \& Faulkner (1992) argued that this led to various absurdities (e.g., that the full mass range of current Hyades G dwarfs were formed with essentially the same mass but then experienced a factor of $\sim$5 variation in mass-loss rate).  Other approaches to solving the G dwarf and Solar Li Problems involve slow mixing induced by rotation (Endal \& Sofia 1976; 1978; 1981; Pinsonneault et~al.\ 1989) or gravity waves (Garcia Lopez \& Spruit 1991).  The rotational models (collectively referred to as the ``Yale'' rotational models) consider a variety of (but not complete list of) instabilities related to rotation.  Stellar models lose angular momentum from the surface due to interactions of their magnetic field with their stellar wind; these models have been calibrated to match the observed spin-down of stars.  A star's outer layers thus slow down, leading to secular-shear instabilities (Zahn 1987) that are relieved by transfer of angular momentum, the act of which is presumed to cause some local mixing.  This approach solves the Solar Li Problem by definition, as the efficiency of rotationally-induced mixing is calibrated to match the solar Li depletion.  The timescales for models with rotationally-induced mixing find that it will not yet play a significant role in young clusters like the Pleiades, but in intermediate-aged clusters (e.g., the Hyades) and older it becomes very important (see C95a).  Another effect of rotation is that faster-rotating stars lose more angular momentum and they lose it more efficiently, which should accordingly induce greater internal mixing and Li depletion.  Therefore, in intermediate-aged and older clusters this can also create scatter in A(Li) at a given T$_{\rm eff}$.

This class of models has enjoyed a variety of successes (see especially the Li-gap discussion, below).  A particularly encouraging success is the ability of these models to lose the vast majority of their initial angular momentum (to match rotation rates of young stars as compared to old halo dwarfs at the turnoff) and thus rotate slowly at the turnoff, but yet retain enough interior angular momentum to explain the rapidly rotating blue horizontal branch stars (Pinsonneault et~al.\ 1991).  Evidence that this type of mixing actually depletes Li in G dwarfs comes from Short Period Tidally Locked Binaries (SPTLBs).  According to tidal circularization theory (Zahn \& Bouchet 1989), stars observed today with sufficiently short periods would have become tidally locked during the early PMS, before their interiors were hot enough to destroy Li; such stars should thus have higher A(Li) than normal single stars.  Such high-Li stars have indeed been observed in open clusters (Hyades: Soderblom et~al.\ 1990; Thorburn et~al.\ 1993; M67: Deliyannis et~al.\ 1994) and other contexts (Ryan \& Deliyannis 1995).  However, these models (and their improved successors, that also include the interaction of rotationally-induced mixing and microscopic diffusion, Chaboyer et~al.\ 1995a, 1995b) might not deplete enough G-dwarf Li at later ages, and they may produce too much scatter in older clusters and little to none of the observed scatter in the young Pleiades.  (The Pleiades scatter is notable because the faster rotators are less depleted rather than more, the reverse of that predicted by rotationally-induced mixing if it began to play a role at such a young age.)  It is also possible that wave-induced mixing plays an important role, particularly at later stages, and that all three mechanisms must be considered together (Talon \& Charbonnel 2005).  Another idea is that a different prescription for the internal angular momentum transport may be more realistic; the G dwarf models of Somers \& Pinsonneault (2016) may provide both more realistic internal rotation curves and surface Li abundances.

Looking directly at the problem of large A(Li) scatter observed in young cluster G dwarfs, most notably in the Pleiades, Somers \& Pinsonneault (2015) provide new rotation models that consider the effects of rotation and inflated stellar radius on PMS Li depletion.  Somers \& Pinsonneault (2015) proposed and Somers \& Stassun (2016) observed a correlation between rotation and radius inflation, where faster rotation leads to greater magnetically-driven radius inflation.  This can recreate the observed Pleiades scatter where at a given T$_{\rm eff}$ faster rotators are the least depleted in Li and the slowest rotators are are $\sim$1 dex more depleted.  These slow rotators fall on the lower envelope of the Pleiades A(Li) scatter, in agreement with the standard model proposed in SP14 (i.e., zero rotation).  Applying this to older clusters, this may effectively counterbalance the subsequent effects of rotationally-induced mixing and angular momentum loss during the MS (as discussed above) and explain why Hyades-aged G dwarfs with likely a distribution of different rotational histories exhibit relatively little scatter in A(Li) compared to \textit{both} younger and older clusters.

Discovery of the Li gap in F dwarfs blatantly contradicted standard theory and resulted in a proliferation of candidate non-standard mechanisms that could create such a Li gap.  These fall into three categories: mass loss (Schramm et~al.\ 1990), microscopic diffusion (Michaud 1986; Richer \& Michaud 1993), and slow mixing induced by various rotation-related mechanisms.  Boesgaard's (1987) discovery of a Li-\textit{v sin i} anticorrelation in cooler Hyades gap stars provided a connection to rotation.  The Yale rotational models also predict the Li gap in terms of increasing initial angular momentum for more massive stars, followed by decreasing rates of angular momentum loss due to the vanishingly small SCZ.  Beryllium and boron, which survive to increasingly deeper depths than does Li, when considered with Li provide critically important diagnostics that identify the responsible mechanisms and potentially constrain the relative contribution of each.  A considerably varied body of evidence favors the Yale rotational models as the dominant Li gap-causing mechanism, and argues against mass loss and diffusion (though diffusion can still play a role, particularly at later stages).  This evidence includes a) the early formation of the Li gap (Steinhauer \& Deliyannis 2004), b) the prototypical Li/Be pattern in 110 Her (Deliyannis \& Pinsonneault 1997) and the strikingly more general Li/Be depletion correlation in F dwarfs (Deliyannis et~al.\ 1998; Boesgaard et~al.\ 2001; Boesgaard et~al.\ 2004), c) the Be/B depletion correlation in F dwarfs (Boesgaard et~al.\ 2005), and d) subgiants in the solar-aged M67 evolving out of the Li gap, whose deepening SCZ reveal the profile of the Li preservation region, and thus the mechanism(s) responsible for its creation.  (For a more detailed discussion, see Section 5.1 of Anthony-Twarog et~al.\ 2009.)

Related to the issues discussed here is that stars in between the Li gap and the (mid-) G (and K) dwarf Li depletions, exhibit the ``Open Cluster Li Plateau''.  Evidence that this plateau is depleted by at least a factor of 2 to 3 from some higher initial value comes from SPTLBs in the Hyades and the much older (solar-aged) M67 (Thorburn et~al.\ 1993; Deliyannis et~al.\ 1994).  Studies of the open cluster Li plateau could conceivably provide insight into the halo Li plateau, interpretation of which remains critically important to the testing of Big Bang theory.  If Big Bang Li is to be consistent with Planck data (Coc et~al.\ 2014), the halo Li plateau must be depleted by a factor of $\sim$3.  Conversely, if the halo Li plateau were shown \textit{not} to be depleted, there could be serious problems with the standard Big Bang model.

A critical component to making further progress is to separate out effects predicted successfully by standard theory from those that need to be explained by non-standard mechanisms.  As discussed, we will adopt that the standard theory can explain the general Li-T$_{\rm eff}$ relation of the Pleiades, but it clearly fails to account for the scatter in Li around this relation and the subsequent Li depletion in F, G, and K dwarfs observed in older clusters.  But what about the other major prediction from standard theory, that Li depletion depends on metallicity?  The P97 and SP14 models predict that the G- and K-dwarf Li-T$_{\rm eff}$ trends depend quite sensitively on metallicity, even if differences in [Fe/H] are as small as 0.1-0.15 dex, which is comparable to the metallicity difference between the Hyades and the Pleiades.  Some evidence that metallicity plays an important role comes from field dwarfs (Ram{\'{\i}}rez et~al.\ 2012), but these standard predictions have never been tested rigorously using open clusters (see in SP14 a discussion of the limited and contradictory tests that have been done).  Unless we can establish how well standard theory predicts the metallicity dependence of Li depletion, we will be hard pressed to quantify the degree of Li depletion required of non-standard mechanisms, and it will be even more difficult to quantify any metallicity dependence that these mechanisms themselves might have.

We thus embark on a program to determine empirically the metallicity dependence of Li depletion in open cluster F, G and K dwarfs, which in turn can be used to test standard (and non-standard) models.  Since age is known to be an important Li-depletion parameter, we have chosen to study the [Fe/H] and A(Li) of five clusters with the \textit{same} age, in particular, the age of the Hyades ($\sim$650 Myr).  To limit systematic effects we observe all of the clusters self-consistently using WIYN/Hydra.\footnote[5]{The WIYN Observatory is a joint facility of the University of Wisconsin Madison, Indiana University, the National Optical Astronomy Observatory and the University of Missouri.}  We begin our program here by comparing the Hyades to Praesepe.  The questions to be addressed include the following.  Is metallicity indeed an important parameter for Li depletion in open clusters?  If yes, is there evidence for parameters other than age and metallicity affecting Li depletion?  How well do the P97/SP14 models match any metallicity dependence we might find in the Li depletion of G and K dwarfs?  In the F dwarfs is the Li-T$_{\rm eff}$ morphology of the Hyades Li gap typical?  Or equivalently -- does the Li-T$_{\rm eff}$ morphology of the Li gap vary from Hyades-aged cluster to Hyades-aged cluster, and if so, what parameters might be relevant?  Do other Hyades-aged clusters also show a Li-\textit{v sin i} anticorrelation, and if yes, does it have a similar slope and zero point?


In this first paper of our Hyades-aged clusters program, we discuss our observations of the Hyades and Praesepe and our data-reduction methods in Section 2, our analysis of radial velocities, \textit{v sin i}, and membership in Section 3, and our atmospheric models and determination of stellar parameters in Section 4.  We discuss our Fe abundances in Section 5, our A(Li) for the Hyades and Praesepe individually in Section 6 where we also compare to and combine with previous studies of these clusters, and we directly compare our Li results from both of these moderately metal-rich clusters in Section 7.  Lastly, results and conclusions are summarized in Section 8. 

\section{Data Observations and Reductions}

The Hyades and Praesepe open star clusters were both observed using the Hydra multi-object spectrograph on the WIYN 3.5-meter telescope using the 316$@$63.4 echelle grating in order 8 with the X19 filter.  The spectra span from 6450 to 6850 \AA.  All Hyades stars and a majority of Praesepe stars were observed with blue cable, which yielded R$\sim$13,600.  The remaining Praesepe stars were observed with the red cable, which yielded a moderately-higher R$\sim$17,600.  These resolutions are based on our own arc-line FWHM measurements.  The Hyades observations were all performed using the STA1 detector, but we binned the spectra by two along the dispersion axis giving a dispersion of $\sim$0.20 \AA\,pixel$^{-1}$.  The Praesepe observations were all performed using the T2KA detector, and without any binning this also gave a dispersion of $\sim$0.20 \AA\,pixel$^{-1}$.  

Most Hyades targets (34) were selected from the final list of members from Perryman et~al.\ (1998, hereafter P98).  P98 began with a preliminary list of members based on a number of proper motion (PM) studies, a large compilation of radial velocities, and their own Hipparcos parallax data.  Detailed consideration of the space motions resulted in P98's final membership list.  We also chose 3 additional targets that do not appear in P98, namely vB 49, vB 59, and vB 93, which are all radial velocity cluster members from Griffin et~al.\ (1988).  The Hyades sample thus comprises a total of 37 highly probable members, all of which also fall on the cluster's narrow MS.  Based on Hipparcos data and a variety of previous studies, P98 list evidence of multiplicity in their Table 2.  31 of our 37 stars are likely single stars as listed in P98, while the remaining six are confirmed binaries.  The Hyades is very nearby (45 pc) with members loosely spread across a large area of the sky; therefore, these 37 Hyades stars had to be observed individually on a single Hydra fiber placed in the center of the field.   The Hyades data were acquired over seven nights from February 2 to 23, 2009.  The total exposures for each target are given in column 11 of Table 6.

For Praesepe, the coordinates and PM analysis of Wang et~al.\ (1995) were used to calculate the current epoch coordinates. PM membership probabilities and consistency with the photometric single-star main sequence were used to select likely single-star dwarf members.  All stars are highly probable PM members of Praesepe: 69 of them have P$_{mu}$=1.00, five have P$_{mu}$=0.99, two have P$_{mu}$=0.98; the remaining two have 0.89 and 0.82.  Praesepe is a moderately nearby cluster (187 pc) that spans a significantly larger area than the degree field of view of Hydra, so multiple Hydra positions and fiber configurations were necessary to broadly analyze the cluster members.  Furthermore, to minimize potentially adverse effects of scattered light on the fainter stars, we limited each Hydra configuration to a range of less than three to four magnitudes.  This required multiple configurations to cover our full target range in V magnitude from 8 to 13.4.  Using two red-cable configurations, we obtained spectra of 34 candidate cluster members on November 16 and 18, 1997.
Using seven blue cable configurations, we obtained spectra of 66 candidate cluster members during seven nights on December 2, 2001; May 1 and 2, 2005; January 25 and 26, 2006; and February 2 and 3, 2006.  Most stars were observed in more than one configuration, giving a wide range of exposure times across our sample.  See column 11 in Table 7 for the total exposure times for each star.  

The difference in characteristics between red cable and blue cable are moderate, with small variations in throughput and the moderate difference in resolutions given above.  For all observations, the comparison lamp spectra were taken in the same configuration as the object data, at least one (if not all) dome flats were taken in configuration, and daytime sky spectra were taken in-configuration when possible.  In those few instances when it was not possible to take calibrations in-configuration, the circle configuration was used.  Taking calibrations in configuration is ideal because it provides consistent fiber throughputs, which can change due to altering a fiber's position or magnitude of curvature.  This is most important when calibrating the throughput of fibers used to determine the sky subtraction; however, all our targets are very bright in this study and the subtracted sky background was very minor.

Several Praesepe stars were observed in both red and blue cable on Hydra.  Instead of co-adding the spectra with slightly differing characteristics, we independently determined the abundances from both observations.  The final stellar abundances are based on a weighted linear-average of the two independent sets of line measurements.  The weighting is based on the relation from Cayrel (1988), as recast in Deliyannis et~al.\ (1993, hereafter the D93 1$\sigma$ relation), which determines an equivalent-width error based on the spectra's signal-to-noise per pixel (hereafter S/N), dispersion, and line broadening.  

Here the steps to create our final spectra are briefly covered, but a more detailed overview is given in Anthony-Twarog et~al.\ (2010) and Cummings et~al.\ (2012).  We processed our images using the standard IRAF\footnote[6]{IRAF is distributed by the National Optical Astronomy Observatory, which is operated by the Association of Universities for Research in Astronomy, Inc., under cooperative agreement with the National Science Foundation.} steps, followed by tracing the apertures, removing cosmic rays with L.A. Cosmic (van Dokkum 2001), and extracting the spectra.  For the nine Praesepe configurations, comparison spectra were taken using the ThAr lamps in eight configurations and the CuAr lamps in one configuration, both of which provide a rich series of lines in the Li spectral region for wavelength calibrating our spectra.  Unlike the data discussed in Cummings et~al.\ (2012), which were observed with the CTIO 4-m using Hydra II, WIYN/Hydra observations do not use etalons and only use comparison lamp spectra.  Praesepe is bright and required only relatively brief (one to two hours) observations each night, but to take into account possible shifts with time, a comparison spectrum was observed in configuration immediately before and after the cluster observations.  Before and after each set of Hyades-dwarf observations, the ThAr lamps were similarly observed with the central fiber.

The Hyades stars are very bright (5.7$\leq$V$\leq$9.4) and the observations were typically only several minutes, so the sky was not subtracted for the Hyades.  Praesepe stars are not quite as bright (8$\leq$V$\leq$13.4) so sky subtraction was performed.  In each Praesepe configuration, due to the low sky density of stars, there were enough available fibers to place 30 to 40 sky fibers at random positions across the field.  To create a uniform sky level in all fibers before it is subtracted from the objects, fiber throughput corrections were applied, which were created using the high-S/N afternoon-sky spectra.  We then applied a Doppler correction to each object image based on the Earth's orbital velocity, which provided a uniform radial-velocity zero point for all stellar spectra and allowed for precise radial-velocity measurements.  Lastly, the individual object spectra were all co-added and continuum-fitted to produce our final spectra.

\section{Radial Velocities, Binarity, and Membership}

\begin{figure*}[htp]
\begin{center}
\subfigure{\includegraphics[scale=0.45]{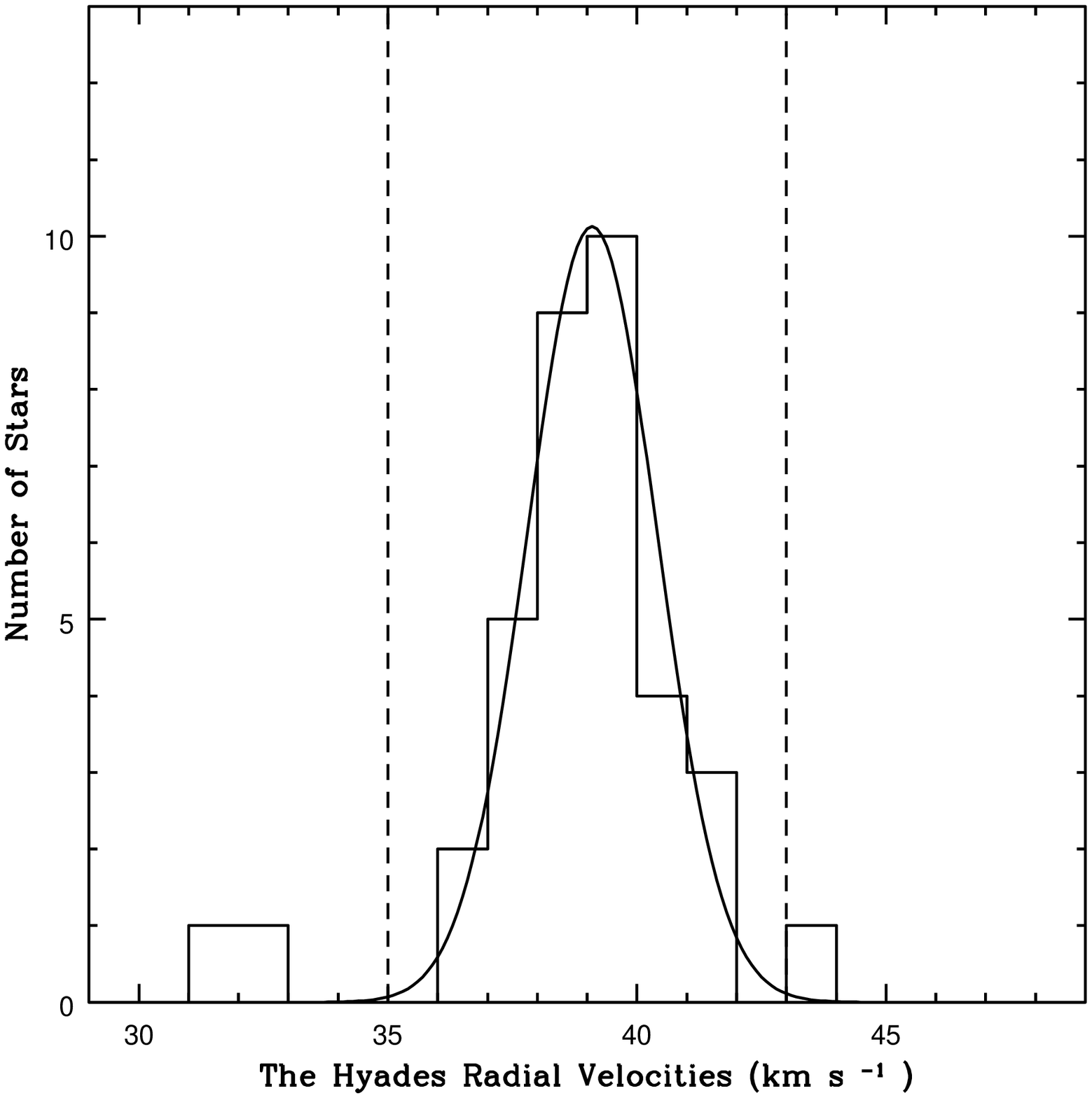}}
\subfigure{\includegraphics[scale=0.45]{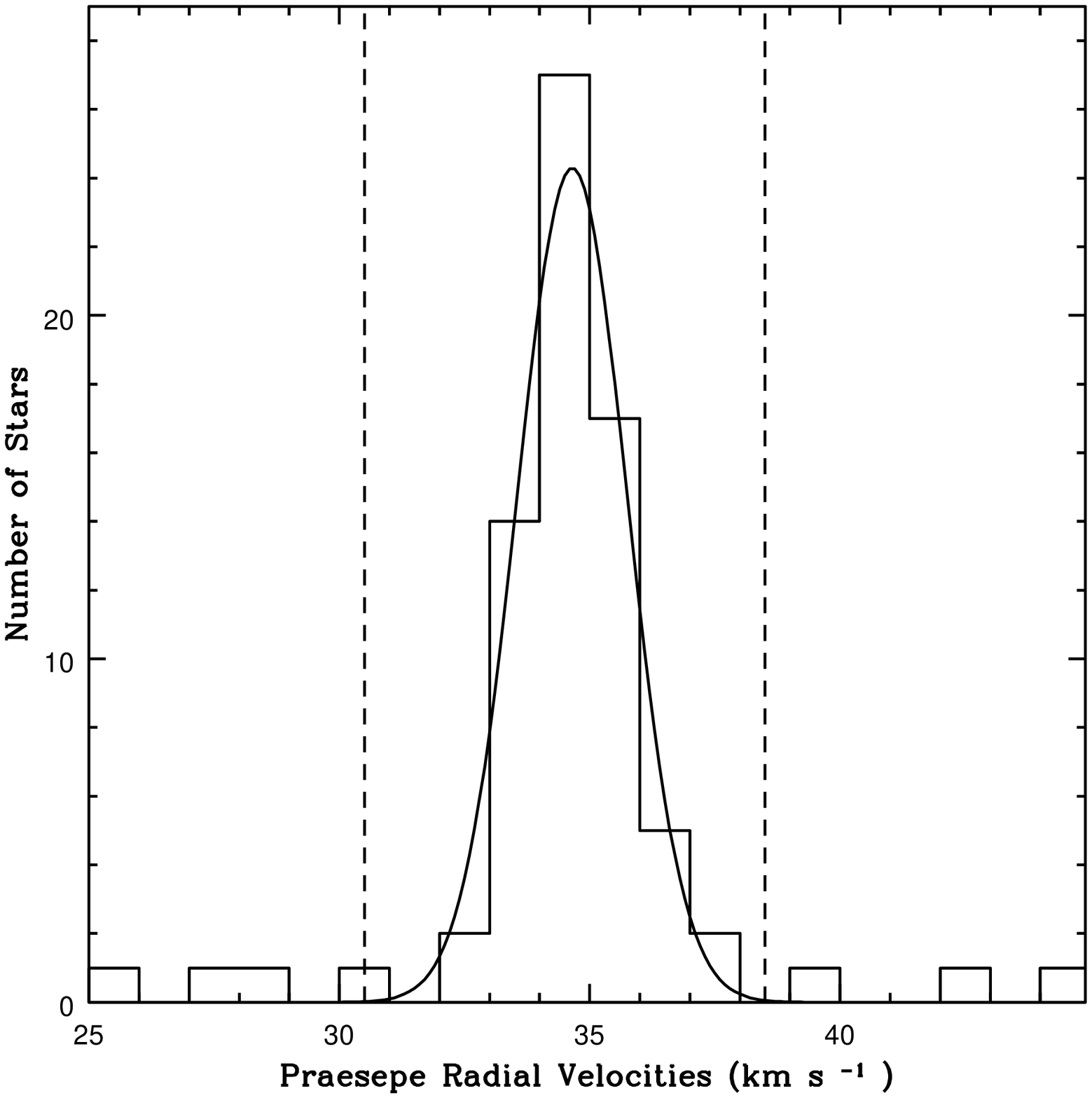}}
\end{center}
\vspace{-0.45cm}
\caption{The histograms of radial velocities for the Hyades and Praesepe, with fit Gaussian distributions.  For each cluster, the dashed lines define the range of stars consistent with single-star membership.}
\end{figure*}

In this study our targets are all highly probable cluster members, based on a variety of criteria for the Hyades, and based on PMs and location on the color magnitude diagram single-star fiducial for Praesepe (Section 2).  However, PM-based membership information is often not available for clusters that we will be reporting on in the future, including some Hyades-aged clusters intended for the present series of studies.  In such cases, stellar targets can often be selected from the photometric single-star fiducial sequence, and membership can be refined to those stars that also have radial velocities consistent with single-star cluster membership.  For the Hyades and Praesepe, the richness of material available affords us to include highly probable members (as judged from PMs, photometry, and overall kinematics) that do not fall inside the single-star radial velocity peak.  This allows us to check to see whether the A(Li) patterns of stars belonging strictly to those within the single-star radial velocity peak differ from those patterns derived when member stars from outside the single-star radial velocity peak are also included.  Such stars are often binary (or multiple) members, and the question then becomes, when can the flux contribution from the secondary be ignored, and when not?  

Not surprisingly, previous studies (e.g., Thorburn et~al.\ 1993) have found that SB2s show a larger scatter in the Hyades Li-T$_{\rm eff}$ relation, even after attempts have been made to correct for the flux contribution of the secondary (see also Boesgaard \& Tripicco 1986).  In part, this is because such flux corrections can be uncertain, but the temperature of the primary can also be uncertain (see Sections 4, 6.1).  This, in turn, creates a corresponding uncertainty in the derived A(Li), even if the flux correction factor were to be very accurate.  Therefore, we will not include SB2s when determining the Li-T$_{\rm eff}$ relation of any given cluster, though we will discuss certain SB2s such as short-period tidally locked binaries (SPTLBs) that have special importance.  The question then becomes, are the A(Li) from probable member SB1s sufficiently reliable so that we can include them in determining the Li-T$_{\rm eff}$ relation?  We will examine these issues in the sections that follow.  

The stellar radial velocities have been measured with the IRAF task \textit{fxcor}, which compares each spectrum to a radial velocity standard star's spectrum of a similar spectral type also observed with WIYN/Hydra.  Stars observed with blue cable were only compared to radial velocity standards observed with the blue cable, and similarly objects observed with the red cable were compared to red-cable standards.  While very high precision radial velocities are not needed for this work, minor ($<$0.5 km s$^{-1}$) systematics can be introduced from applying a standard observed in a central fibre to all fibres.  For narrow lined spectra (\textit{v sin i} $<$ 30 km s$^{-1}$), we used the spectral comparison range of 6600 to 6800 \AA, which is broad but avoids H$\alpha$ ($\sim$6563 \AA) and the complications that its inclusion would introduce.  In increasingly broadened spectra (\textit{v sin i} $>$ 30 km s$^{-1}$), the moderate-to-weak strength lines in this chosen spectral range become increasingly challenging for \textit{fxcor} to match.  Therefore, in fast rotators a narrow region (10 to 15 \AA) centered on the strong H$\alpha$ line was used as the comparison wavelength to determine radial velocities.

The \textit{fxcor} comparison of the two spectra creates an approximately Gaussian cross-dispersion profile, which provides both the Doppler shift of the star and for slower rotators (\textit{v sin i} $<$ 30 km s$^{-1}$) the width of the profile gives a direct measurement of the rotational broadening (\textit{v sin i}).  The precision with which \textit{fxcor} can fit the center of the profile provides our 1$\sigma$ radial velocity errors.  Furthermore, a significantly asymmetric or double peaked cross-dispersion profile is strong evidence for binarity, where the Gaussian profile for each star of the system is offset and superimposed.  None of our Hyades dwarfs and only two Praesepe dwarfs (KW 181 and KW 367\footnote[7]{Our Praesepe IDs are from Klein Wassink (1927) when available.  Otherwise, they are followed by a W and are taken from Wang \& Jiang (1991).}, clear SB2s) show clear evidence for binarity based on this criterion.  For faster rotators (\textit{v sin i} $>$ 30 km s$^{-1}$), when only using H$\alpha$ for the \textit{fxcor} reference, we similarly derive reliable radial velocities, but we cannot determine \textit{v sin i} because the very strong H$\alpha$ is also heavily broadened through other mechanisms.  Determining \textit{v sin i} for fast rotators is very desirable because of the potential effects of varying rotation rates on A(Li).  Therefore, we compared the observed spectral features near Li to rotationally broadened synthetic lines that were convolved with our spectral resolution (see Section 6).  Our \textit{v sin i}, radial velocities, and 1$\sigma$ errors in the radial velocities are shown in Columns 12-13 of Tables 6 and 7 for the Hyades and Praesepe, respectively.

Single-star cluster members have similar radial velocities with only moderate dispersion, so we use the radial velocities of our PM and photometrically selected dwarfs to define a set of stars consistent with single-star membership.  We will refer to these as ``radial-velocity members''.  The left panel of Figure 1 shows that the Hyades radial velocities produce a narrow Gaussian distribution.  We note that Figure 1 includes known SB1s.  This may increase the width slightly of the observed distribution, but this is appropriate for illustrating our methods, which in future publications will be applied to other clusters without previous published binary analyses.  Stars inside the vertical dashed lines are defined to be consistent with single-star membership; this includes 33 of our 37 program stars.  These 33 stars give a mean cluster radial velocity of 39.1 with a standard deviation of the mean (hereafter $\sigma_\mu$) of $\pm$0.2 km s$^{-1}$ and a standard deviation ($\sigma$) of 1.3 km s$^{-1}$, which is remarkably consistent with this sample's mean cluster radial velocity from P98 of 39.1$\pm$0.3 km s$^{-1}$ ($\sigma_\mu$), they also have a consistent radial velocity distribution width with ours.  The four Hyads with inconsistent radial velocities in comparison to the primary group are vB 2, vB 4, vB 38, and vB 127.  Since all four of these stars are well-established PM members, and since Hyads have relatively large PMs that easily separate them from the field, these discrepant velocities may be explained by binarity.  For example, there is previous evidence that vB 2 and vB 38 are binaries (see P98 and references therein).  However, these four stars have no direct evidence of binarity from \textit{fxcor} and are all photometrically consistent with the single-star sequence.  Therefore, any flux contributed to the spectrum by a possible companion is likely small and will not cause significant errors in the abundances that we derive for the primary. In the cases of vB 2, vB 4, and vB 127 the velocity discrepancy is less than 8 km s$^{-1}$ from the cluster mean.  But in the case of vB 38, its discrepancy is 25 km s$^{-1}$.  vB 38 is further singled out by being the only one of these four stars that has significant T$_{\rm eff}$ dispersion (see Section 4), which suggests the presence of a faint companion that we cannot ignore or some other issue with the atmosphere that likely makes its derived abundance unreliable.  Therefore, we will continue to include vB 2, vB 4, and vB 127 in our final sample, but their slight radial velocity inconsistencies should be noted and in Section 6.1.1 we will test their and vB 38's A(Li) consistency with the Hyades radial velocity members.  

The right panel of Figure 1 shows that the Praesepe radial velocities also produce a narrow Gaussian distribution.  In our Praesepe sample of 78 stars, 67 are radial-velocity members (defined as those stars within the dashed lines in Figure 1).  These 67 stars give a mean cluster radial velocity of 34.7$\pm$0.2 km s$^{-1}$ ($\sigma_\mu$) with a $\sigma$ of 1.1 km s$^{-1}$, which is remarkably consistent with the mean cluster radial velocity determined by Mermilliod \& Mayor (1999) of 34.54$\pm$0.12 km s$^{-1}$ ($\sigma_\mu$).  The 11 stars lying outside the dashed lines are KW 38, KW 45, KW 124, KW 154, KW 181, KW 183, KW 268, KW 367, KW 375, KW 416, and KW 434.  However, due to the relatively large PM of Praesepe, it is likely that many of these stars are Praesepe member binaries.  Indeed, KW 181 and KW 367 have clear double-lined spectra.  KW 416 and KW 434 have separate observations in the red and blue cable, with KW 416 showing a 6 km s$^{-1}$ difference in each epoch's radial velocity and KW 434 showing a 55 km s$^{-1}$ difference.  Another possible signature of binarity is a high T$_{\rm eff}$ dispersion (see Section 4), and nine of these radial-velocity discrepant stars have high dispersion.  Based on the binarity analysis of Mermilliod et~al.\ (2009) and Patience et~al.\ (2002), roughly half of these radial velocity discrepant stars are known binaries (KW 181, KW 268, KW 367, KW 416, and KW 434).  Only KW 45 and KW 183 do not show any signatures of binarity in our observations or elsewhere beyond their disagreement with the cluster radial velocity.  Consistent with our procedures for the Hyades, we present Li results for all radial velocity discrepant stars that are not double-lined spectra, and in Section 6.2.1 we test their A(Li) consistency with the Praesepe radial velocity members.  Lastly, we include KW 45 and KW 183 in our final sample because of their small T$_{\rm eff}$ dispersion.

\section{Stellar Parameters and Atmospheres}

\subsection{Photometry}

Near the Li I resonance line (6708 \AA) there are no measurable Fe II lines or Fe I lines with a large enough variation in excitation potential to determine spectroscopically either T$_{\rm eff}$ or log g.  Fortunately, both Praesepe and the Hyades are very well studied photometrically and have near-zero reddening (see discussion below).  Therefore, fairly precise, accurate, and self-consistent temperatures can be derived photometrically with a color-temperature relation.  For our Praesepe photometry we used the UBV photometry of Johnson (1952, hereafter J52) and the UBVRI photometry of Mendoza (1967, hereafter Mend67).  To test for consistency, we compared 31 stars that span a broad range of V magnitude (6.78 to 11.99) and B-V color (0.22 to 0.82) from both J52 and Mend67.  Figure 2 shows this comparison and that the differences in B-V have no significant systematic offset or slope with magnitude.  We combine these two independent photometric studies (J52 and Mend67) confident that we are not introducing any significant systematic effects.  The eight faintest Praesepe stars observed in our sample are beyond the J52 or Mend67 magnitude limits.  Therefore, for seven of these fainter stars we used the BV photometry of Upgren et~al.\ (1979, hereafter U79) and for the case of 624W we used the BV photometry of Weis (1981).  U79 compared their data to those of J52 for the 36 stars common to both samples, and they found negligible offsets of -0.002$\pm$0.004 in V and +0.002$\pm$0.004 in B-V.  Lastly, for our adopted Hyades photometry we similarly combined the photometry of Johnson \& Knuckles (1955) and Mend67.

\begin{figure}[htp]
\begin{center}
\includegraphics[scale=0.44]{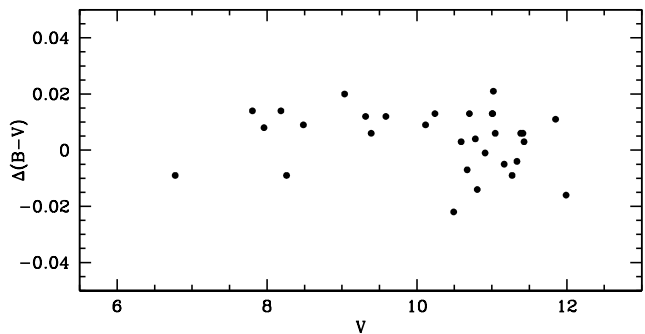}
\end{center}
\vspace{-0.45cm}
\caption{Comparison of the B-V photometry versus the V magnitude for the 31 Praesepe members that have photometric observations in both J52 and Mend67.  There is no significant systematic offset or trend with magnitude for the difference between the two studies.  Therefore, there are no systematic corrections that need to be considered when combining the Praesepe photometry of these two studies.}
\end{figure}

\begin{figure*}[htp]
\begin{center}
\includegraphics[scale=0.95]{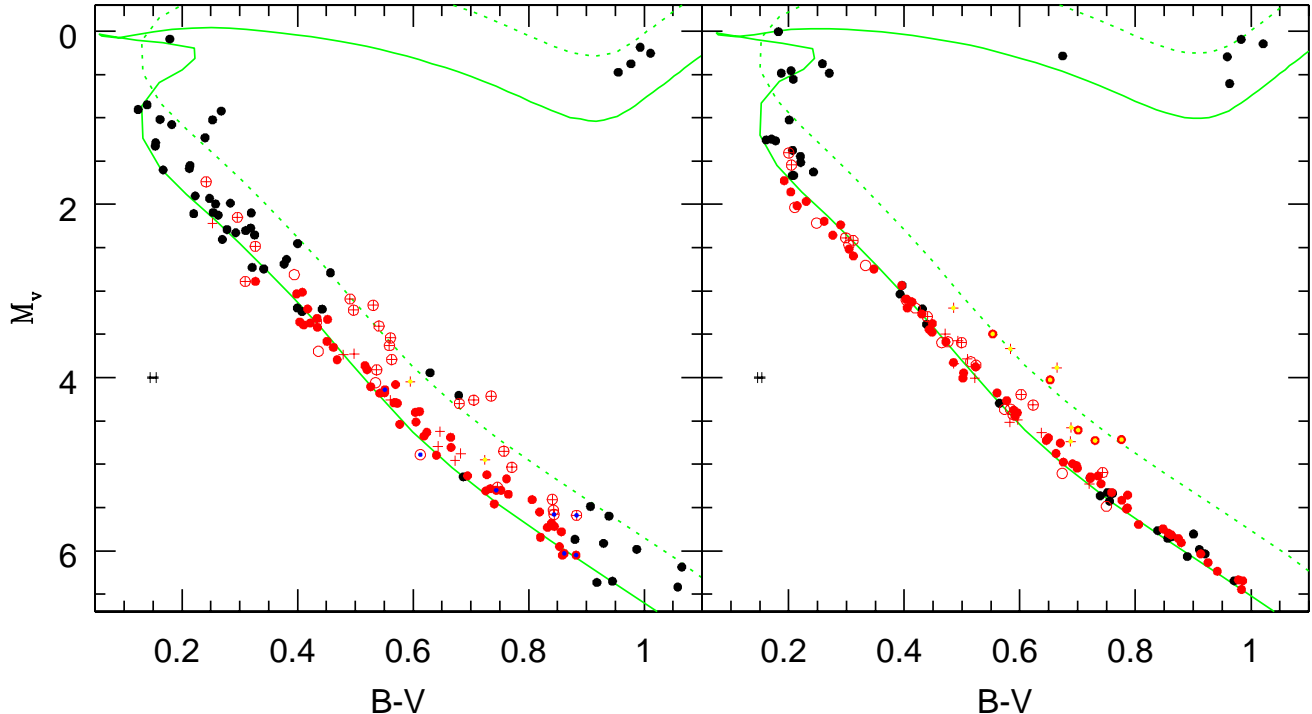}
\end{center}
\vspace{-0.45cm}
\caption{The left panel shows the color magnitude diagram of the Hyades.  Data in red have had their Li analyzed in this paper.  These stars are further differentiated as single stars with low $\sigma_{\rm Teff}$ (solid red) and high $\sigma_{\rm Teff}$ (open red) and binaries with low $\sigma_{\rm Teff}$ (red plus) and high $\sigma_{\rm Teff}$ (circled red plus).  Stars that do not have individual Hipparcos distances are further overlaid with blue data points.  Stars with low $\sigma_{\rm Teff}$ that photometrically deviate from the single star fiducial are further overlaid with yellow data points.  In solid green a Yale-Yonsei isochrone (Yi et~al.\ 2001, hereafter Y$^2$ isochrone) of 635 Myr is fit to the data by eye with a corresponding dashed-green isochrone representing equal-mass binaries.  The right panel shows the color magnitude diagram of Praesepe using the same format, and in green a Y$^2$ isochrone of 670 Myr is fit to the data by eye with a corresponding dashed-green isochrone representing equal-mass binaries.  Lastly, in both diagrams we plot on the left representative 1$\sigma$ error bars based on the published magnitude and color errors in Johnson \& Knuckles (1995) and J52, respectively.}
\end{figure*}

To determine our photometrically based T$_{\rm eff}$, we took advantage of the fact that besides the eight faintest Praesepe stars, all program stars have measurements in all five UBVRI filters.  We used all 10 color combinations of these filters, rather than the B-V color alone.  This should result in a more precise T$_{\rm eff}$ and perhaps smaller errors as well.  To perform this multicolor analysis, for Praesepe we began by cutting the full published photometric sample to a narrow MS fiducial of stars that have up to 10 UBVRI-based colors.  Each of the nine non-(B-V) colors were plotted against B-V, and a low-order fit (typically 3rd order) provided an \textit{empirical} relation to convert each color to an effective B-V.  This gives up to 10 effective B-V colors for each star, which have been averaged to give a final mean B-V for each Praesepe dwarf.  Maderak et~al.\ (2013) applied these same methods to the Hyades photometry, and we adopt their mean Hyades B-V colors.

This method provides several advantages.  A T$_{\rm eff}$ derived from a single color may be affected by high surface activity or a binary companion of different color that contributes a significant fraction to the total combined flux.  A cluster member's color can also be affected by a reddening deviant from the cluster mean or a poor measurement or large error in at least one filter magnitude.  Non-uniform reddening or photometric error are not a concern in either the Hyades or Praesepe, but can be in many other clusters that we will analyze in the future.  In comparison, multiple color analysis will also be affected by these concerns but non-uniformly, which can be used to identify stars with peculiar colors.  Therefore, our method identifies complications that even the most precise \textit{single} color photometry could not detect.

There are also several advantages to our method over direct application of these multiple colors to their own independent color-temperature relation from, for example, Ram{\'i}rez \& Mel{\'e}ndez et~al.\ (2005; which presents relations for only 4 of our 10 total colors).  First, with most clusters a reddening is critical for deriving photometrically based T$_{\rm eff}$, and precise transformations between different reddening factors for each color would be necessary.  With our method the \textit{empirical} comparisons of the observed colors directly fits the relative effects of reddening between each color.  Second, because we work directly with the standardized colors to derive the color transformations independently for each cluster, this corrects for any potential systematic issues introduced in the analysis or observations between the multiple observed colors.  Because with our method we put everything into the B-V system, any systematics in the observed B-V color can cause systematic offsets in our final T$_{\rm eff}$, but photometric systematics in any color will not artificially increase the resulting dispersion in the 10 derived effective B-V values.  Lastly, similar systematics within the photometry used to derive these multiple color-temperature relations will not affect our results because we only need to adopt one single color-temperature relation. 

To test the reliability of these final combined B-V, we calculated the standard deviation in T$_{\rm eff}$ based on each of the 10 B-V colors (see color-temperature relation discussion below).  The $\sigma_{\rm Teff}$ of our spectroscopically observed Hyades and Praesepe sample ranges from only 6 K to nearly 400 K, with a typical $\sigma_{\rm Teff}$ of $\sim$40 K.  The derived A(Li) depend strongly on the assigned T$_{\rm eff}$ (roughly 0.1 dex per 100 K, depending on T$_{\rm eff}$).  The very high precision of the Hyades and Praesepe photometry indicates that these large $\sigma_{\rm Teff}$ are not simply the result of photometric errors and are truly indicative of color peculiarities across some part of the full UBVRI color spectrum.  Thus, stars with a standard deviation in T$_{\rm eff}$ (hereafter $\sigma_{\rm Teff}$) greater than 75 K were considered to be problematic and removed from our final sample.  In Praesepe, for example, 9 of the 11 stars not consistent with the cluster radial velocity have high $\sigma_{\rm Teff}$, while conversely only 7 out of 59 of the stars that are both radial velocity and PM members with full photometric analysis have high $\sigma_{\rm Teff}$.  We note that for future analyses of any clusters with moderately variable reddening or higher photometric error our high $\sigma_{\rm Teff}$ definition of $>$75 K will be appropriately adjusted.

The A(Li) for these high $\sigma_{\rm Teff}$ stars in both the Hyades and Praesepe are still of interest, nonetheless, and they remain in Tables 6 and 7.  Similar to the stars with discrepant radial velocities, in Section 6 we will test if these high $\sigma_{\rm Teff}$ stars have distinct A(Li).  For the eight faintest Praesepe stars we only adopted their B-V color from U79 and Weis (1981) because the R and I photometry from both of these studies used different filter sets than Mend67, and our color transformations are not applicable.  Therefore, these eight stars similarly have larger uncertainty in their parameters, but they are so heavily depleted in Li that only upper limits are presented in Table 7. 

\subsection{Color Magnitude Diagrams}

In Figure 3 we plot the color magnitude diagrams (CMDs) of the Hyades (left panel) and Praesepe (right panel) to further analyze their photometric characteristics and to self-consistently fit isochronal ages.  In the Hyades, we plot the final averaged B-V colors versus absolute magnitude (M$_V$) based on their individual Hipparcos distances (van Leeuwen 2007).  This is because variations in distance for individual Hyades members are a large enough fraction of the Hyades distance itself.  Therefore, apparent magnitude does not as accurately illustrate the relative luminosities.  (Seven Hyades stars are marked in Figure 3 with overlain blue data points; these have no published Hipparcos distances and we applied a uniform distance of 46.7 pc from van Leeuwen 2009).  For the more distant Praesepe, accounting for the detailed variations in each star's distance is not required, but we apply a uniform distance of 181.5 $\pm$ 6.0 pc (van Leeuwen 2009) to place it on the same scale as the Hyades. 

In both figures we have marked in red the members that have had their Li analyzed in our WIYN/Hydra data or in our supplemental data.  We further differentiate these members as single stars with low $\sigma_{\rm Teff}$ (solid red) and high $\sigma_{\rm Teff}$ (open red) and binaries with low $\sigma_{\rm Teff}$ (red plus) and high $\sigma_{\rm Teff}$ (circled red plus).  As expected, we see that in the Hyades the single stars with low $\sigma_{\rm Teff}$ create a reasonably tight MS fiducial.  Similarly, the binaries with low $\sigma_{\rm Teff}$ are consistent with this fiducial.  This supports the idea that their secondary companions are not contributing significant light.  In contrast to this, we see that the single stars with high $\sigma_{\rm Teff}$ and most clearly the binaries with high $\sigma_{\rm Teff}$ deviate significantly from the single star fiducial.  The stars overlain with blue data points, which have had a uniform distance applied, appear photometrically consistent with their stellar types.  We note that the Hyades main sequence appears to be somewhat broad, and for our analyzed single and binary stars with low $\sigma_{\rm Teff}$ we find that they have a median absolute deviation from the isochrone of 0.135 magnitudes.\footnote[8]{We adopt the median absolute deviation from the isochrone because it is not sensitive to the outliers.}  We define stars with low $\sigma_{\rm Teff}$ that are more than three times this median absolute deviation from the isochrone as photometrically discrepant stars.  The two binaries vB 102 and vB 114, which have low $\sigma_{\rm Teff}$, are discrepant and overlaid with a yellow data point to illustrate them clearly.  

In the Praesepe CMD we see consistent characteristics with a very tight fiducial of single stars and binaries with low $\sigma_{\rm Teff}$, which gives a median absolute deviation from the isochrone of 0.097 magnitudes.  For Praesepe we find ten single and binary stars with low $\sigma_{\rm Teff}$ that meaningfully deviate from the fiducial (KW 31, KW 90, KW 182, KW 257, KW 275, KW 322, KW 334, KW 365, KW 536, and KW 540).  These photometrically discrepant stars are again overlaid with yellow data points.  In contrast to the Hyades, in Praesepe there are several photometrically discrepant binaries that have low $\sigma_{\rm Teff}$ and are consistent with the plotted equal-mass binary sequence.  Binaries with two components of nearly equal mass will not have peculiar colors (based on $\sigma_{\rm Teff}$), but their high luminosities will be apparent in the CMD.  In our Li analysis of both the Hyades and Praesepe we will identify these low $\sigma_{\rm Teff}$ stars that are photometrically discrepant from the single star fiducial and test if they have distinct A(Li).

Before we can derive our T$_{\rm eff}$ from a color-temperature relation and fit our isochrones in Figure 3, it is necessary to consider the cluster reddenings.  (See Taylor 2006 for a history of reddening determinations toward the Hyades and Praesepe.)  While early work (1975-1981) on the Hyades suggested a E(B-V) of at most a few to several mmag, some studies of the 1980s and 1990s claimed E(B-V)'s as high as a few times 0.01 mag.  Based on polarization data, Taylor (2006) finds E(B-V) $<$ 0.001 mag at the 95\% confidence level; this is a refinement of his earlier work where he found a limit of E(B-V)= 0.003 $\pm$ 0.002 (Taylor 1980).  We have adopted E(B-V)=0.00 for the Hyades.  When applying no reddening to the Hyades CMD we curiously found that the Yi et~al.\ (2001) isochrones with the derived metallicity of [Fe/H]=+0.146 (Deliyannis et~al.\ in prep.) did not match the Hyades single-star main sequence.  The Hyades M$_V$ were 0.08 magnitudes fainter than the isochrones.  Applying this correction and focusing on matching the left-edge of the turnoff by-eye derives an isochronal age 635$\pm$25 Myr.  

Many researchers have used the same reddening value for Praesepe as for the Hyades (usually 0.00) on the basis that both clusters show similar trends in the (U-B) versus (B-V) plane (e.g., Johnson \& Knuckles 1955), though other arguments for identical reddenings have also appeared.  There also exist at least several claims of small reddening values up to a few times 0.01 (see summary in Taylor 2006).  Taylor (2006) derives a value of 0.027$\pm$0.004 based on polarimetric and photometric data, but also discusses some caveats and cautions that more work is required before a definitive value can be established.  We have adopted E(B-V)=0.00, but it could be as high as 0.03 and throughout the paper we comment on the general effects that adopting E(B-V)=0.03 would have on our results.  Applying a reddening of E(B-V)=0.00 to the Praesepe CMD fits by eye an isochronal age of of 670$\pm$25 Myr using the isochrones of Yi et~al.\ (2001) and our derived metallicity of [Fe/H]=+0.156 (see Section 5).  Here we acknowledge that from an M$_V$ of approximately 2 to 3 magnitudes, the observed Praesepe stars are systematically fainter than the isochrone.  It is unclear what may be causing this systematic, but it is not observed in the Hyades and tests based on adopting differing reddenings and metallicities that are within reason for Praesepe or differing model isochrones (e.g., Bressan et~al.\ 2012; Choi et~al.\ 2016) do not improve its match.

Unlike with the Hyades, no correction was necessary for the isochrone to match the Praesepe main sequence (besides the brighter stars below the turnoff).  There are several possibilities for this systematic inconsistency we should briefly discuss.  First we note that P98 found a similar mismatch in their analysis of the Hipparcos photometry for the Hyades.  Therefore, it is not a systematic in the photometry but may be a systematic in the distances, but it would have to be quite large to create this offset.  It is not indicative of issues with our adopting Hyades reddening because it would require a negative reddening (E(B-V)$\sim$-0.01).  Another possibility is that while the isochrones already match well with most of Praesepe, if we adopt a solar Z of 0.016 (instead of the 0.018 adopted in Yi et~al.\ 2001 isochrones) this corrects the systematic offset observed in the Hyades and requires only a minor reddening of 0.01 (including the corresponding [Fe/H] shift) in Praesepe for its isochrone to identically match its main sequence.  A lower solar Z is argued for in the results of Asplund et~al.\ (2005; 2009) but helioseismology argues for the higher solar Z (see review in Chaplin \& Basu 2008).  This isochrone adjustment has no effect on the age fit for the Praesepe, but it requires a 30 Myr older (665 Myr) isochrone to match the Hyades turnoff.  This removes the already relatively minor isochronal age difference between these clusters, but the increased Praesepe reddening makes their spectroscopic metallicity difference significant.

\subsection{Atmospheric Parameters}

For our model atmospheres we used the Kurucz (1992) models with convective overshoot.  Effective temperatures were derived from the final averaged B-V by applying the color-temperature relation described in Deliyannis et~al.\ (2002):
\small
\begin{multline}
T_{\rm eff} = 8575 - 5222.27 (B-V)_0+ 1380.92 (B-V)_0\,^2 \\ 
        + 701.7 (B-V)_0([Fe/H]_* - [Fe/H]_{Hyades}) 
\end{multline}
\normalsize

This uses the Carney (1983) (B-V)-T$_{\rm eff}$ relation, the Cayrel et~al.\ (1985) Hyades zero point, and the Saxner \& Hammarbach (1985) metallicity dependence.  (See also Thorburn et~al.\ 1993; Deliyannis et~al.\ 1994; and Maderak et~al.\ 2013.)  Because this relation is based on the Hyades, the metallicity term is zero for the Hyades itself.  We adopted [Fe/H]$_{Hyades}$ = +0.15 based on a review of a number of high-resolution studies (see discussions in Maderak et~al.\ 2013 and Deliyannis et~al.\ in prep.), most of which fall in the very narrow range +0.13 to +0.17, and to be consistent with our previous studies.  Furthermore, using an identical [Fe/H] analysis as in the present paper, Deliyannis et~al.\ (in prep.) find a consistent [Fe/H]=+0.146$\pm$0.004.  If we had adopted +0.146 for our T$_{\rm eff}$ analysis, our temperatures for Praesepe would have differed by only +2.8 (B-V)$_0$ K, an insignificant difference.  Because T$_{\rm eff}$ depends on [Fe/H] and vice-versa, for Praesepe an initial guess is used for [Fe/H]$_*$, and the final determination of both T$_{\rm eff}$ and [Fe/H] is an iterative process.

\begin{figure}[htp]
\begin{center}
\includegraphics[scale=0.44]{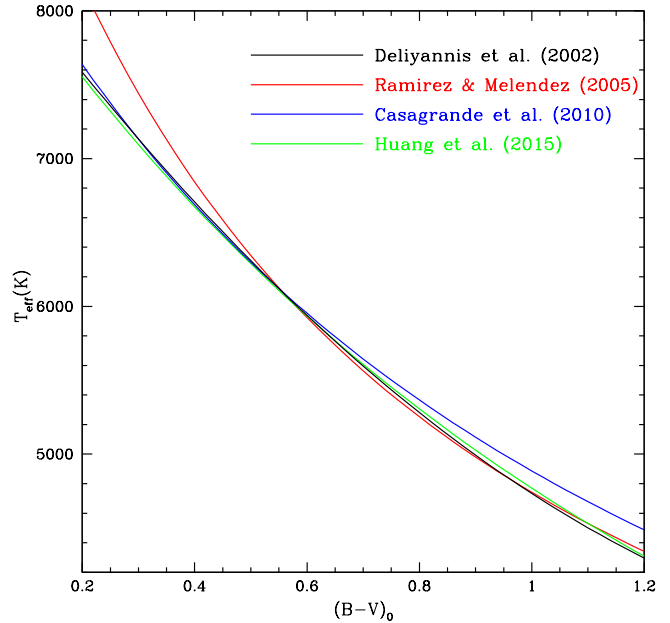}
\end{center}
\vspace{-0.45cm}
\caption{A direct comparison of the B-V color-temperature relations from Deliyannis et~al.\ (2002; black), Ram{\'i}rez \& Mel{\'e}ndez et~al.\ (2005; red), Casagrande et~al.\ (2010; blue), and Huang et~al.\ (2015; green).  All are plotted for a metallicity of [Fe/H]=+0.15, appropriate for both the Hyades and Praesepe.}
\end{figure}

The adoption of Equation 1's color-temperature relation will give us important consistency with all of our previous studies, but T$_{\rm eff}$ is critical for abundance analysis and we will compare it to the color-temperature relations from Ram{\'i}rez \& Mel{\'e}ndez (2005), Casagrande et~al.\ (2010), and Huang et~al.\ (2015).  Figure 4 plots these four relations for B-V from 0.2 to 1.2 and for a metallicity of [Fe/H]=+0.15, appropriate for both the Hyades and Praesepe.  This illustrates that for these parameters there is generally strong agreement between all four relations.  The Huang et~al.\ (2015) relation and Equation 1 never deviate more than 30 K from each other, but for hotter stars (B-V $<$ 0.50) the Ram{\'i}rez \& Mel{\'e}ndez (2005) relation becomes increasingly hotter relative to the other three relations while for cooler stars (B-V $>$ 0.70) the Casagrande et~al.\ (2010) relation becomes increasingly hotter than the other three relations.  This comparison shows that while potential systematics remain, our adoption of the Deliyannis et~al.\ (2002) relation remains appropriate in comparison to more recent color-temperature relations.  (See Huang et~al.\ 2015 for an in depth review of the different methods and measurements that introduce these systematics.)  Throughout the rest of the paper, when appropriate, we will comment on the effects of adopting the Casagrande et~al.\ (2010) relation for cooler stars and the Ram{\'i}rez \& Mel{\'e}ndez (2005) relation for hotter stars.

The log g of each star was determined from the Y$^2$ isochrones, adopting an age of 650 Myr for both the Hyades and Praesepe.  Lastly, the microturbulence was calculated using the empirical relation for MS stars of Edvardsson et~al.\ (1993), which has both temperature and log g dependence.  A lower limit of 0.8 km s$^{-1}$ was used for the microturbulence in the coolest stars.

\section{Praesepe Iron Abundance}

To determine the [Fe/H] of Praesepe, we focused only on stars consistent with the cluster radial velocity and that had $\sigma_{\rm Teff}$ less than 75 K.  Additionally, only stars with a $\textit{v sin i}$ of less than 25 km s$^{-1}$ were used because of the challenges of measuring heavily broadened lines and that they are likely subject to increased neighboring line contamination.  For our measurements we used 16 Fe I lines (see Table 1) that at our resolution we expected to have minimal contamination from neighboring lines, as evaluated using the high-resolution Delbouille et~al.\ (1989) solar atlas.  For each Praesepe spectrum, we measured the equivalent width of as many of the 16 Fe I lines as possible using the IRAF task \textit{splot}.  In certain stars (with minor rotational broadening, high T$_{\rm eff}$, or low S/N), several of the weaker Fe I lines could not be measured reliably, but typically 13 to 16 Fe I lines were measured in each star.  Fe lines with an equivalent width greater than 150 m\AA\, were rejected because of the increasing difficulty of fitting their wings properly.  Cutting all equivalent widths greater than 100 m\AA\, was considered, but no systematic differences were found in the resulting abundances.  Lastly, we require that all line measurements are stronger than the 3$\sigma$ equivalent width determined by the D93 $\sigma$ relation.  With our high S/N observations, however, only one of our Fe-line measurements was weaker than 3$\sigma$.  Therefore, only one line was excluded on the basis of Poisson noise.

\vspace{-0.5cm}
\begin{center}
\renewcommand{\baselinestretch}{1.1}
{\small \begin{longtable}{c c c}
\multicolumn{3}{c}%
{{\bfseries \tablename\ \thetable{} – Fe I Line Parameters}} \\
\hline
$\lambda$ (\AA) & Excitation Potential (eV) & log gf\\
\hline
\hline
6597.560 &  4.80 & -1.04 \\
6608.044 &  2.28 & -4.02 \\
6609.118 &  2.56 & -2.67 \\
6627.540 &  4.55 & -1.57 \\
6653.910 &  4.15 & -2.44 \\
6677.997 &  2.69 & -1.22 \\
6703.576 &  2.76 & -3.13 \\
6710.320 &  1.49 & -4.77 \\
6725.364 &  4.10 & -2.30 \\
6726.673 &  4.61 & -1.12 \\
6733.153 &  4.64 & -1.52 \\
6750.164 &  2.42 & -2.48 \\
6752.716 &  4.64 & -1.30 \\
6806.856 &  2.73 & -3.24 \\
6810.267 &  4.61 & -1.12 \\
6820.374 &  4.64 & -1.27 \\
\hline
\end{longtable}}
\end{center}

\begin{figure*}[htp]
\begin{center}
\includegraphics[scale=0.74]{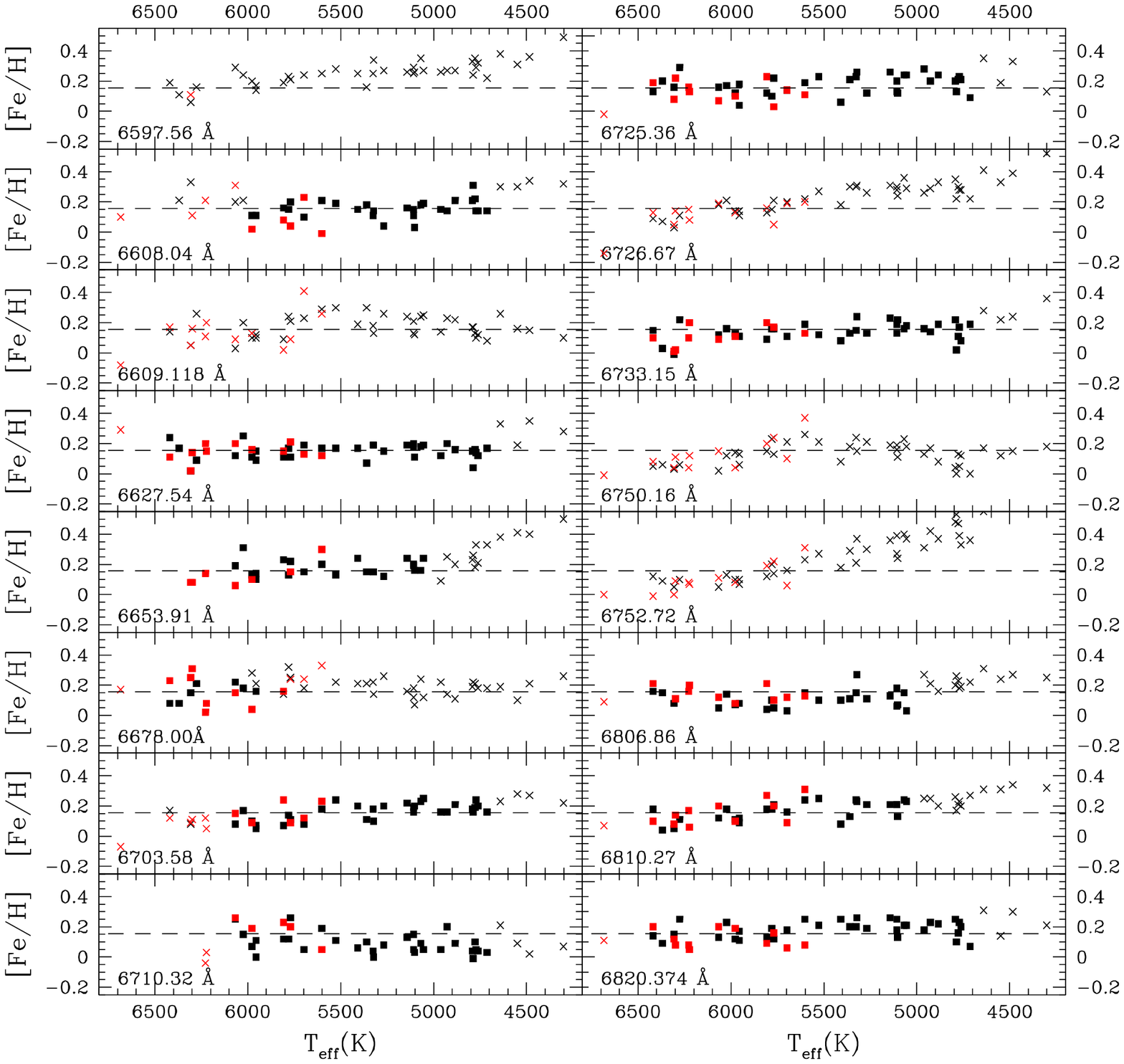}
\end{center}
\vspace{-0.45cm}
\caption[Each Praesepe line's {[Fe/H]} versus T$_{\rm eff}$]{All Praesepe [Fe/H] measurements from each individual line plotted against the stellar T$_{\rm eff}$.  Red and blue cable measurements are differentiated as red and black data points, respectively.  Solid squares are measurements that were kept, and X's are measurements that have been rejected either because they were above the 150 m\AA\, cutoff or because they were part of a clear systematic trend with T$_{\rm eff}$, as in the case of 6753 \AA.  Stars above 6500 K and below 4700 K are rejected because they have systematically lower and higher metallicities.  Lastly, the dashed horizontal line represents our weighted linear average of [Fe/H]=+0.156.}
\end{figure*}

\begin{figure}[htp]
\begin{center}
\includegraphics[scale=0.4]{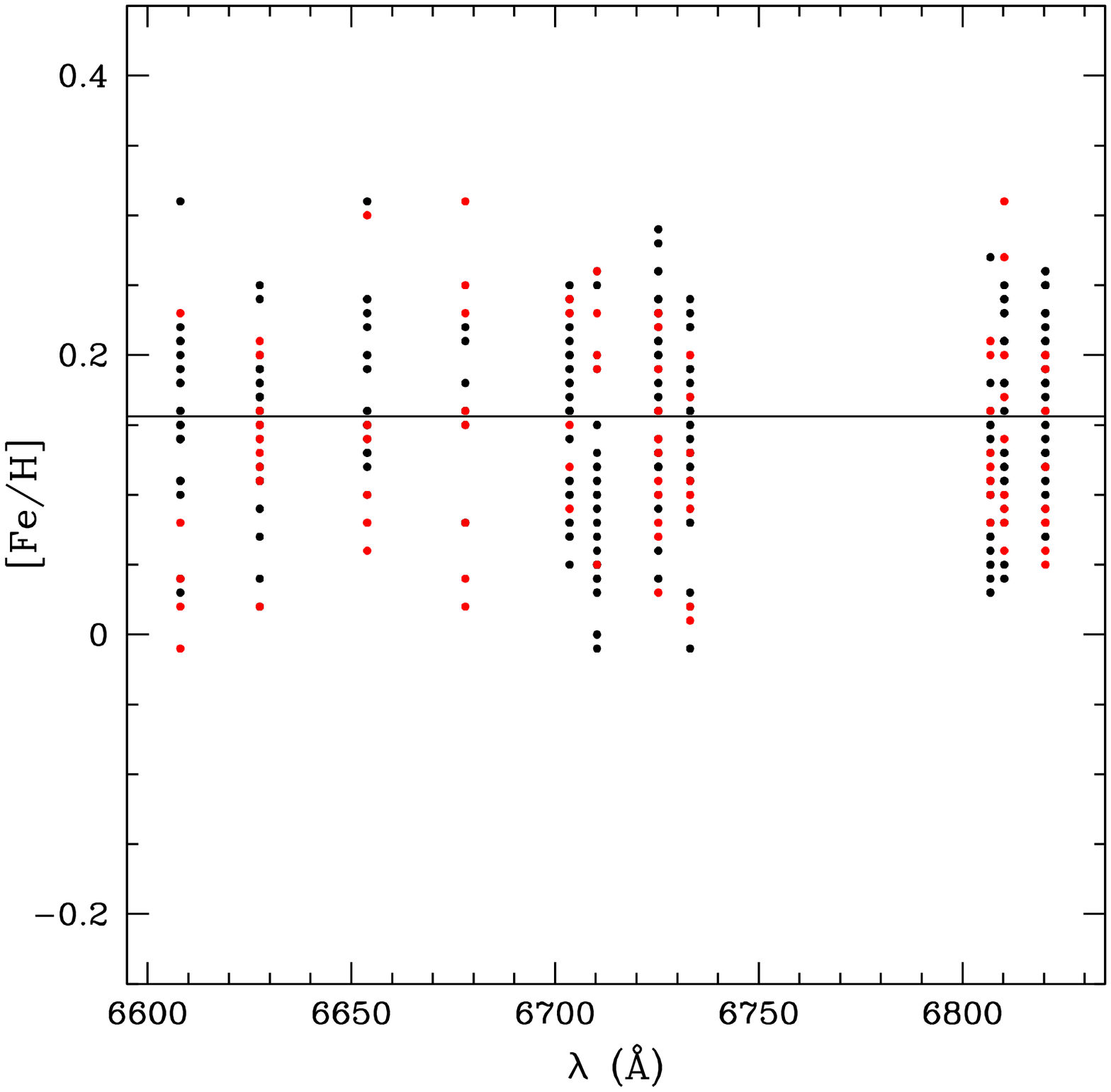}
\end{center}
\vspace{-0.45cm}
\caption[Praesepe {[Fe/H]} measurements versus wavelength of Fe line]{The [Fe/H] for of all the uncut lines versus the wavelength of the line for Praesepe.  This shows that in comparison to the spread of abundances measured from each line, there are no significant systematic abundance differences for the different Fe I lines.}
\end{figure}

\begin{figure}[htp]
\begin{center}
\includegraphics[scale=0.4]{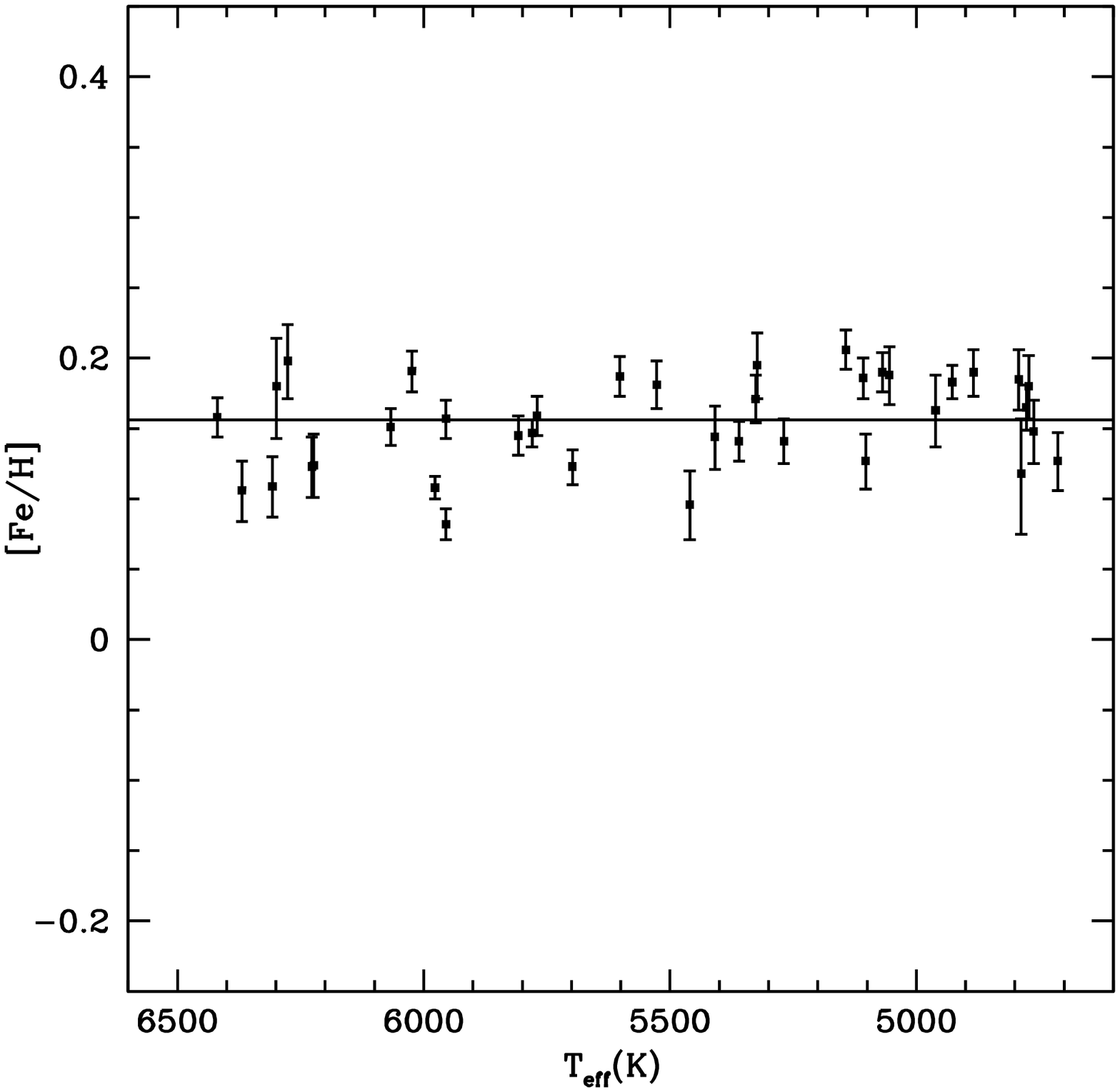}
\end{center}
\vspace{-0.45cm}
\caption[Praesepe stellar {[Fe/H]} versus T$_{\rm eff}$]{The final [Fe/H] for all of the slow-rotating Praesepe members ($<$25 km s$^{-1}$).  There is no significant scatter or trends seen with T$_{\rm eff}$.  The vertical solid-black line represents the weighted linear-average of all line measurements and is our final Praesepe metallicity of [Fe/H]=0.156$\pm$0.004.}
\end{figure}

\vspace{-0.8cm}
With our final line measurements, we used the MOOG spectral analysis program (Sneden 1973) and its \textit{abfind} routine with our stellar atmospheres to determine each line's absolute Fe/H abundance.  For both red and blue cable, we observed daytime solar spectra with high S/N and co-added all the available Hydra apertures to achieve S/N$\sim$4000.  These provided each line's consistently determined absolute solar Fe/H abundance, and direct comparison of the stellar and solar absolute abundances gave the [Fe/H] for that line measurement.  Therefore, we effectively employed the method of solar gf values and avoided the uncertainties associated with laboratory gf values.  This procedure also minimizes possible effects of weak-line blends contributing to the line strengths.  This analysis was carried out separately for both blue and red cable and applied to Praesepe stars observed with the corresponding cable.

We performed several tests on our data before finalizing our cluster [Fe/H].  Our sample of Praesepe stars span a T$_{\rm eff}$ range of nearly 2500 K.  Figure 5 shows the individual abundances for each measurement of all 16 Fe I lines versus T$_{\rm eff}$, which provides an important test for trends with T$_{\rm eff}$.  Red cable measurements are shown in red and blue cable measurements are shown in black.  Reassuringly, there are no systematic differences between the two cable measurements.  Additionally, measurements that have been cut are shown as X's and our final adopted measurements are shown as solid squares.  While there are no significant trends with temperature for most of the individual line abundances, the line at 6752.72 \AA\, and to a lesser extent the 6597.56 and 6726.27 \AA\, lines have significant trends with T$_{\rm eff}$.  These trends are likely the effect of subtle neighboring blends that we missed in our initial line selection.  As observed these blends would increase the abundance in increasingly cooler stars.  Other factors may also be important, such as deficiencies in the model atmospheres or opacities.  The 6609.18 and 6750.16 \AA\, lines have more complicated trends, which also concern us but are difficult to explain.  We have cut all measurements of these five lines in our [Fe/H] analysis.  This bases our final [Fe/H] on the remaining 11 lines.  

Considering specific T$_{\rm eff}$ ranges, we found that our highest T$_{\rm eff}$ star (KW 439) and the four stars cooler than 4700 K (KW 299, 288W, 624W, and 792W) have significantly lower and higher [Fe/H] in nearly every Fe I line, respectively, indicating problems with deriving Fe abundances at these extreme T$_{\rm eff}$.  (Schuler et~al.\ 2006 and 2009; and Maderak et~al.\ 2013 have discussed possible reasons for this effect.)  In regard to our adopted color-temperature relation, the hot KW 439 would increase by 130 K with the Ram{\'i}rez \& Mel{\'e}ndez (2005) relation, giving it an [Fe/H] of 0.07 higher but it would still be systematically Fe-poor relative to the cooler Praesepe dwarfs.  Conversely, for the coolest stars that are systematically too [Fe/H] rich, adopting the color-temperature relation from Casagrande et~al.\ (2010) would increase their derived T$_{\rm eff}$ and exacerbate the measured [Fe/H] discrepancy further.  To be conservative, we removed these five stars from our [Fe/H] analysis.  Beyond these discussed cuts, we have also cut all measurements below 5000 K for the 6653.91, 6806.86, and 6810.27 \AA\, lines due to consistently observed systematics.  Lastly, we remind the reader that all lines stronger than 150 m\AA\, have been cut, which is why the strong 6678 \AA\, line is cut in all but the hottest stars.  

Overall, given various uncertainties in the model atmospheres and the color-T$_{\rm eff}$ calibration, it is remarkable that the majority of our lines are so well-behaved over such a large range in T$_{\rm eff}$.

\vspace{-0.1cm}

\begin{center}
\tablefontsize{\footnotesize}
\begin{deluxetable}{c c c c c c c c}
\multicolumn{8}{c}%
{{\bfseries \tablename\ \thetable{} - Stellar Metallicity data for Praesepe}} \\
\hline
ID & T$_{\rm eff}$ & N & [Fe/H] & +STD & -STD & S/N & \textit{v sin i}\\
\hline
KW &     (K)       &   &        &      &      &     &   (km s$^{-1}$) \\
\endfirsthead
\hline
\multicolumn{8}{l}{{Blue Cable}}  \\
\hline
 27  &  5527  & 10  & 0.181 & 0.051 & -0.058 &  400 & 6.4 \\
 32  &  5323  & 10  & 0.195 & 0.069 & -0.081 &  250 & 6.4 \\
 48  &  4961  &  7  & 0.163 & 0.062 & -0.073 &  240 & 5.8 \\
 79  &  5103  & 10  & 0.127 & 0.057 & -0.065 &  310 & 5.0 \\
100  &  6024  & 10  & 0.191 & 0.043 & -0.048 &  500 & 6.0 \\
162  &  6067  & 10  & 0.146 & 0.057 & -0.065 &  450 & 8.2 \\
172  &  4884  &  7  & 0.190 & 0.042 & -0.046 &  210 & 6.4 \\
183  &  4772  &  7  & 0.180 & 0.056 & -0.065 &  260 & 6.0 \\
198  &  4792  &  7  & 0.185 & 0.054 & -0.062 &  210 & 6.4 \\
208  &  5977  & 10  & 0.107 & 0.019 & -0.020 &  550 & 9.3  \\
209  &  4777  &  7  & 0.165 & 0.040 & -0.044 &  240 & 5.2 \\
213  &  5326  & 10  & 0.171 & 0.050 & -0.057 &  290 & 8.2 \\
222  &  6369  &  7  & 0.106 & 0.054 & -0.062 &  310 & 8.2 \\
237  &  4762  &  7  & 0.148 & 0.056 & -0.064 &  300 & 7.6 \\
238  &  6307  &  7  & 0.095 & 0.056 & -0.065 &  450 & 19.3  \\
263  &  5269  & 10  & 0.141 & 0.048 & -0.054 &  277 & 8.2 \\
272  &  4713  &  7  & 0.127 & 0.051 & -0.057 &  120 & 6.4 \\
282  &  6276  &  6  & 0.198 & 0.060 & -0.070 &  326 & 23.5 \\
288  &  5955  & 10  & 0.133 & 0.033 & -0.036 &  500 & 7.6 \\
313  &  5055  & 10  & 0.188 & 0.062 & -0.072 &  280 & 6.4 \\
326  &  5602  & 10  & 0.199 & 0.029 & -0.031 &  700 & 7.6 \\
335  &  5780  & 10  & 0.147 & 0.029 & -0.031 &  500 & 8.2 \\
344  &  5108  & 10  & 0.155 & 0.040 & -0.044 &  280 & 6.0 \\
349  &  5108  & 10  & 0.186 & 0.043 & -0.048 &  270 & 5.2 \\
363  &  4927  &  7  & 0.183 & 0.031 & -0.033 &  250 & 8.2 \\
399  &  5808  & 10  & 0.123 & 0.048 & -0.055 &  388 & 8.2 \\
403  &  5360  & 10  & 0.141 & 0.042 & -0.047 &  230 & 8.2 \\
430  &  5143  & 10  & 0.206 & 0.043 & -0.047 &  280 & 6.4 \\
432  &  5698  & 10  & 0.122 & 0.049 & -0.055 &  325 & 6.4 \\
448  &  5069  & 10  & 0.190 & 0.042 & -0.047 &  320 & 6.0 \\
454  &  6419  &  7  & 0.151 & 0.047 & -0.052 &  440 & 17.0 \\
466  &  5770  & 10  & 0.165 & 0.059 & -0.068 &  340 & 6.4 \\
476  &  5409  & 10  & 0.144 & 0.066 & -0.078 &  240 & 6.4 \\
508  &  5955  &  9  & 0.082 & 0.031 & -0.033 &  520 & 8.2 \\
899W &  4787  &  7  & 0.118 & 0.097 & -0.126 &  180 & 10.0 \\
\hline
\hline
\multicolumn{8}{l}{{Red Cable}}  \\
\hline
162  &  6067  & 10  & 0.160 & 0.051 & -0.058 &  280 & 6.5 \\
208  &  5977  & 11  & 0.110 & 0.050 & -0.056 &  350 & 8.5 \\
217  &  6299  &  8  & 0.180 & 0.091 & -0.115 &  530 & 15.2 \\
238  &  6307  &  7  & 0.125 & 0.084 & -0.105 &  420 & 22.3 \\
288  &  5955  & 11  & 0.182 & 0.066 & -0.077 &  450 & 5.0  \\
326  &  5602  & 10  & 0.160 & 0.091 & -0.116 &  320 & 5.0 \\
341  &  6227  &  8  & 0.123 & 0.056 & -0.065 &  491 & 10.1 \\
399  &  5808  & 10  & 0.190 & 0.052 & -0.059 &  140 & 5.0  \\
421  &  6223  &  7  & 0.124 & 0.056 & -0.064 &  210 & 8.3 \\
432  &  5698  &  7  & 0.126 & 0.047 & -0.053 &  131 & 5.0 \\
454  &  6419  &  7  & 0.170 & 0.051 & -0.058 &  230 & 24.3 \\
466  &  5770  & 10  & 0.144 & 0.055 & -0.063 &  140 & 5.0 \\
\hline
\hline
\multicolumn{8}{l}{{Final Combined [Fe/H]}}  \\
\hline
162 &   6067 &  20  & 0.151 & 0.055 & -0.063 & - & - \\
208 &   5977 &  21  & 0.108 & 0.036 & -0.039 & - & - \\
238 &   6307 &  14  & 0.109 & 0.073 & -0.088 & - & - \\
288 &   5955 &  21  & 0.157 & 0.058 & -0.067 & - & - \\
326 &   5602 &  20  & 0.187 & 0.058 & -0.067 & - & - \\
399 &   5808 &  20  & 0.145 & 0.058 & -0.067 & - & - \\
432 &   5698 &  17  & 0.123 & 0.048 & -0.054 & - & - \\
466 &   5770 &  20  & 0.159 & 0.058 & -0.067 & - & - \\
\hline
\caption[Praesepe stellar metallicity data]{Our Praesepe IDs are from Klein Wassink (1927) when available.  Otherwise, they are followed by a W and are taken from Wang \& Jiang (1991).}
\end{deluxetable}
\end{center}

\vspace{-0.85cm}
Figure 6 shows the line abundances plotted against wavelength and illustrates that there are no significant offsets in average abundance from line to line.  To calculate each individual star's final [Fe/H] we used the weighted linear-average of its individual line measurements.  The [Fe/H] error is given by the weighted standard deviation of the mean calculated in linear space.  These weights are linear and are based on the 1$\sigma$ abundance error of each line using the D93 1$\sigma$ relation and the local S/N at each line's wavelength.  Figure 7 shows each star's average [Fe/H] versus T$_{\rm eff}$.  Across the nearly 2000 K range, [Fe/H] is uniform with no temperature trends. Table 2 lists the T$_{\rm eff}$, number of lines used, stellar average [Fe/H], errors, S/N, and \textit{v sin i} for each of our 39 Praesepe stars selected for Fe analysis.

For the total cluster average, we similarly based it on the weighted linear-average of the entire sample of line abundances.  This average gives a Praesepe [Fe/H] of +0.156$\pm$0.004 ($\sigma_\mu$) +0.061/-0.071 ($\sigma$, per line not per star).  After all of the cuts discussed above, this result is based on 415 Fe I line measurements from the spectra of 39 slowly-rotating (\textit{v sin i}$<$25 km s$^{-1}$) Praesepe members, where eight of these members had independent measurements of Fe I lines from both the red and blue cable.

Before continuing, we will discuss the various methods for determining a stellar average and a cluster average.  The star average is more straightforward because the uncertainty that creates the abundance dispersion in [Fe/H] is dominated by measurement errors in equivalent width due to spectral noise.  When propagating this equivalent width error to an abundance error, the resulting abundance error is more symmetric in linear rather than log space.  Therefore, averaging the individual Fe I line abundances from a single star in linear space to determine its ``average" abundance provides a better statistical representation of the mean than averaging in log space does.  For a cluster average, equivalent width error is no longer the only factor because the error in each stellar T$_{\rm eff}$ further increases abundance dispersion.  Errors in T$_{\rm eff}$ propagate into abundance errors that are more symmetrically distributed in log rather than linear space.  Therefore, an argument can be made that a weighted logarithmic-average of the individual Fe I line abundances would provide a better representation of the distribution and of the cluster's true [Fe/H].  For the sake of comparison and completeness, we will present the result for Praesepe from four different averaging methods.  As discussed above, a weighted linear-average for Praesepe gives an [Fe/H] of +0.156$\pm$0.004 ($\sigma_\mu$) +0.061/-0.071 ($\sigma$, per line not per star).  A weighted logarithmic-average gives an [Fe/H] of +0.151$\pm$0.004 ($\sigma_\mu$) $\pm$0.066 ($\sigma$).  As expected, the average in logarithmic space gives a lower abundance, but the difference is not meaningful.  However, we note that observations of most clusters will have larger measurement dispersions, and the difference between these two methods increases rapidly with increasing dispersion.

\begin{center}
\tablefontsize{\footnotesize}
\begin{deluxetable*}{c c c c}
\multicolumn{4}{c}%
{{\bfseries \tablename\ \thetable{} - Cluster Metallicity data for Praesepe}} \\
\hline
Type of Average & [Fe/H] & Standard Deviation ($\sigma$) & Standard Deviation of Mean ($\sigma_\mu$)\\
\hline
\textbf{Weighted Linear-Average} &  \textbf{+0.156} & \textbf{+0.061,-0.071} & \textbf{$\pm$0.004} \\
Linear Average & +0.154 & +0.061,-0.071 & $\pm$0.004\\
Weighted Logarithmic-Average & +0.151 & $\pm$0.065 & $\pm$0.004 \\
Logarithmic Average & +0.149 & $\pm$0.066 & $\pm$0.004 \\
\hline
\multicolumn{4}{l}%
{{Total Number of Stars = 39; Total Number of Lines = 415}} \\
\hline
\caption[Praesepe mean cluster {[Fe/H]} data]{}
\end{deluxetable*}
\end{center}

\vspace{-0.96cm}
Another averaging method is to not weight the measurements at all.  Since there is a variation in abundance errors from the Fe I line measurements and from stars with differing S/N, this method is not as robust, but we include it for comparison.  For Praesepe, the linear average with no weighting gives an [Fe/H] of +0.154$\pm$0.004 ($\sigma_\mu$) +0.061/-0.071 ($\sigma$).  A logarithmic average with no weighting gives an [Fe/H] of +0.149$\pm$0.004 ($\sigma_\mu$) $\pm$0.066 ($\sigma$).  This shows that with Praesepe and its high S/N spectra, weighting has no significant effect, but we will not have this advantage with other typically fainter clusters where weighting will provide a more rigorous metallicity.  Table 3 lists these final Praesepe cluster [Fe/H] averages and errors using the four different methods discussed.  

A potentially larger systematic error is due to the possibility that Praesepe has a non-zero reddening.  In the extreme case of a E(B-V)=0.03, the T$_{\rm eff}$ are increased (for example, by 115 K near 6000 K and by 83 K near 5000 K), which for our sample increases the overall cluster [Fe/H] by 0.057 dex to +0.213.

\subsection{Previous Iron Studies}

There exist a number of previous high-resolution studies of the metallicity of Praesepe, including the seven studies described in Table 4.  The cluster averages have a rather large range from solar to very metal rich, but half are similar to our metallicity of [Fe/H]=+0.156$\pm$0.004 ($\sigma_\mu$).  The extremely metal-rich (+0.40) value from
Burkhart \& Coupry (1998) is derived from Am stars, which are chemically peculiar stars that often show significantly higher metallicities than other stars from the same cluster, so this large discrepancy is likely due to this phenomenon.  

The discrepancy with the high metallicity of Pace et~al.\ (2008, hereafter P08), who studied similar stellar types as we did, will be examined in greater detail.  We observed five of their seven dwarfs, but it should be noted that two of these five have very high $\sigma_{\rm Teff}$ and were removed from our sample (see Section 4).  Detailed comparison of the remaining three dwarfs shows that while the log g, microturbulence, and stellar atmospheres of P08 are consistent with ours, their spectroscopically determined T$_{\rm eff}$ values are significantly higher than ours.  For these three stars, Table 5 shows our T$_{\rm eff}$, the metallicity we derive from our spectra using our T$_{\rm eff}$, their T$_{\rm eff}$, the metallicity they derived, and the metallicity we would derive from our spectra using their T$_{\rm eff}$.  It is clear that for these three stars the discrepancies between our [Fe/H] and their [Fe/H] can primarily be explained by the differences in the adopted stellar T$_{\rm eff}$.  This higher spectroscopic T$_{\rm eff}$ may suggest that, as discussed above, the reddening of Praesepe is non-zero.  Adopting the extreme reddening of 0.03 for these three dwarfs would explain a majority, but not all, of the difference found for their T$_{\rm eff}$. 

\begin{center}
\tablefontsize{\footnotesize}
\begin{deluxetable*}{l c c c}
\multicolumn{4}{c}%
{{\bfseries \tablename\ \thetable{} - Praesepe Cluster Metallicity Comparisons}} \\
\hline
Study & [Fe/H]$\pm\sigma$ & Stellar Types\\
\hline
\hline
Boesgaard \& Budge (1988) & +0.13$\pm$0.07 & 5 F dwarfs and 1 binary\\ 
Friel \& Boesgaard  (1992) & +0.038$\pm$0.039 & 6 F dwarfs and sub-giants\\ 
Burkhart \& Coupry (1998) & +0.40$\pm$0.14 & 10 Am stars\\
Pace et~al.\ (2008) & +0.27$\pm$0.10 & 7 F and G dwarfs\\
Carrera \& Pancino (2011) & +0.16$\pm$0.05 & 3 Giants\\
An et~al.\ (2007) & +0.11$\pm$0.03 & 4 G dwarfs\\
Our analysis & +0.156$\pm$0.066 & 39 F, G, and K dwarfs\\
\hline
\end{deluxetable*}
\end{center}

\vspace{-0.4cm}
\begin{center}
\tablefontsize{\footnotesize}
\begin{deluxetable*}{c c c c c c}
\multicolumn{6}{c}%
{{\bfseries \tablename\ \thetable{} - Parameter and Abundance Comparisons to Pace et~al.\ 2008}} \\
\hline
ID & Our T$_{\rm eff}$ & P08 T$_{\rm eff}$ & Our [Fe/H] & P08 [Fe/H] & Our [Fe/H] using P08 T$_{\rm eff}$\\
\hline
\hline
100 & 6024 & 6150  & +0.191$\pm$0.045 & +0.27$\pm$0.10 & +0.265$\pm$0.048\\
208 & 5977 & 6280  & +0.107$\pm$0.020 & +0.28$\pm$0.10 & +0.257$\pm$0.024\\
326 & 5602 & 5800  & +0.199$\pm$0.030 & +0.29$\pm$0.10 & +0.294$\pm$0.039\\
\hline
\caption{Comparison of our derived T$_{\rm eff}$ and the spectroscopic T$_{\rm eff}$ from
P08, and the comparison of our final [Fe/H], that published in P08, and our rederived
[Fe/H] using the P08 T$_{\rm eff}$.}
\end{deluxetable*}
\end{center}

\vspace{-1.5cm}
The cluster average [Fe/H] derived by Carrera \& Pancino (2011), Boesgaard \& Budge (1988, hereafter BB88), and An et~al.\ (2007) are all very similar to ours.  Nonetheless, it is of interest to analyze any possible systematic differences between their chosen parameters and ours. Carrera \& Pancino (2011), however, only observed giants so we cannot directly compare their stellar parameters with ours.  While we did not observe any of the six dwarfs observed by BB88, J52 and Mend67 did observe all six of these dwarfs photometrically.  Therefore, we determined their photometric T$_{\rm eff}$ in the same manner as our own sample, and we found no significant difference between our values and the T$_{\rm eff}$ values adopted in BB88.  The comparison of parameters with An et~al.\ (2007) is more complex.  We did not observe any of their dwarfs ourselves, but again J52 and Mend67 did observe photometrically all four of their dwarfs, so we derived the parameters of these stars using our methods.  While the log g's are similar, An et~al.\ (2007) use T$_{\rm eff}$ values that are roughly 75 to 175 K hotter than ours and microturbulences that are roughly half of ours.  Again, adopting a small non-zero reddening could explain their systematically hotter T$_{\rm eff}$, but their significantly lower microturbulences are difficult to explain.  Although we did not observe their stars, we can examine the effects of applying their parameters to one of our G dwarfs (KW 335) that is similar to their stars.  Increasing the T$_{\rm eff}$ from 5781 K to 5900 K and decreasing the microturbulence from 1.08 km s$^{-1}$ to 0.5 km s$^{-1}$ increases the derived [Fe/H] by 0.09 dex and 0.06 dex, respectively, for a total change of 0.15 dex.  Furthermore, there are no differences that were introduced by the adopted models because our analysis and that of An et~al.\ (2007) used identical models.  Thus, it is difficult to understand why their [Fe/H] measurement is so similar to ours. 

Lastly, Friel \& Boesgaard (1992, hereafter FB92) give the lowest [Fe/H] at +0.038$\pm$0.039.  We observed five of their six dwarfs, but found one of these five (KW 416) to have a radial velocity significantly different than the cluster and with high $\sigma_{\rm Teff}$.  Therefore, we do not consider it a single-star member, but based on its PM and its variable radial velocity it is likely a binary member.  Similarly, Mermilliod et~al.\ (2009) find it to be a binary member.  For the other four stars, FB92's choice of parameters is very similar to ours.  While their adoption of the older Kurucz (1979) atmospheric models may play an important role in this difference, BB88 also used the same models with similar parameters yet they derived a higher [Fe/H] that is much closer to our value; therefore, this is not the likely explanation for the difference between our result and that of FB92. The cause of this difference thus remains mysterious.

\subsection{Comparison of Praesepe and the Hyades}

For the Hyades, Deliyannis et~al.\ (in prep.) find [Fe/H]=+0.146$\pm$0.004 ($\sigma_\mu$) using identical methods, and for Praesepe we have reported [Fe/H]=+0.156$\pm$0.004 ($\sigma_\mu$).  It is noteworthy that these values are consistent to high precision, using data taken with the same telescope and instrument and analyzed in the same way.  These clusters also have nearly indistinguishable ages with our own isochrones age measurement (see Figure 3) using Yi et~al.\ (2001) isochrones for the Hyades of 635$\pm$25 Myr and for Praesepe of 670$\pm$25 Myr.

Among others, Eggen (1992) has noted the similarities (and subtle differences) in the space motions of these two clusters, in the context of a possible common origin.  Our finding of identical super-solar metallicities is consistent with this idea, especially since super-solar metallicities are less common than slightly sub-solar ones (Edvardsson et~al.\ 1993).  That said, we also remind the reader about the possibility that Praesepe is slightly reddened, which would lead to both a higher [Fe/H] (+0.213) and a younger isochronal age (570 Myr) than that of the Hyades, potentially contradicting the common origin hypothesis. 

\section{Lithium Abundance Analysis and Results}

Spectral synthesis was used to fit the Li line region, including the Li line near 6707.8 \AA, the nearby blended Fe I line (6707.44 \AA), and the continuum level. This is advantageous compared to direct equivalent-width fits because the Hyades and Praesepe have moderately super-solar metallicity and many of the hotter stars have moderate to large rotational broadening.  For such rotators, the increased blending of the neighboring lines makes it challenging to judge the continuum level and account for neighbor contamination.  But a synthetic reference allows for more proper fits of moderately weak Li lines and the neighboring lines alike.  

The MOOG spectral-analysis program (Sneden 1973) was used to create and compare the synthetic and observed spectra.  The line list for the Li region was taken from Hiltgen (1996; see King et~al.\ 1997 for a discussion), which includes both fine and hyperfine structure for the Li feature.  All input abundances for the synthetic spectra are scaled-solar relative to our cluster [Fe/H] average.  Once the continuum level has been matched in the observed spectrum, a \textit{v sin i} broadening parameter convolved with a spectral resolution of 0.5 \AA\, is used to match the observed line broadening.  The input A(Li) is then changed at increments of 0.005 dex until the best fit by eye is found.  This \textit{v sin i} value based on our Li synthesis is further tested by matching the width of nearby isolated Fe I lines.  This provides a more reliable determination of \textit{v sin i} in the fastest rotators versus \textit{fxcor} (see Section 3) and is used as our published value for when \textit{v sin i} is greater than 30 km s$^{-1}$.

\begin{figure}[htp]
\begin{center}
\includegraphics[scale=0.41]{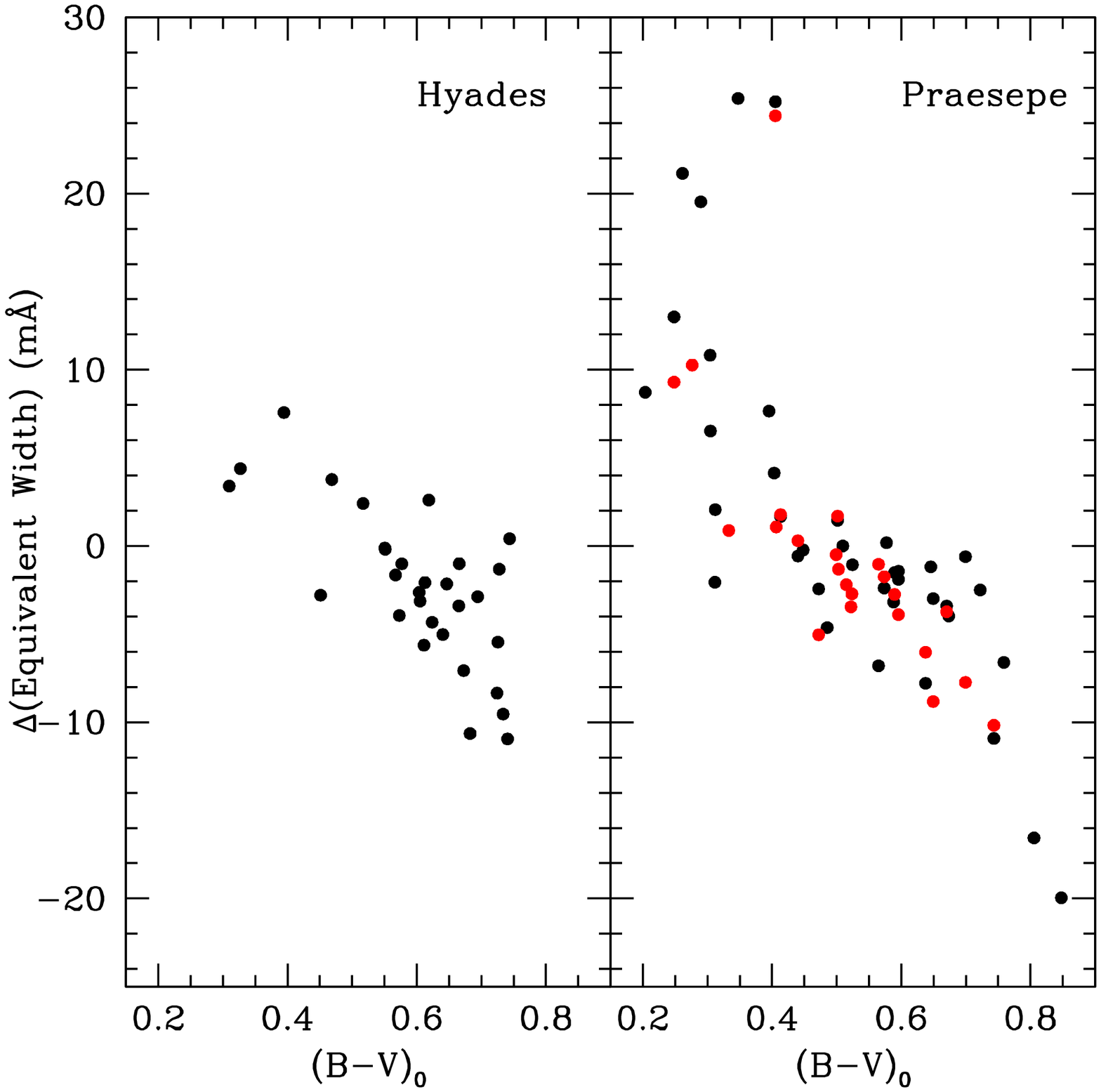}
\end{center}
\vspace{-0.45cm}
\caption{A direct comparison of fit Li equivalent widths minus the synthetically derived Li equivalent widths versus a star's (B-V)$_0$.  The left panel focuses on the WIYN/Hydra observations of Hyades members, and the right panel focuses on our WIYN/Hydra observations of Praesepe members with blue cable observations in black and red cable observations in red.}
\end{figure}

The synthetic Li measurements were applied to our manually created curves of growth (hereafter COGs) for the Li feature (Steinhauer et~al.\ in prep.), which provided the equivalent width corresponding to the synthetic abundance.  We stress that our equivalent widths are not measured directly but are derived from the synthetic A(Li) and the stellar parameters applied to our COGs.  These manually created COGs are based on the Li feature and exclude the Fe I line at 6707.44 \AA, but they also consider the contributions from the weak neighboring lines when applying the resolution of WIYN/Hydra (R$\sim$13,500).  This creates a more realistic synthetic equivalent width, and we have created a series of COGs at differing metallicity to account for the increasing neighbor contribution (besides that of the 6707.44 \AA\, Fe I line) with increasing metallicity, which becomes important at the metallicity of the Hyades and Praesepe.  

In Figure 8 we plot $\Delta$(Equivalent Width) (measured minus synthetically derived) in m\AA\, versus (B-V)$_0$ for our observations of both the Hyades and Praesepe.  As with Figures 5 and 6, in the right panel for Praesepe measurements we plot the blue cable observations in black and the red cable observations in red.  Some notable systematics are consistently observed in both clusters.  In Praesepe, at the bluest colors (B-V $\leq$ 0.4) there is a broad range of systematic differences ($\sim$0 to 25 m\AA) in equivalent widths, and this large scatter is a direct result of the large variation in \textit{v sin i} in these hotter stars.  As we discussed earlier, direct attempts at equivalent width measurements in fast rotators will not be able to deblend the series of neighboring lines contributing to the increasingly broad Li feature.  The smaller dispersion in the Hyades (left panel) is due solely to the Hyades having a smaller range of observed \textit{v sin i}.  In stars of intermediate color (0.4 $<$ B-V $<$ 0.7) the systematic differences are minor but still important, with both clusters illustrating that direct equivalent width measurements are $\sim$2 to 3 m\AA\, weaker.  In the reddest stars (B-V $\geq$ 0.7) this systematic increases rapidly.  These weaker equivalent widths are primarily due to the challenge of defining the continuum near the Li region, in particular in metal-rich clusters, and as the neighboring line strengths increase in cooler stars this systematic difference increases.  For both of these challenges, increases in spectral resolution will provide only marginal improvements; spectral synthesis provides the most precise Li abundances.

The D93 relation again provided Poisson-noise 1$\sigma$ Li equivalent width errors, which were translated to A(Li) errors using our COGs.  These are shown in Column 8 of Tables 6 and 7 for the Hyades and Praesepe, respectively.  Additionally, the Li line is sensitive to T$_{\rm eff}$ (but not log g or microturbulence).  Application of the $\sigma_{\rm Teff}$ (from the 10 empirically derived colors) to our COGs provided the T$_{\rm eff}$-based error for A(Li).  These are shown in Column 9 of Tables 6 and 7 for the Hyades and Praesepe, respectively.  These two errors added in quadrature give our final A(Li) errors.  Lastly, we note the possible systematic errors based on our choice of reddening.  For the Hyades, the upper limit in E(B--V) of 0.001 has negligible effects in the derived T$_{\rm eff}$ and A(Li), and for Praesepe, an upper limit in E(B--V) of 0.03 would give relatively small increases in the derived A(Li) of 0.09 dex for our hottest stars and 0.14 dex for our coolest stars.

Several of our observed members of Hyades and Praesepe have undergone far too much Li depletion to make a reliable detection at the 3$\sigma$ level or greater.  In these cases, the D93 relation was used to calculate the 3$\sigma$ equivalent width for each spectrum, and this was converted using our COGs to a 3$\sigma$ based A(Li) upper limit.   In most of our temperature range, the additional neighboring lines near the Li line at 6707.8 \AA\, have negligible strength (e.g., compared to the Fe I feature at 6707.44 \AA).  However, at cooler T$_{\rm eff}$ and in view of our clusters' high (super-solar) metallicity, some of these neutral features begin to gain strength at the same time that the Li line vanishes (due to stellar Li depletion).  Therefore, at cooler T$_{\rm eff}$ each metallicity-dependent COG has a lower limit set by when the neighboring lines begin to dominate the Li feature.  (In this line comparison the moderately strong Fe I line at 6707.44 \AA\, is ignored.)  If a star's 3$\sigma$ equivalent width is below this lower limit of the COG, the lower limit is adopted as the A(Li) upper limit.

\subsection{Hyades}

A major goal in re-observing the Hyades is to ensure that our WIYN/Hydra A(Li) are on the same scale as those in previous studies of the Hyades (e.g., Boesgaard \& Tripicco 1986, hereafter BT86; BB88; Thorburn et~al.\ 1993, hereafter T93; Takeda et~al.\ 2013, hereafter T13).  We will address this in Section 6.1.2, but first in Section 6.1.1 we use our WIYN/Hydra abundances to examine what criteria are required to best define the Li-T$_{\rm eff}$ trend.  We take it as an axiom that highly probable single star members with small $\sigma_{\rm Teff}$ (here, $<$75 K) are ideal for this purpose.  We will also examine whether high $\sigma_{\rm Teff}$ stars, known member SB1s, PM members that fall outside the radial velocity single-star peak, or members that photometrically deviate from the single star cluster fiducial might also be suitable.  

\subsubsection{New Hyades Lithium Observations}

The left panel of Figure 9 shows the A(Li) we derived for our WIYN/Hydra sample for the Hyades cluster, which we have found to be moderately metal-rich ([Fe/H]=0.146$\pm$0.004 ($\sigma_\mu$); Deliyannis et~al.\ in prep.).  Solid circles (detections) and solid inverted triangles (upper limits) are Hyades PM members (including both stars within and outside the radial velocity peak from 35 to 43 km s$^{-1}$) that have small $\sigma_{\rm Teff}$ ($<$ 75 K).  Open circles (detections) and open inverted triangles (upper limits) are members that have large $\sigma_{\rm Teff}$. Plus signs (detections) and plus signs with downward arrows (upper limits) are for stars identified by P98 as binaries that have small $\sigma_{\rm Teff}$, but the addition of an open circle to a plus sign signifies that it has a large $\sigma_{\rm Teff}$.  Lastly, our three observed radial velocity members that are not in P98 (vB 49, vB 59, and vB 93) are treated as single stars because none show any evidence of binarity in the photometry or in our spectra (from either \textit{v$_{rad}$} or \textit{fxcor}).

\begin{figure}[htp]
\begin{center}
\includegraphics[scale=0.43]{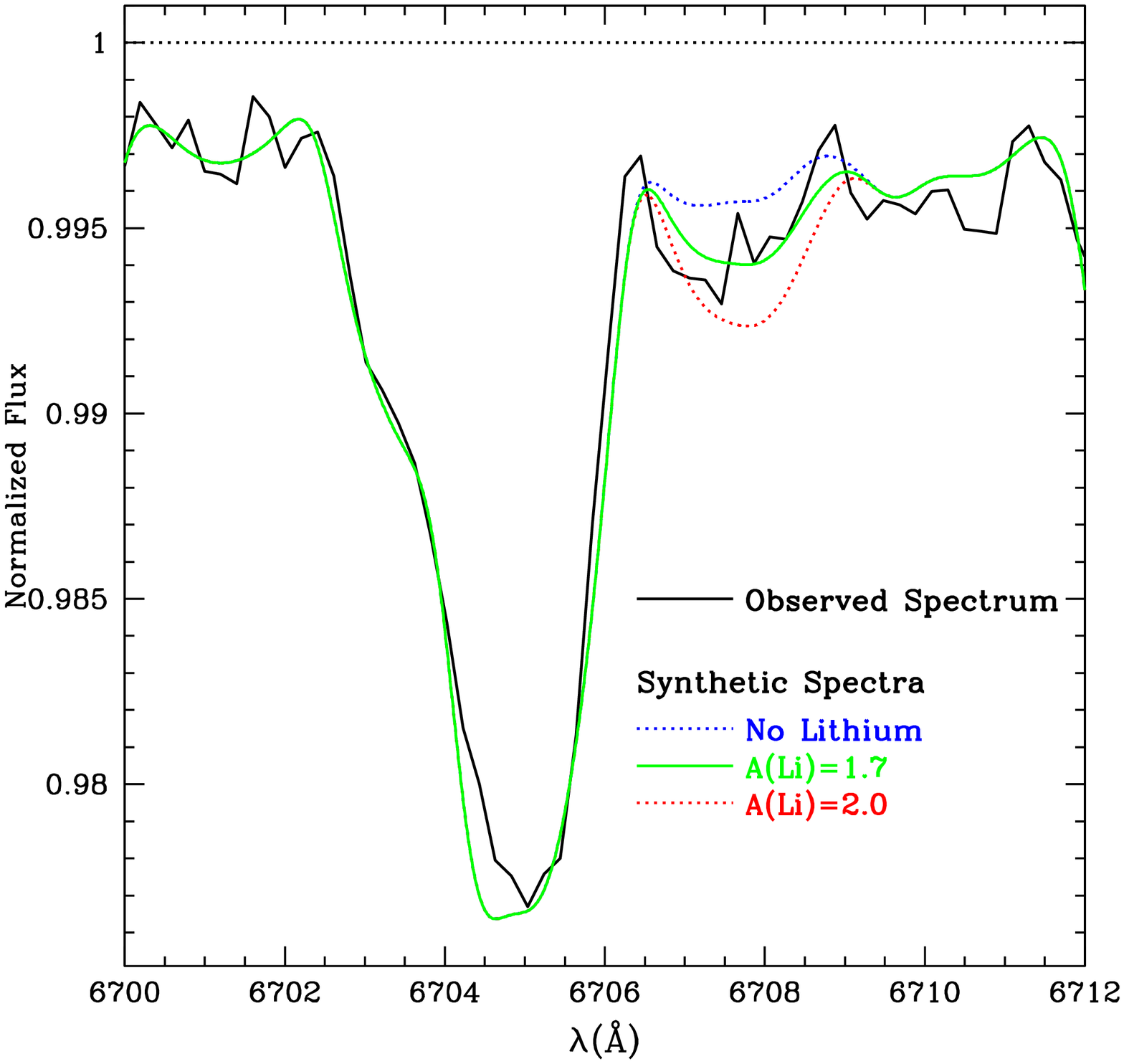}
\end{center}
\vspace{-0.45cm}
\caption{The matching of the very high S/N spectrum of vB 90 and three synthetic spectra of varying Li abundance at no Li (dotted blue), A(Li)=1.7 (solid green), and A(Li)=2.0 (dotted red).  A(Li)=1.70 represents our best match to the observed spectrum.  The straight dotted black line at 1.0 illustrates the true synthetic continuum in this spectrum with large rotational broadening (\textit{v sin i}=50 km s$^{-1}$) and at the rich metallicity of the Hyades.  We also note a few mismatches in the synthesis that likely cannot be explained by noise and are discussed in more detail in the text.}
\end{figure}

\begin{figure*}[htp]
\begin{center}
\subfigure{\includegraphics[scale=0.95]{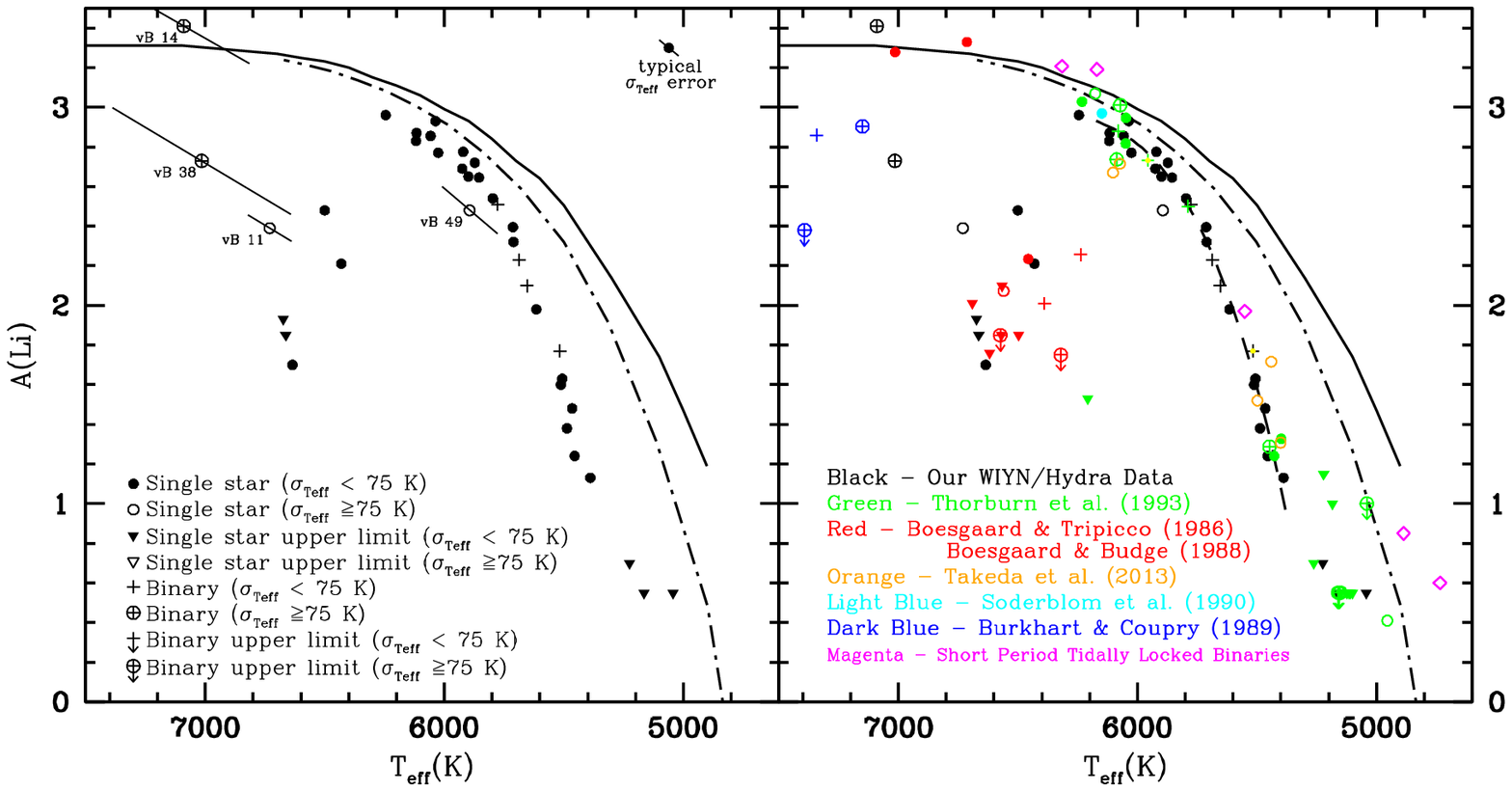}}
\end{center}
\vspace{-0.75cm}
\caption{The left panel shows the Li synthesis results from our observations with WIYN/Hydra.  The symbols are explained in the figure.  The solid curve and dashed-dot curve are Li isochrones from the standard (no rotation, diffusion, mass loss, etc.) models of P97 and SP14, respectively, for the Hyades age and metallicity.  The right panel shows the full set of Hyades A(Li), including our own WIYN/Hydra data, and the data from T93, BT86, BB88, BC89, T13, and S90 adjusted to our T$_{\rm eff}$ and A(Li) scales as described in the text.  The dashed curve is the fit to the observed G-dwarf Li-depletion trend, fitting only the solid circles and the pluses; that is, only detections with small $\sigma_{\rm Teff}$, that are single stars or stars with no evidence of contamination from a secondary in the spectra.}
\end{figure*}

We see that there is a broad range of roughly 3 dex in A(Li) spanning over 2000 K in temperature.
The well-known general features of the Hyades Li-T$_{\rm eff}$ relation are evident: substantial A(Li) near 7000 K, a severe Li depletion in mid-F dwarfs (6800 to 6300 K, known as the Li ``dip" or ``gap", among other terms found in the literature), higher A(Li) in the Li plateau in late-F/early-G dwarfs (near 6200 K), and the increasingly steep decline in Li with lower T$_{\rm eff}$ in G dwarfs until the Li line becomes undetectable near 5000 K.  These features have been observed repeatedly in the Hyades (e.g., Cayrel et~al.\ 1984; T93) and in older open clusters (for example, in NGC 752: Hobbs \& Pilachowski 1986; in NGC 3680: Anthony-Twarog et~al.\ 2009; and the much older M67: Pasquini et~al.\ 2008; Jones et~al.\ 1999).  We stress that the six Li gap stars with T$_{\rm eff}$ between 6400 and 6800 K show very clear depletion of Li relative to the cooler Li plateau and to the hotter star vB 14.  Our low (and unprecedented) Li detection in vB 90 in the middle of the Li gap, which is a greater than 3$\sigma$ detection with A(Li)=1.70, required extremely high S/N ($\sim$1500) to produce a reliable detection of the weak and rotationally broadened Li line.  In Figure 9 we illustrate vB 90's Li detection through synthesis.  We show three syntheses: one with no Li, A(Li)=1.7, and 2 dex, with 1.70 dex representing our by-eye fit.  This shows the advantages of very high S/N but also the significant challenge of measuring such weak Li in rapid rotators.  In particular, the horizontal line shown at the top that represents the true synthetic continuum of the syntheses, which based on the observed spectrum (that shown and in the surrounding region) cannot be reliably determined without a synthetic reference.  We also note a few apparent mismatches in the spectrum, in particular the too strong synthetic feature near 6704.5 \AA, the possibly too weak synthesis of the blend in the blue wing of the broadening Li feature (near 6707.0 \AA), and the too weak synthesis of the weak features from 6710.5 to 6711.0 \AA.  Most of these cannot be explained by noise but may be the result of T$_{\rm eff}$ errors, minor errors in our adopted linelist, or limitations of our scaled-solar abundance assumption.  Consistent synthetic analyses of our observations of vB 11, vB 13, and vB 37 at similar T$_{\rm eff}$ but with weaker rotational broadening suggest that our synthesis of the blend blueward of Li (near 6707.0 \AA, including the Fe I 6707.45 \AA\, feature) is reliable, but the comparison is more limited by noise.  In either case, for our syntheses of the typically much stronger Li feature in our other stars, spectral noise will dominate these possible issues in our final errors but we acknowledge their importance in vB 90.  This detection, however, encourages re-observation of other Li gap stars that have only upper limits in Li using extremely high S/N.  

Before delving into more detail about the features of the Li morphology, we conduct a preliminary examination of what criteria stars must satisfy to be included as definers of the Li morphology.  We first focus on our WIYN/Hydra sample but we will revisit this issue after we expand the total sample using previously published data.  We begin by assuming that all (WIYN/Hydra) highly probable members with no evidence of binarity, that are radial velocity members, and have small $\sigma_{\rm Teff}$ are included.  We first compare to the three PM members that fall outside the radial velocity peak but show no further evidence for binarity and have small $\sigma_{\rm Teff}$ (vB 2, vB 4, and vB 127).  These are strongly consistent with the very tight G-dwarf Li-T$_{\rm eff}$ trend, so even though these PM members fall outside the Hyades radial velocity peak, and the numbers are limited, it appears that when such stars have small $\sigma_{\rm Teff}$ they can be included as definers of the Li morphology.  The left panel of Figure 10 marks both of these prime data types with solid circles (detections) and solid inverted triangles (upper limits).  We next look at our four observed stars that are listed as binaries in P98 but show no evidence for binarity in our observed spectra, are radial velocity members, and have small $\sigma_{\rm Teff}$ (vB 39, vB 106, vB 114, and vB 142).  Nonetheless, these stars also fall right on top of the very tight G-dwarf Li-T$_{\rm eff}$ trend (the left panel of Figure 10 marks these with pluses).  These data suggest that if known binaries show no evidence of binarity in their spectra from Fourier analysis and $\sigma_{\rm Teff}$ is small, then the effects (if any) of the secondary are negligible in determining T$_{\rm eff}$ and A(Li).  To help illustrate the relatively minor T$_{\rm eff}$ based errors for these G dwarfs with small $\sigma_{\rm Teff}$, the left panel of Figure 10 shows the typical $\sigma_{\rm Teff}$ of 40 K and its resulting affect on A(Li).

In contrast to these G-dwarf types with a tightly consistently Li depletion trend, the one G dwarf with large $\sigma_{\rm Teff}$ falls off of the Li-T$_{\rm eff}$ trend (see vB 49; open circle in the left panel Figure 10).  Although it does not fall very far off, the trend is so tight that this star appears to be clearly discrepant from the trend; even the large $\sigma_{\rm Teff}$ and the resulting effects on $\sigma$A(Li) (illustrated in the left panel of Figure 10 by the diagonal line passing through vB 49) are not enough to bring this star to the trend.  This and the illustrated typical error show that errors in A(Li) that result from errors in T$_{\rm eff}$ go fairly parallel to the Li-T$_{\rm eff}$ trend in this region, so changes in T$_{\rm eff}$ must be very large to bring stars like this back to the Li-T$_{\rm eff}$ trend.  Furthermore, the inclusion of such stars in other clusters might lead to overestimates of the intrinsic scatter around the Li-T$_{\rm eff}$ trend.\footnote[9]{We note the caveat that while vB 49 is both a photometric and radial velocity member, it is one of our three observed Hyades stars without a Hipparcos membership analysis.  However, we are confident in its membership and many G-dwarfs with high $\sigma_{\rm Teff}$ in both our supplemental Hyades analysis (see Section 6.1.3) and Praesepe (see Section 6.2) also show peculiar A(Li).}  This underscores the importance of using multiple color indices to estimate T$_{\rm eff}$ and $\sigma_{\rm Teff}$.  Lastly, note the large A(Li) uncertainties of the other high $\sigma_{\rm Teff}$ stars (the single star vB 11 and the binaries vB 14, and vB 38; similarly illustrated in the left panel of Figure 10 by diagonal lines passing through each data point), and thus the difficulties -- which have often been overlooked -- in determining the precise Li morphology in that region.  The hot side of the Li gap is sometimes portrayed as a steep function of T$_{\rm eff}$ (e.g., BT86; AT09).  Being sure that this is the case requires using multiple stars that have more accurate T$_{\rm eff}$ than vB 38 (and more accurate Li as well).  We will revisit these issues and make a final determination of which stars define the Li morphology after we have supplemented the sample with previous studies.

\subsubsection{Supplemental Hyades Lithium Abundances}

We have supplemented our Hyades A(Li) with previous observations, and our first step was to ensure that all A(Li) we use are on the same scale.  We compared our A(Li) of stars in common with some key previous studies, namely T93, BT86 and BB88, Soderblom et~al.\ (1990, hereafter S90), Burkhart \& Coupry (1989, hereafter BC89), and T13.  For all comparison stars we used the same methods for deriving stellar parameters as we employed for our WIYN/Hydra spectra.  In the case of T93, we used the actual T93 spectra, provided kindly by J. Thorburn (2005, private communication), to perform spectral synthesis using our methods and employing our [Fe/H]-dependent Li COGs.  The T93 spectra have higher resolution (R $\approx$ 32,000) than our WIYN/Hydra spectra, providing us with an opportunity to test for possible systematics introduced through the different spectrographs and spectral reductions; it is for this reason that we intentionally observed 24 of the T93 stars using WIYN/Hydra.  For BT86, BB88, and S90, and BC89 we used the Li 6708 \AA\, equivalent widths from these studies and have rederived A(Li) using our COGs, stellar parameters, and correction of the blended Fe I line at 6707.44 \AA\, in the cases when this was not already done in the published equivalent widths.  Three special cases in BT86 (vB 36, vB 51, and vB 124) were treated differently and will be discussed in more detail in the following paragraph.  We also note that we did not reanalyze stars from the BC89 sample for which we derived a T$_{\rm eff}$ of greater than 7750 K.  Lastly, T13 performed spectral synthesis but did not give corresponding equivalent widths.  Therefore, we used our method of deriving synthetic equivalent widths described at the beginning of Section 6: we applied their abundances and T$_{\rm eff}$ to our COGs to give effective equivalent widths.  Adopting these equivalent widths and \textit{our} stellar parameters gave our rederived T13 A(Li).  

Three special cases in BT86 are vB 36, vB 51, and vB 124.  All three of these are heavily 
depleted stars in the Li gap with large rotational broadening, but BT86 observed high S/N spectra
for all three stars and gives Li detections.  Their characteristics are very similar in nature to vB 
90 (Figure 9), which has comparably weak A(Li) and similar T$_{\rm eff}$ and \textit{v sin i}.  vB 90 is a 
good a reference because at these high \textit{v sin i} ($>$40 km s$^{-1}$) the increase in resolution in BT86 
does not meaningfully improve Li's clarity.  In Figure 9, using the method applied in BT86 of direct 
integration of the Li region yields in the synthetic spectrum with no Li an equivalent width of 9 m\AA.  For such 
weak and broadened Li (spanning more than 2 \AA) this begins to illustrate the challenges with direct 
application of the published equivalent width measurements in BT86 to derive self-consistent A(Li) on 
our scale.  When creating comparable synthetic spectra for vB 36, vB 51, and vB 124 by applying our 
derived atmospheric parameters and each star's rotational broadening convolved with the resolution of 
BT86, the non-Li contributions are even stronger at 10 to 11 m\AA.  These are larger than both BT86's 
estimate that the Fe contribution here is only $\sim$2 m\AA\, and their reported Li+Fe(6707.45 \AA) 
measurements of 5.9, 6.5, and 9 m\AA, respectively.  We suspect BT86 placed their continua below those 
given by our syntheses.  This is very easy to do because the highest peaks of, for example, vB 90's 
synthetic spectrum (between 6700.0 and 6702.5 \AA) are at absorption levels $\sim$0.998 and thus lie 
below the synthetic continuum (of 1.000); in other words, no portions of the metal-rich and 
rotationally-broadened spectra (near the Li feature) are pure continuum.  Without direct synthesis of 
the actual BT86 spectra, it is very difficult to know what A(Li) we would derive on our scale.

Nevertheless, we attempt to derive what information we can from the published detections.
If in each star's synthetic spectrum we instead measure the equivalent widths relative to the nearby 
spectral peak ($\sim$6706.1 \AA) and not the real continuum, we find non-Li components for the Li 
regions in vB 36, vB 51, and vB 124 at 3, 6.3, and 8 m\AA, respectively.  This large variation is the 
result of the differences in rotational broadenings between these three stars.  The reference continuum 
peak approaches the real continuum as the rotational broadening decreases, resulting in a larger apparent 
measurement.  All three of these non-Li components are below the measurements of Li+Fe published in BT86 of 5.9, 
6.5, and 9 m\AA, respectively.  Subtracting the non-Li components from these published equivalent widths 
gives the Li-only component.  Applying the D93 formula to the Li-only component, and taking into account 
each star's rotational broadening, results in none of the three having detections at the 3$\sigma$ level, 
though vB 36 comes close at 2.7$\sigma$.  (A similar analysis of our observations of vB 90 finds a Li-only 
component of 3.9 m\AA, a 6.6$\sigma$ detection).  Therefore, while BT86 publishes detections for these 
three stars, we are forced to adopt appropriate 3$\sigma$ upper limits.

\begin{figure}[htp]
\begin{center}
\includegraphics[scale=0.41]{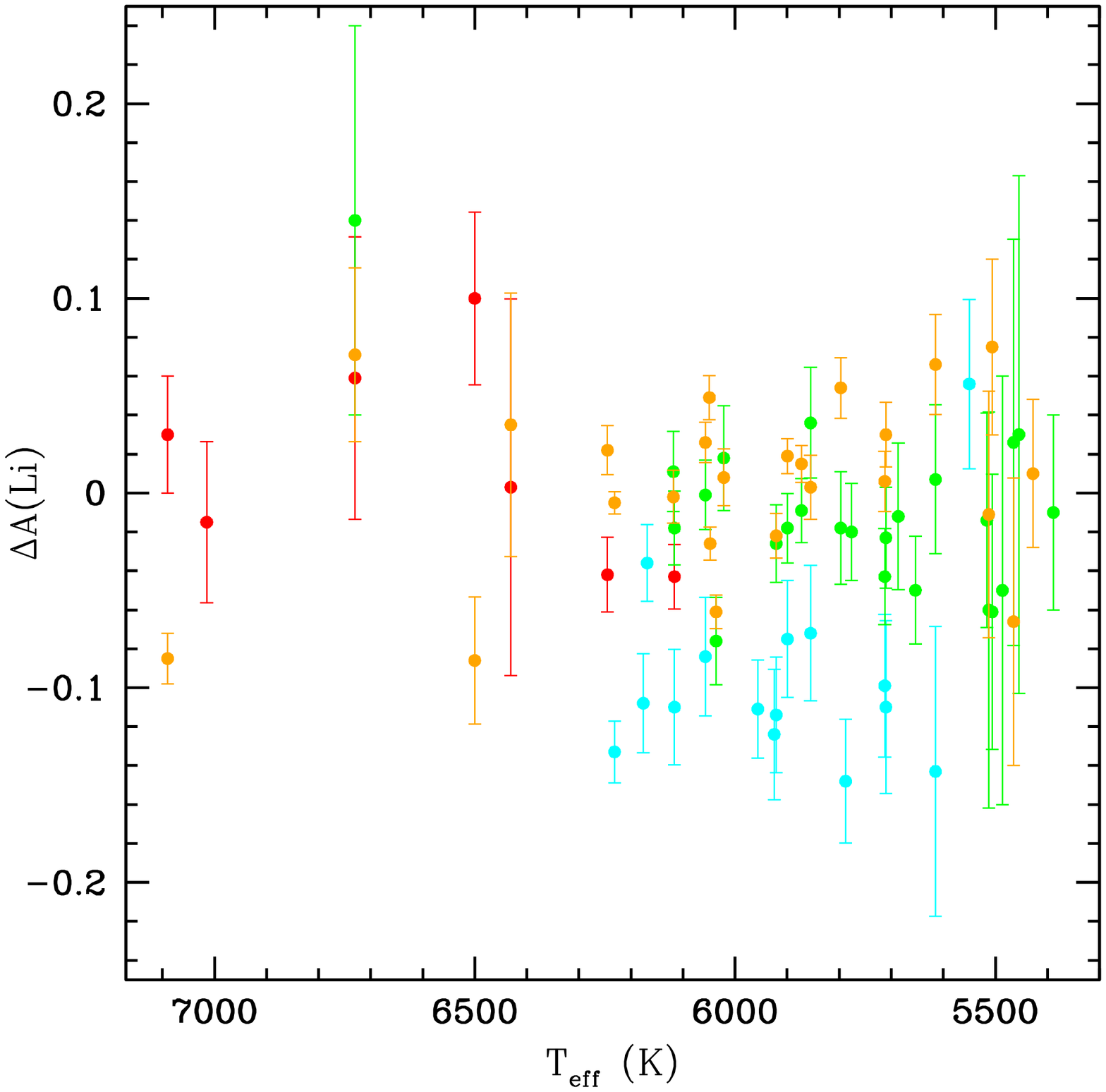}
\end{center}
\vspace{-0.45cm}
\caption{Comparisons of our Hyades A(Li) from WIYN synthesis to the A(Li) from our synthesis of T93's spectra are shown in green.  Similarly, comparisons of our WIYN synthesis to abundances that we rederived from the published equivalent widths of BB88 and BT86 are shown in red, for S90 are shown in light blue, and for T13 in orange.  The differences shown are the abundances rederived from their data minus our new abundances.  Error bars are based on the S/N based $\sigma$s for both compared measurements added in quadrature.  Point types are the same as those used in the right panel of Figure 10.  There is no significant offset between our results and those of T93, BB88, BT86, and T13.  The systematic offset for S90, however, is significant.}
\end{figure}

Figure 11 shows a comparison of A(Li) for stars in common from our WIYN/Hydra sample versus the A(Li) rederived from the above studies using our parameters and synthesis as described above.  The vertical axis is their (rederived by us) A(Li) minus our WIYN/Hydra A(Li), and the horizontal axis is the T$_{\rm eff}$ derived using our methods.  Comparison with T93 (green) shows no significant systematic difference, with our A(Li) being 0.010$\pm$0.009 dex higher, and with no significant trends with either A(Li) or T$_{\rm eff}$.  This strongly suggests that no significant systematic effects have been introduced by the use of different telescopes and instrumental configurations.  Comparison with BT86 and BB88 (red) again shows no significant systematic difference, with their rederived A(Li) being 0.013$\pm$0.022 dex higher, and with no significant trends with either A(Li) or T$_{\rm eff}$.  Similarly, in comparison to T13 (orange) our A(Li) and including also our reanalysis of the T93 spectra (to increase the number of stars compared) there is no trend and the difference between our abundances are not significant with their rederived A(Li) being 0.005$\pm$0.009 dex higher.  Comparison with S90 (light blue) to our A(Li) and again including our reanalysis of the T93 spectra does show a systematic difference of 0.094$\pm$0.013 dex with theirs being lower than ours, with no significant trends with either A(Li) or T$_{\rm eff}$.  This is a result of S90's equivalent widths being systematically 9.2$\pm$1.3 m\AA\, lower.  Lastly, our comparison to the abundances from BC89 is not shown because we only have vB 38 in common.  Therefore, while there is no meaningful difference in the measured equivalent widths for vB 38, we cannot reliably test for systematics.  Possible systematics are not a major concern, however, because the BC89 sample only includes A dwarfs that are predominantly binaries with large $\sigma_{\rm Teff}$, which will not affect our conclusions.  

Based on these comparisons, we thus compile our augmented sample by taking the \textit{rederived} A(Li) from the BT86, BB88, T93, BC89, and T13 studies as is, but by increasing the rederived S90 abundances by 0.09 dex.  We are thus confident that the augmented sample is on the same self-consistent A(Li) and T$_{\rm eff}$ scale (ours), and can be used reliably in interpreting the Hyades A(Li).  For stars observed in more than one study, we give priority to the A(Li) from our own WIYN/Hydra data/analysis first, followed by T93, BB88, BT86, T13, S90, and BC89 in that order. 

Table 6 shows the vB star ID, the mean B-V, the derived T$_{\rm eff}$, the $\sigma_{\rm Teff}$, the Li equivalent width (EqW), and $\sigma$ equivalent width based on Poisson noise in Columns 1-6.  For stars from our WIYN/Hydra data or the data from T93 and T13, these equivalent widths are based on our method of deriving them synthetically.  For all other sources the directly measured equivalent widths are given with subtraction of the Fe I line at 6707.44 \AA, if this was not already performed in the original publication.  The measured systematic between our analysis and S90 was also applied to the equivalent width of vB 88.  For the several stars that have no given equivalent width, this is because their synthetic A(Li) was below the lower limit of our COGs.  Similarly, stars with no equivalent width $\sigma$ had no published S/N.  Following these are the derived A(Li), the $\sigma$A(Li) based on Poisson noise only, and the $\sigma$A(Li) based on $\sigma_{\rm Teff}$ (Columns 7-9); the S/N per pixel (placed on a uniform scale), the total exposure time, the derived \textit{v sin i}, and the derived radial velocity and $\sigma$\textit{v$_{rad}$} (Columns 10-13).  Next is the source of the data used to derive the abundance (Column 14), whether that be our own WIYN/Hydra analysis or our rederived A(Li) from a previous study.  

In our analysis of Li depletion in the Li gap, a star's \textit{v sin i} is a key aspect, but BT86 does not give \textit{v sin i} values for their observed stars.  Therefore, we have looked at the \textit{v sin i} values from Kraft (1965), which is also the source of the given \textit{v sin i} values in BB88.  To test for systematic differences we have compared the 10 Hyades stars that were analyzed both in our WIYN/Hydra observations and in Kraft (1965).  We found that there are meaningful differences that can be well defined by the relation of \textit{v sin i}=\textit{v sin i}$_{Kraft}$$\times$1.12+6.0 km s$^{-1}$.  The source of this systematic is unknown, but for our analysis of the broad range of \textit{v sin i} in the Li gap, the relative velocities are more important than the absolute velocities.  Therefore, our final adopted \textit{v sin i} values (see Table 6) for the stars from both BT86 and BB88 are based on those from Kraft (1965) but shifted to the same scale as ours using this relation.

Lastly, we note three photometric limitations for a small number of these supplemental Hyades stars.  For all five T13 supplemental stars there was no available photometry from Johnson \& Knuckles (1955) or Mend67.  Therefore, we adopted their Tycho II B-V colors (H{\o}g et~al.\ 2000).  To place the Tycho II B-V colors, which are not equivalent to Johnson B-V, on the same scale as our photometry, we matched the full Hyades sample's colors from Johnson \& Knuckles (1955) to their corresponding Tycho II colors.  This created a simple quadratic relation with low scatter for stars of Johnson B-V color of 0.9 and bluer, well within the color range for these six.  However, due to this limited color information, the $\sigma_{\rm Teff}$ in Table 6 for these six stars are marked as '-'.  Similarly, for one T93 star (vB 12) there is only published Cousins R and I photometry (Taylor et~al.\ 2008).  Comparison of the Taylor et~al.\ (2008) Cousins R-I colors to the Johnson R-I colors of Mend67 finds a quadratic relations with low scatter between the two color sets, and using the same transformations as in Section 4 we can consistently convert this to an effective B-V.  Lastly, we note that 13 of the stars we analyzed from T93, BT86 and BB88 have clearly discrepant photometry in comparison to the Hyades MS fiducial in M$_V$ space.  Therefore, their T$_{\rm eff}$ was not determined using our derived color relations because they would not be appropriate.  Their final T$_{\rm eff}$ are thus based only on their Johnson \& Knuckles (1955) B-V and they have increased uncertainty, and in Table 6 their $\sigma_{\rm Teff}$ are marked as '*'. These 19 stars will be grouped together with those that have $\sigma_{\rm Teff}$ $>$ 75 (``large" $\sigma_{\rm Teff}$) for interpretation purposes. 

\subsubsection{Final Hyades Lithium Sample}

The right panel of Figure 10 shows the final augmented sample, using the same symbols as in the left panel, but differentiating between studies using color.  Our WIYN/Hydra data are in black, T93 are in green, BT86 and BB88 are in red, S90 is in light blue, BC89 are in dark blue, and T13 are in orange.  SPTLBs (discussed below) are shown as magenta open diamonds.  We also remind the reader that in the left panel of Figure 3 we identified stars with low $\sigma_{\rm Teff}$ that photometrically deviated from the Hyades single star fiducial and marked them with overlaid yellow data points.  These two stars (vB 102 and vB 114) are also marked here with overlaid yellow data points.  

A few stars that appeared in the earlier Li studies, namely vB 9 (T93), vB 61 (BB88), and vA 771 (T93), are not shown because they are considered to be non-members by P98.  When considering errors, vB 61 would have been on the high side of or slightly above the Li plateau, whereas vA 771 would have been on the low side of or slightly below the Li plateau, giving the appearance that there is more intrinsic spread in the Li plateau than is really there.  vB 9 would have been well below the tight G-dwarf Li-T$_{\rm eff}$ trend.  The inclusion of these non-members can obfuscate what might otherwise be clear trends.  We will no longer consider these three stars.  We have also excluded stars previously identified as SB2s in T93 (vB 29, vB 57, vB 58, vB 75, vB 120, vB 122, vB 140, and vB 162), except for the special cases of SPTLBs (vB 22, vB 62, vB 121, BD+22$^{o}$669, and BD+22$^{o}$635).  In SB2s, the secondary can contribute a significant flux to the total observed spectrum.  T93 studied 13 SB2s and tried to determine A(Li) for the primaries after correcting for the flux contributed by the secondary.  Not surprisingly, many of the SB2s in T93 deviated from the Li-T$_{\rm eff}$ trend, possibly reflecting how difficult this flux-correction process is, but also how difficult it is to assign an appropriate temperature to the primary.  We also note that all of the T93 SB2s have large $\sigma_{\rm Teff}$, further establishing the effectiveness of this detailed photometric temperature analysis in identifying binaries. 

\LTcapwidth=\textwidth
\begin{center}
\renewcommand{\baselinestretch}{1.1}
{\footnotesize \begin{longtable*}{p{1.1cm} c c c c c c c c c c c c c c}
\multicolumn{15}{c}%
{{\bfseries \tablename\ \thetable{} - Stellar Lithium Data for The Hyades}} \\
\hline
ID & B-V & T$_{\rm eff}$ & $\sigma_{\rm Teff}$ & EqW & $\sigma$EqW & A(Li) & $\sigma$Li$_{S/N}$ & $\sigma$Li$_{\rm Teff}$ & S/N &  Exp. & \textit{v sin i} & \textit{v$_{rad}$} & Ref & Prime\\
\hline
\tiny vB &  &   \tiny(K)       &       \tiny(K)           &\tiny(m\AA)& \tiny(m\AA)     &       &                    &                         &     &  \tiny(s)  & \tiny(km s$^{-1}$)    &  \tiny(km s$^{-1}$)             &     &             \\
\endfirsthead
\multicolumn{15}{c}%
{{\bfseries \tablename\ \thetable{} -- continued from previous page}} \\
\hline
ID & B-V & T$_{\rm eff}$ & $\sigma_{\rm Teff}$ & EqW & $\sigma$EqW & A(Li) & $\sigma$Li$_{S/N}$ & $\sigma$Li$_{\rm Teff}$ & S/N &  Exp. & \textit{v sin i} & \textit{v$_{rad}$} &  Ref & Prime\\
\hline
\tiny vB &  &   \tiny(K)       &      \tiny(K)           &\tiny(m\AA)& \tiny(m\AA)     &       &                    &                         &     &  \tiny(s)  & \tiny(km s$^{-1}$)    &  \tiny(km s$^{-1}$)              &     &             \\
\hline
\hline
\endhead
\hline
\hline
  1 & 0.567 & 6048.1  &  55.9 & 102.23  &  1.16  &  2.945  &  0.008  &  0.052 & 325 & -  & 7  & 32.44$\pm$0.21 & 2 & \checkmark \\
  2 & 0.619 & 5872.6  &  20.3 &  93.71  &  1.35  &  2.72   &  0.009  &  0.020 & 396 & 360& 11 & 32.78$\pm$0.42 & 1 & \checkmark \\
  6 & 0.327 & 7013.7  &  65.6 &  52.30  &  1.85  &  3.28   &  0.018  &  0.058 & 535 & -  & 62 & 35.0$\pm$2.5  & 3 & \checkmark \\
 10 & 0.576 & 6025.5  &  68.5 &  80.94  &  1.83  &  2.77   &  0.014  &  0.075 & 287 & 350& 10 & 37.80$\pm$0.28 & 1 & \checkmark \\ 
 11 & 0.394 & 6730.1  &  92.3 &  13.62  &  1.15  &  2.39   &  0.042  &  0.066 & 610 & 960& 30 & 36.52$\pm$1.09 & 1 & \\
 12 & 0.914 & 4954.2  &    -  &    -    &  1.33  &  0.41   &    -    &    -   & 283 & -  & 6  & 41.49$\pm$0.30 & 2 & \\
 15 & 0.665 & 5712.5  &  46.8 &  72.30  &  1.72  &  2.395  &  0.015  &  0.050 & 330 & 400& 15 & 38.16$\pm$0.26 & 1 & \checkmark \\
 17 & 0.694 & 5614.9  &  56.4 &  44.48  &  1.76  &  1.98   &  0.025  &  0.072 & 311 & 450& 13 & 38.60$\pm$0.27 & 1 & \checkmark \\
 18 & 0.640 & 5797.0  &  20.3 &  80.06  &  1.91  &  2.54   &  0.015  &  0.020 & 272 &1200& 9  & 38.78$\pm$0.36 & 1 & \checkmark \\ 
 19 & 0.517 & 6245.0  &  53.6 &  80.68  &  1.56  &  2.96   &  0.012  &  0.039 & 368 & 200& 16 & 37.97$\pm$0.90 & 1 & \checkmark \\
 20 & 0.398 & 6713.6  &  40.4 &  83.97  &  5.93  &  3.33   &  0.045  &  0.041 & 176 & -  & 68 & 36.4$\pm$1.2  & 3 & \checkmark \\
 26 & 0.741 & 5465.1  &  15.2 &  30.33  &  1.79  &  1.48   &  0.071  &  0.031 & 297 & 420& 11 & 38.88$\pm$0.33 & 1 & \checkmark \\
 27 & 0.728 & 5506.3  &  21.8 &  33.08  &  1.70  &  1.63   &  0.044  &  0.030 & 307 & 450&$<$6& 38.54$\pm$0.28 & 1 & \checkmark \\
 31 & 0.567 & 6057.2  &  32.4 &  87.87  &  1.35  &  2.855  &  0.010  &  0.033 & 403 & 350& 13 & 37.71$\pm$0.30 & 1 & \checkmark \\
 42 & 0.765 & 5388.7  &  43.0 &    -    &  1.96  &  1.13   &  -      &  0.043 & 276 &1700& 12 & 38.30$\pm$0.27 & 1 & \checkmark \\
 44 & 0.462 & 6456.7  &  22.7 &  15.32  &  1.40  &  2.235  &  0.049  &  0.030 & 570 & -  & 40 & 35.9$\pm$0.5  & 3 & \checkmark \\
 48 & 0.520 & 6231.7  &  10.5 &  91.40  &  2.15  &  3.03   &  0.015  &  0.021 & 198 & -  & 12 & 38.94$\pm$0.13 & 2 & \checkmark \\
 49 & 0.613 & 5893.7  & 116.0 &  61.46  &  1.86  &  2.48   &  0.018  &  0.118 & 283 & 300& 10 & 39.68$\pm$0.36 & 1 & \\
 50 & 0.604 & 5924.9  &  53.7 &  82.73  &  1.37  &  2.69   &  0.011  &  0.052 & 390 &2340& 11 & 41.55$\pm$0.46 & 1 & \checkmark \\
 52 & 0.611 & 5899.4  &  40.1 &  80.67  &  1.04  &  2.65   &  0.008  &  0.039 & 528 &1200& 13 & 38.18$\pm$0.33 & 1 & \checkmark \\
 59 & 0.551 & 6116.4  &  22.5 &  82.30  &  1.51  &  2.87   &  0.011  &  0.021 & 354 & 200& 11 & 39.06$\pm$0.78 & 1 & \checkmark \\
 64 & 0.666 & 5710.5  &  34.0 &  64.24  &  1.68  &  2.32   &  0.016  &  0.037 & 318 &1200& 11 & 39.14$\pm$0.34 & 1 & \checkmark \\
 65 & 0.535 & 6176.3  &   *   & 103.36  &  2.10  &  3.07   &  0.014  &    -   & 198 & -  & 10 & 39.32$\pm$0.24 & 2 & \\
 66 & 0.551 & 6118.3  &  39.8 &  75.70  &  1.64  &  2.82   &  0.013  &  0.037 & 328 & 400& 12 & 39.31$\pm$0.74 & 1 & \checkmark \\
 73 & 0.605 & 5921.0  &  32.2 &  94.93  &  1.59  &  2.775  &  0.011  &  0.031 & 354 & 350& 15 & 39.89$\pm$0.44 & 1 & \checkmark \\
 76 & 0.752 & 5427.8  &  50.9 &  14.79  &  0.89  &  1.24   &  0.038  &  0.049 & 354 & -  & 3  & 39.02$\pm$0.17 & 2 & \checkmark \\
 78 & 0.451 & 6500.2  &  28.6 &  22.74  &  1.34  &  2.48   &  0.030  &  0.024 & 533 &1800& 32 & 39.18$\pm$1.02 & 1 & \checkmark \\
 86 & 0.469 & 6430.7  &  28.6 &  15.32  &  1.87  &  2.21   &  0.066  &  0.029 & 378 & 200& 31 & 39.90$\pm$1.02 & 1 & \checkmark \\
 87 & 0.734 & 5486.5  &  44.5 &  26.48  &  1.90  &  1.38   &  0.078  &  0.057 & 275 & 600& 7  & 40.00$\pm$0.37 & 1 & \checkmark \\
 88 & 0.542 & 6148.7  &  13.6 &  83.09  &  -     &  2.97   &  -      &  0.024 & 120 & -  &  - & 40.98$\pm$0.31 & 4 & \checkmark \\
 90 & 0.418 & 6635.0  &  70.0 &   3.90  &  0.59  &  1.70   &  -      &  0.061 & 1500&10945&50 & 39.76$\pm$0.90 & 1 & \checkmark \\
 92 & 0.744 & 5454.7  &  16.4 &  13.00  &  2.47  &  1.24   &  0.094  &  0.033 & 212 & 300& 10 & 40.29$\pm$0.39 & 1 & \checkmark \\
 97 & 0.624 & 5855.1  &  32.8 &  85.80  &  2.04  &  2.645  &  0.016  &  0.032 & 269 & 400& 13 & 40.82$\pm$0.47 & 1 & \checkmark \\
105 & 0.569 & 6049.2  &  30.2 &  84.42  &  1.38  &  2.815  &  0.010  &  0.035 & 240 & -  & 6  & 39.86$\pm$0.29 & 2 & \checkmark \\
118 & 0.573 & 6036.4  &  41.0 & 101.63  &  1.29  &  2.93   &  0.008  &  0.044 & 407 &1440& 10 & 41.09$\pm$0.46 & 1 & \checkmark \\
127 & 0.726 & 5512.1  &  65.1 &  31.54  &  2.19  &  1.6    &  0.062  &  0.114 & 269 & 920& 18 & 43.16$\pm$0.27 & 1 & \checkmark \\
128 & 0.436 & 6559.8  &  87.1 &  12.26  &  0.88  &  2.205  &  0.037  &  0.073 & 838 & -  & 34 & 42.5$\pm$1.5  & 3 & \\ 
187 & 0.762 & 5398.7  &   7.6 &  18.26  &  1.90  &  1.325  &  0.059  &  0.020 & 198 & -  & 5  & 43.51$\pm$0.19 & 2 & \checkmark \\
\tiny HD35768  & 0.555 & 6102.0  &  - & 60.41 & -&  2.670  &      -  &     -  &  -  & -  & 6  & 42.10$\pm$0.43 & 6 \\
\tiny HD14127  & 0.563 & 6074.0  &  - & 68.13 & -&  2.715  &      -  &     -  &  -  & -  & 7  & 26.40$\pm$0.32 & 6 \\
\tiny HD240648 & 0.748 & 5440.0  &  - & 42.85 & -&  1.716  &      -  &     -  &  -  & -  & 5  & 42.31$\pm$0.18 & 6 \\
\tiny HD19902  & 0.730 & 5498.0  &  - & 30.58 & -&  1.520  &      -  &     -  &  -  & -  & 4  & 27.27$\pm$0.22 & 6 \\
\tiny HD242780 & 0.761 & 5401.0  &  - & 30.21 & -&  1.308  &      -  &     -  &  -  & -  & 5  & 43.70$\pm$0.23 & 6 \\
\hline
\hline
\multicolumn{15}{l}{{Spectroscopic Binaries based on Perryman et~al.\ (1998; and references therein)}}  \\
\hline
 14    & 0.310 & 7090.1  & 273.8 &  60.59  &  1.22  &  3.41   &  0.012  &  0.190 & 455  & 720 & 14 & 37.25$\pm$0.79 & 1 &\\
 22$^b$& 0.714 & 5550.3  &   *   &  56.00  &  1.28  &  2.049  &  0.017  &    -   & 220  & -   & 9  & 38.18$\pm$0.13 & 2 &\\ 
 38    & 0.327 & 7015.0  & 373.9 &  17.61  &  1.51  &  2.73   &  0.015  &  0.269 & 467  & 360 & 30 & 64.53$\pm$1.36 & 1 &\\
 39    & 0.683 & 5653.7  &  48.8 &  49.81  &  1.06  &  2.1    &  0.014  &  0.058 & 500  & 1800& 11 & 38.75$\pm$0.27 & 1 & \checkmark \\
 40    & 0.563 & 6072.6  &   *   & 105.00  &  1.66  &  3.01   &  0.013  &    -   & 226  & -   & -  & 37.4$\pm$2.9   & 2 &\\ 
 45    & 0.296 & 7150.2  &   *   &  22.89  &  1.37  &  2.95   &  0.029  &    -   & 400  & -   & 13 & 37.7$\pm$0.3   & 5 &\\
 62$^b$& 0.537 & 6168.9  &   *   & 121.06  &  1.18  &  3.18   &  0.008  &    -   & 220  & -   & 6  & 38.77$\pm$0.14 & 2 &\\ 
 63    & 0.643 & 5787.7  &  29.9 &  76.21  &  1.76  &  2.50   &  0.014  &  0.035 & 212  & -   & 8  & 39.39$\pm$0.31 & 2 &\checkmark \\ 
 69    & 0.746 & 5447.7  &   *   &  15.26  &  2.04  &  1.285  &  0.077  &    -   & 184  & -   & 7  & 39.91$\pm$0.08 & 2 &\\ 
 77    & 0.519 & 6238.1  &  46.7 &  23.00  &  1.92  &  2.255  &  0.045  &  0.046 & 385  & -   & 34 & 39.90$\pm$0.11 & 3 &\checkmark \\
 81    & 0.479 & 6389.2  &  48.4 &  10.04  &  0.98  &  2.01   &  0.040  &  0.047 & 670  & -   & 26 & 38.0$\pm$2.5   & 3 &\checkmark \\
 83    & 0.253 & 7342.6  &  58.6 &  15.91  &  1.68  &  2.915  &  0.051  &  0.042 & 400 &  -   &26.7& 39.56$\pm$0.23 & 5 &\checkmark \\
102$^a$& 0.595 & 5956.3  &  48.3 &  84.99  &  1.87  &  2.73   &  0.014  &  0.050 & 198  & -   & 8  & 42.00$\pm$0.33 & 2 & \\ 
106    & 0.646 & 5776.0  &  70.2 &  78.66  &  2.17  &  2.51   &  0.017  &  0.071 & 239 & 480  & 9  & 39.89$\pm$0.38 & 1 &\checkmark \\
113    & 0.561 & 6079.5  &  39.2 &  88.08  &  1.77  &  2.88   &  0.013  &  0.041 & 212  & -   & 8  & 42.47$\pm$0.11 & 2 &\checkmark \\ 
114$^a$& 0.724 & 5516.6  &  18.6 &  38.99  &  1.87  &  1.77   &  0.035  &  0.027 & 311 & 1200 & 17 & 40.01$\pm$0.37 & 1 & \\
119    & 0.559 & 6087.3  &   *   &  69.52  &  2.61  &  2.735  &  0.022  &    -   & 170  & -   & 11 & 41.40$\pm$0.16 & 2 &\\ 
121$^b$& 0.498 & 6316.0  &  17.8 & 108.77  &  0.87  &  3.205  &  0.008  &  0.026 & 670  & -   & 19 & 42.74$\pm$0.17 & 3 &\\
142    & 0.673 & 5686.5  &  43.8 &  57.85  &  2.11  &  2.23   &  0.023  &  0.049 & 256  & 360 & 12 & 42.00$\pm$0.38 & 1 &\checkmark \\
182    & 0.844 & 5151.1  &   *   &  11.45  &  0.98  &  0.55   &    -    &    -   & 270  & 382   & 7  & 40.8$\pm$0.2  & 2 &\\ 
\tiny BD+22$^{o}$669$^b$&0.940 & 4886.2&-&23.0&  1.25  &  0.85   &  0.030  &    -   & 220     & -  &   -    &  -   & 2 &\\
\tiny BD+22$^{o}$635$^b$&1.000 & 4733.6&-&13.0&  2.80  &  0.60   &  0.070  &    -   & 130     & -  &   -    &  -   & 2 &\\
\hline
\hline
\multicolumn{15}{l}{{3$\sigma$ Upper Limits}}  \\
\hline 
  4 & 0.840 & 5163.8  &  13.1 &  6.31   &  -     &  0.55  &    -    &    -   & 247  & 550 & 10 & 31.79$\pm$0.38 & 1 & \checkmark \\
  8 & 0.404 & 6691.0  &  39.5 &  7.31   &  -     &  2.01  &    -    &    -   & 408  & -   & 62 & 39.1$\pm$1.1  & 3 & \checkmark \\
 13 & 0.411 & 6663.5  &  34.2 &  4.73   &  -     &  1.85  &    -    &    -   & 404  & 150 & 24 & 36.67$\pm$0.96 & 1 & \checkmark \\
 21 & 0.819 & 5224.5  &  18.0 &  5.04   &  -     &  0.70  &    -    &    -   & 297  & 600 &$<$6&37.88$\pm$0.23 & 1 & \checkmark \\
 36 & 0.434 & 6568.6  &  51.9 &  3.36   &  -     &  1.85  &    -    &    -   & 804  & - & 51   & 40.8$\pm$2.4  & 3 & \checkmark \\
 37 & 0.408 & 6673.0  &  45.0 &  5.43   &  -     &  1.93  &    -    &    -   & 332  & 150& 20  & 38.90$\pm$0.77 & 1 & \checkmark \\
 46 & 0.857 & 5114.6  &  20.7 &  4.00   &  -     &  0.55  &    -    &    -   & 170  & - & 5    & 40.29$\pm$0.06 & 2 & \checkmark \\
 51 & 0.452 & 6496.7  &  57.7 &  3.76   &  -     &  1.85  &    -    &    -   & 637  & - & 40   & 40.1$\pm$0.6  & 3 & \checkmark \\
 79 & 0.820 & 5220.4  &  28.2 &  4.05   &  -     &  1.15  &    -    &    -   & 340  & - & 6    & 40.7$\pm$0.06 & 2 & \checkmark \\
 85 & 0.435 & 6566.6  &  49.2 &  10.02  &  -     &  2.10  &    -    &    -   & 312  & - & 68   & 40.9$\pm$1.3  & 3 & \checkmark \\
 93 & 0.882 & 5042.9  &  36.1 &  5.81   &  -     &  0.55  &    -    &    -   & 269  & 600 & 10 & 40.38$\pm$0.37 & 1 & \checkmark \\
 94 & 0.422 & 6617.5  &  22.3 &  3.22   &  -     &  1.64  &    -    &    -   & 838  & - & 51   & 36.9$\pm$0.9  & 3 & \checkmark \\
 99 & 0.861 & 5101.2  &  20.1 &  4.00   &  -     &  0.55  &    -    &    -   & 382  & - & 5    & 41.5$\pm$0.1  & 2 & \checkmark \\
109 & 0.806 & 5262.9  &  31.4 &  5.00   &  -     &  0.70  &    -    &    -   & 198  & - & 6    & 41.34$\pm$0.16 & 2 & \checkmark\\
116 & 0.833 & 5184.5  &  40.0 &  6.19   &  -     &  1.00  &    -    &   -    & 249  & - & 6    & 41.62$\pm$0.15 & 2 & \checkmark \\
143 & 0.527 & 6207.9  &  36.4 &  5.00   &  -     &  1.53  &    -    &    -   & 198  & - & 10   & 42.92$\pm$0.19 & 2 & \checkmark \\
153 & 0.853 & 5124.3  &  12.5 &  3.07   &  -     &  0.55  &    -    &    -   & 354  & - & 6    & 26.62$\pm$0.21 & 2 & \checkmark\\
178 & 0.845 & 5149.6  &  58.7 &  3.28   &  -     &  0.55  &    -    &    -   & 354  & - & 4    & 40.94$\pm$0.08 & 2 & \checkmark \\
180 & 0.858 & 5109.7  &  13.2 &  6.61   &  -     &  0.55  &    -    &    -   & 156  & - & 5    & 40.97$\pm$0.06 & 2 & \checkmark \\
\hline
\hline
\multicolumn{15}{l}{{Spectroscopic Binary 3$\sigma$ Upper Limits}}  \\
\hline
 91 & 0.883 & 5040.4  &   *   &   4.00  &  -     &  1.00  &    -    &    -   & 297  & - & -    &  -     & 2 &\\
 96 & 0.841 & 5159.8  &   *   &   6.00  &  -     &  0.55  &    -    &    -   & 212  & - & 9    & 37.6$\pm$1.2  & 2 &\\ 
101 & 0.433 & 6573.5  &  86.6 &   5.92  &  -     &  1.85  &    -    &    -   & 456  & - & 51   & 33.7$\pm$1.2  & 3 &\\ 
115 & 0.843 & 5154.0  &   *   &   3.00  &  -     &  0.55  &    -    &    -   & 382  & - & 5    & 41.84$\pm$0.44 & 2 &\\ 
124 & 0.497 & 6320.6  &   *   &   8.69  &  -     &  1.75  &    -    &    -   & 256  & - & 34   & 39.83$\pm$0.24 & 3 &\\
130 & 0.242 & 7392.1  &   *   &   4.10  &  -     &  2.24  &    -    &    -   & 400  & - & 13   & 44.16$\pm$0.14 & 5 &\\
\hline
\caption[Hyades stellar Li data]{An (a) superscript marks stars that deviate from the single-star fiducial (see Figure 3).  A (b) superscript marks SPTLBs.  Ref: (1) Our WIYN/Hydra sample, (2) Thorburn et~al.\ (1993), (3) Boesgaard \& Budge (1988) and Boesgaard \& Tripicco (1986), (4) Soderblom et~al.\ (1990), (5) Burkhart \& Coupry (1989), and (6) Takeda et~al.\ (2013).  The source for radial velocities of stars we did not observe is Perryman et~al.\ (1998) (and references therein) and for vB 99 and vB 182 it is Griffin et~al.\ (1988).  The source for \textit{v sin i} of the Boesgaard \& Tripicco (1986) and Boesgaard \& Budge (1988) stars (3) are Kraft (1965) but scaled to our measurements.  The source for \textit{v sin i} of the Burkhart \& Coupry (1989) stars (5) are Debernardi et~al.\ (2000).}
\end{longtable*}}
\end{center}

\vspace{-0.8cm}
The same general Li morphology that was evident in our WIYN/Hydra sample is also reflected in the augmented total sample: a substantial A(Li) for late A/early F dwarfs, a Li gap for mid-F dwarfs, a Li plateau for late F/early G dwarfs, and a Li decline for later G (and K) dwarfs.  Taking these in turn: two more late A/early F detections with low $\sigma_{\rm Teff}$ help define the high-Li high side of the Li gap, while additional detections and upper limits in the Li gap, most of which have low $\sigma_{\rm Teff}$, help define the Li gap morphology.  It remains unclear whether the hot side of the Li gap has as clearly defined of a steep Li-T$_{\rm eff}$ relation as previously suggested in BT86 and AT09.  It is also curious that the three P98 binaries on the cool side of the Li gap (red plus signs) all significantly deviate from the trend suggested by the single stars (solid circles); all five stars have low $\sigma_{\rm Teff}$ but the Hyades sample alone remains too small to draw meaningful conclusions about the nature of Li depletion in the gap for well behaved single stars versus binaries.  

Three stars that have been previously used to help define the Li-T$_{\rm eff}$ morphology of the Li gap have high $\sigma_{\rm Teff}$.  These three are vB 128 (an apparently single star), vB 34 (an SB2), and vB 124 (a triple system).  BT86 and BB88 derived A(Li) for both components of vB 34, and assumed a T$_{\rm eff}$ for the primary of vB 124 based on its spectral type, but the assumed T$_{\rm eff}$ are rather uncertain; thus, their usefulness in defining the morphology of the Li gap is limited.  For example, using the B-V color of vB 124, we find T$_{\rm eff}$=6320 K, which is 310 K lower than BT86's value of 6630 K.  BT86's value may be more accurate if the redder companions make the overall B-V color redder than that of the primary alone.  However, the primary's T$_{\rm eff}$ remains uncertain, and in a region where the Li-T$_{\rm eff}$ trend is a steep function of T$_{\rm eff}$, it is best to exclude stars with high $\sigma_{\rm Teff}$.  For similar reasons we also do not analyze the two components of vB 34.  In Section 7.2 we look more closely at the complex Hyades Li gap in comparison to the similar Praesepe Li gap.

Very striking is the enormous range in A(Li) of nearly 2 dex (or perhaps more) at the hot edge of the Li plateau near 6250 K.  At the high end are the two SPTLBs (see discussion below) with A(Li) near 3.2, the Li plateau itself with
A(Li) near 3.0, and well below that is vB 77 (red plus) with A(Li)=2.31, and lastly, well below that is vB 143 with A(Li)$<$1.53.  If this enormous range in A(Li) at virtually identical T$_{\rm eff}$ is real, it presents quite a challenge for models to explain.  The high reliability of the SPTLBs and the Li plateau are discussed below and above, respectively.  vB 77 is a radial velocity member that meets some of our high-reliability criteria such as small $\sigma_{\rm Teff}$ but P98 labeled it a binary; a Fourier cross-correlation analysis to detect a possible companion would be useful.  T93 discuss various membership indicators for vB 143 and conclude that this puzzling star is indeed a Hyades member, as does P98.

The conclusions that can be drawn about G-dwarf Li from the combined sample are striking, and they corroborate those drawn from the WIYN/Hyades sample alone.  The stars with small $\sigma_{\rm Teff}$ fall right on the very tight G-dwarf Li-T$_{\rm eff}$ relation found by the WIYN/Hydra sample.  Furthermore, stars with small $\sigma_{\rm Teff}$ that are P98 binaries also fall right on the very tight Li-T$_{\rm eff}$ relation defined by the WIYN/Hydra sample.  So \textit{all} stars in this region that have small $\sigma_{\rm Teff}$ fall on a very tight Li-T$_{\rm eff}$ relation, regardless of whether they are bona fide single stars, suspected binaries with no evidence of contamination from the secondary in the spectra, radial velocity members, or those that fall outside the radial velocity single-star membership peak (we reiterate that all stars are highly probable PM members and photometric members).  Conversely, most other stars seem to exhibit more scatter around the tight Li-T$_{\rm eff}$ trend:  The T93 single stars and binaries with large $\sigma_{\rm Teff}$ exhibit far more scatter around the G-dwarf Li trend than the stars with low $\sigma_{\rm Teff}$.  Changes in Li due to errors in T$_{\rm eff}$ move a star parallel to the Li-T$_{\rm eff}$ trend, rather than bring it back to the trend.  Therefore, T$_{\rm eff}$ errors alone cannot explain this increased scatter.  Lastly, for the otherwise well behaved stars that photometrically deviate from the single star fiducial, there are only two of these stars in the Hyades.  These stars appear to be consistent with the observed Li trend but we will not consider these stars in our final Hyades sample for the sake of consistency because these types of stars are shown to be problematic in the Praesepe sample (see Section 6.2.3).

One of the main motivations of this work is to study G-dwarf Li depletion, and since the very tight Li-T$_{\rm eff}$ is a well-behaved function of T$_{\rm eff}$, we can define this trend by fitting a third-order polynomial between T$_{\rm eff}$ = 6300 and 5300 K (dashed curve in the right panel of Figure 10).  For the fit, we used only solid circles and pluses; that is, we used only Li detections with small $\sigma_{\rm Teff}$ that are consistent with the Hyades single star fiducial and that are single stars or stars with no evidence of contamination from a secondary in the spectra.  Figure 10 also shows an appropriate Li isochrone from the standard models of P97 (solid line) and SP14 (dashed dot) assuming an age of 700 Myr and [Fe/H]=+0.15, and initial A(Li) of A(Li)$_{\rm init}$=3.31, which is equal to the meteoritic abundance (the solar A(Li)$_{\rm init}$, Anders \& Grevesse 1989).  The Hyades A(Li)$_{\rm init}$ must be assumed because it remains unknown.  Both the Li gap and G-dwarf Hyads have depleted far more Li than the P97/SP14 standard models predict, and the slope of the G-dwarf depletion is also greater than that in the models.  The standard model fails to explain any of the observations and the discrepancy becomes even larger if we adopt Galactic Li production, giving that the super-solar-metallicity Hyades formed with even more Li than the Sun did.  In fact, evidence for the Galactic Li production might be present in the right panel of Figure 10: we could argue that the P97/SP14 isochrones ought to be at or above vB 6, vB 20, and the two SPTLBs vB 62 and vB 121.  Moving the isochrones upward by roughly 0.1--0.2 dex to an A(Li)$_{\rm init}$ of about 3.4--3.5 would accomplish that, but this remains speculative. 

Rotationally-induced mixing related to stellar angular momentum loss and internal redistribution has been proposed as a primary agent that might explain the differences between the standard model and observations (Pinsonneault 1988; Pinsonneault et~al.\ 1990; C95a), and SPTLBs have provided key evidence supporting the general precepts of these types of models (S90; DDK; T93; Deliyannis et~al.\ 1994; Ryan \& Deliyannis 1995).  We re-evaluate the key role of the SPTLB A(Li) in light of our self-consistent analysis of the Hyades Li-T$_{\rm eff}$ trend.  The rotational models suggest that rotation-related mixing occurs \textit{after} standard Li depletion is done, toward the end of the pre-MS and during the MS, when angular momentum is lost, and the slowing down of the surface coupled with the spinning up of the interior sets off shear instabilities that cause mixing.  However, tidally locked binaries cooler than $\sim$6300 K with periods observed today that are less than $\sim$8 to 9 days would have completed most of their spinning down during the early pre-MS (Zahn \& Bouchet 1989), before the stellar interior was hot enough to destroy Li.  They are thus predicted to have higher (though not necessarily perfectly preserved) Li than normal, single stars do.  As noted in previous studies and in support of this scenario, this is indeed what SPTLBs in the Hyades and elsewhere exhibit.  That said, mixing induced by gravity waves may also play a role in the Li depletion of G dwarfs (Garcia Lopez \& Spruit 1991).

Five SPTLBs have been identified, so far, in the Hyades (vB 22, vB 62, vB 121, BD+22$^{o}$669, and BD+22$^{o}$635) and are included in the right panel of Figure 10.  The first two appear to lie above the Li plateau, and since both are SB1s, there is the potential that their A(Li) are reliable.  vB 121 also has small $\sigma_{\rm Teff}$ (only 17 K!), so according to the various arguments presented above, its A(Li) is robustly above the (now) very tight Li-T$_{\rm eff}$ trend that defines the Li plateau.  Although vB 62 has a large $\sigma_{\rm Teff}$, a) errors in Li due to errors in T$_{\rm eff}$ move it A(Li) parallel to the Li-T$_{\rm eff}$ trend, and thus vB 62 remains above the Li plateau, and b) if the secondary is contributing non-negligible flux to the total spectrum, then flux-correcting the primary's Li line strength would result in an even \textit{higher} A(Li).  Therefore, vB 62 also lies robustly above the Li plateau.  The other three SPTLBs are SB2s, making it difficult to assess their T$_{\rm eff}$ and A(Li).  We derived their A(Li) from the flux-corrected equivalent widths in T93, rather than from direct synthesis of the spectra, as was the case with the other T93 stars.  vB 22 does seem to lie above the G-dwarf Li-T$_{\rm eff}$ trend, but the large error in T$_{\rm eff}$ (and in A(Li)) means this conclusion is rather tentative; it is possible vB 22 actually lies on the trend, farther above the trend, or even below the trend.  The two BD stars are the Li detections cooler than 5000 K.  Extension of the Li-T$_{\rm eff}$ G-dwarf trend would suggest that these two stars lie far above the trend.  There is no doubt that these stars contain Li, so uncertainties in their A(Li) are unlikely to bring down their A(Li) sufficiently to join the Li-T$_{\rm eff}$ trend at their T$_{\rm eff}$ shown in the right panel of Figure 10.  And although their T$_{\rm eff}$ are more limited and only based on B-V, these stars would have to be 600 to 800 K hotter to join the Li-T$_{\rm eff}$ trend, which seems unlikely.  It would appear that these stars really do lie above the Li-T$_{\rm eff}$ trend.  Overall, the SPTLBs in the Hyades do exhibit larger-than-normal A(Li) than the Li-T$_{\rm eff}$ trend, especially in the Li plateau region, still providing evidence supporting rotationally-induced mixing as an important agent of Li depletion for solar-type stars.  It has also been suggested that the SPTLBs lying above the Li plateau provide circumstantial evidence that the halo Li plateau (observed at similar T$_{\rm eff}$) is itself depleted, which is of relevance to the cosmological Li problem and the testing of Big Bang theory.

To summarize, we have presented evidence that probable members with small $\sigma_{\rm Teff}$ as derived from multiple colors, that are either bona fide single stars or show no evidence of contamination in our spectra from a secondary (e.g., using \textit{fxcor}), exhibit a tight Li-T$_{\rm eff}$ relation, which is especially tight for later F/G dwarfs (the Li plateau and G-dwarf Li depletion).  This is true whether we examine our WIYN/Hydra sample only, or whether we combine with several previous studies to form the largest self-consistently analyzed sample to date of A(Li) in Hyades dwarfs.  This is also most encouraging for our future WIYN/Hydra studies of clusters that have much less PM and binarity information available, if any.  We should be able to define a reliable Li-T$_{\rm eff}$ relation for a cluster as long as we use multiple colors to determine T$_{\rm eff}$ and $\sigma_{\rm Teff}$, we use only stars that have no evidence of binarity in our spectra, and by necessity we confine the sample to be within the radial velocity single star membership peak because that plus the photometric single-star fiducial sequence will be our primary membership indicators.  For the Hyades, the stars included (as described above) in our most reliable Li-T$_{\rm eff}$ relation, which we will refer to as our ``prime sample'', are indicated by a checkmark in the last column of Table 6. 

\vspace{-0.3cm}
\subsection{Praesepe}

\subsubsection{New Praesepe Lithium Observations}

Table 7 shows data for our WIYN/Hydra sample of the Hyades-aged Praesepe, which we found to have indistinguishable metallicity compared to the Hyades (Section 4).  Table 7 has identical column formatting as Table 6, and again we break the data into single star detections, single star upper limits, binary detections, and binary upper limits.  This binarity is based on the analysis of Praesepe in Mermilliod et~al.\ (2009), Patience et~al.\ (2002), and Burkhart \& Coupry (1998). The left panel of Figure 12 shows the A(Li) for Praesepe and reveals the same general patterns as in the Hyades: a broad range of over 3 dex in A(Li) across a range exceeding 3000 K in temperature; well-populated and substantial A(Li) near and above 7000 K, a severe Li depletion in mid-F dwarfs (6700 to 6300 K, the Li ``gap"), higher A(Li)s in the Li plateau in late-F/early-G dwarfs (near 6200 K), and the increasingly steep decline in Li with lower T$_{\rm eff}$ in G dwarfs until the Li line becomes undetectable near 5000 K.  

As with the Hyades, binaries that show no clear binary signatures such as a high $\sigma_{\rm Teff}$ or double lines still have A(Li) consistent with the Li-T$_{\rm eff}$ trend defined by single stars.  Stars with low $\sigma_{\rm Teff}$ that have radial velocity inconsistent with the cluster were found in the Hyades to have A(Li) in agreement with the Li-T$_{\rm eff}$ trends defined by radial-velocity members.  In Praesepe, KW 45 and KW 183 were similarly found to have radial velocity inconsistent with Praesepe but show no other signs of binarity.  However, we cannot perform a detailed comparison because both are cool and heavily depleted with only Li upper limits.  Lastly, binaries and single stars with high $\sigma_{\rm Teff}$ are again found to have significant scatter relative to the well behaved stars with low $\sigma_{\rm Teff}$.

\begin{figure*}[htp]
\begin{center}
\includegraphics[scale=0.95]{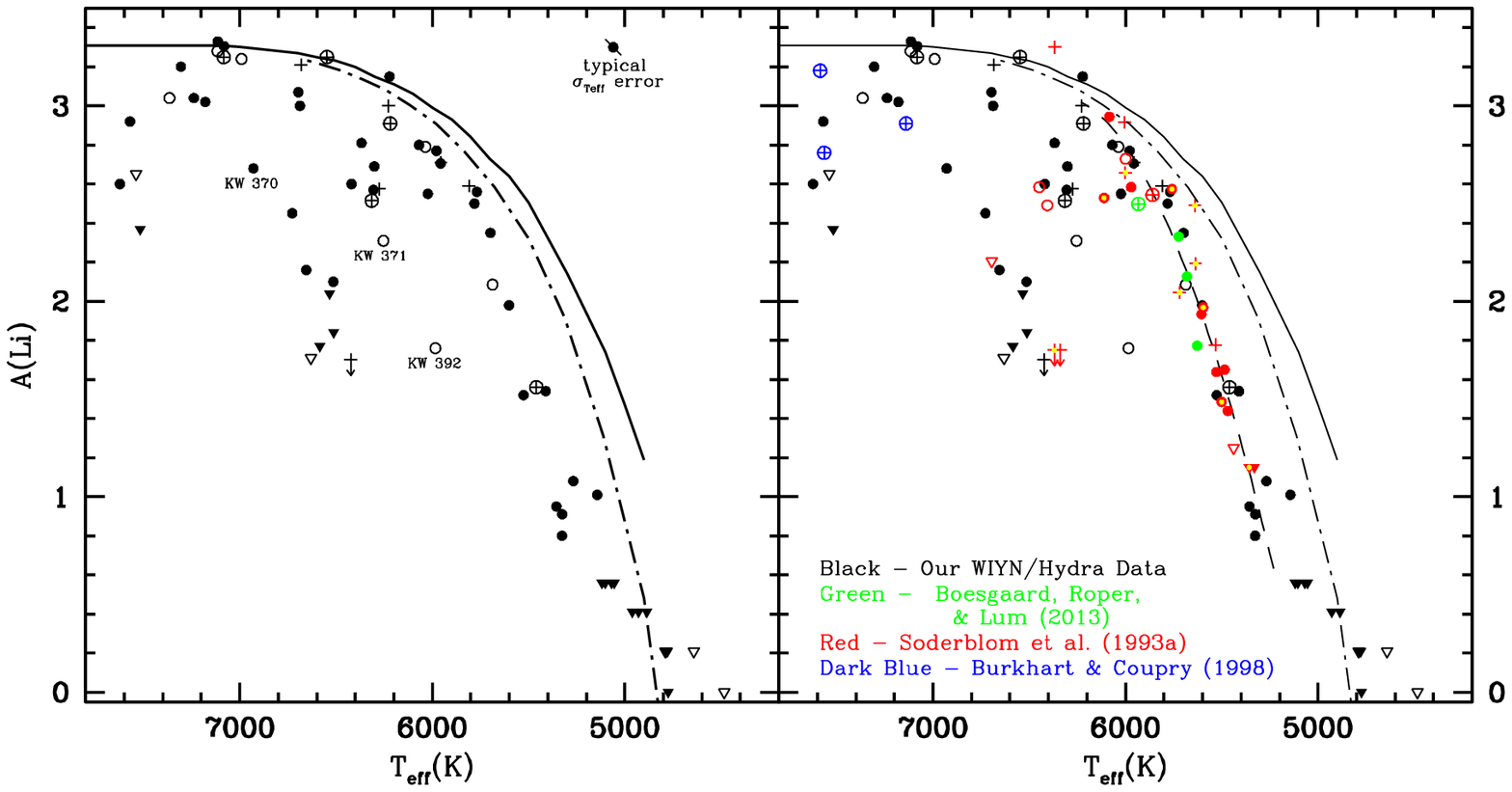}
\end{center}
\vspace{-0.45cm}
\caption[Praesepe A(Li) versus T$_{\rm eff}$]{The left panel shows the Praesepe A(Li) from our WIYN/Hydra analysis.  The data point types are equivalent to those described in Figure 10.  Again, the solid curve and dashed-dot curves represent the standard Li depletion models of P97 and SP14, respectively, for a cluster of this age and metallicity.  The right panel shows the full set of Praesepe A(Li), including our own WIYN/Hydra data, and the data from B13, S93a, BB88, and BC98.  The dashed curve is the fit to the observed G-dwarf Li-depletion trend, fitting only the solid circles and the pluses; that is, only detections with small $\sigma_{\rm Teff}$, that are single stars or stars with no evidence of contamination from a secondary in the spectra.}
\end{figure*}

Looking at the Li gap in more detail, the five detections and two upper limits near the hot edge of the Li gap in the very narrow T$_{\rm eff}$ range 6600 to 6700 K raise again the possibility that the hot edge is indeed a \textit{very} steep function of T$_{\rm eff}$ (BT86; AT09; Section 7.2).  The depleted A(Li)=2.68 in KW 370, which is just slightly hotter at T$_{\rm eff}$=6925 K, however, brings this view into question, and raises a rather different possibility.  Perhaps the hot side of the Li gap rises more gradually between 6600 and 7100 K, but at the same time, the three Li-rich stars with A(Li) near 2.95 to 3.3 dex and at $\sim$6750 K perhaps reflect a huge variation in the behavior of Li-gap stars: some deplete their Li very severely, whereas some are able to preserve their Li almost intact!  It would clearly be of interest to know which of these two rather different scenarios (if either) is more realistic.  The cool side of the Li gap is well populated at higher A(Li) than in the Hyades.  Two stars seem to fall below the cool side of the Li gap and Li plateau, namely KW 371, and especially KW 392 with its very low A(Li)=1.76, which is also somewhat reminiscent of the Hyades dwarf vB 143 in a similarly peculiar location significantly below the Li plateau; however, both of these Praesepe dwarfs have high $\sigma_{\rm Teff}$, so we will ignore them.  

Just as in the Hyades, the G-dwarf Li depletion trend is a tight Li-T$_{\rm eff}$ relation, though not quite as tight as that of the Hyades; a third-order polynomial fit to solid points and pluses is again shown by the dashed line.  The solid curve and the dashed-dot curve are the same P97 and SP14 Li isochrones, respectively, shown in Figure 10.  Lastly, as discussed in Section 4, there are eight cool Praesepe stars that have more limited photometry from U79.  These stars are distinguished in Table 7 by having $\sigma_{\rm Teff}$ set to 0.  None of these stars have Li detections, and all are substantially cooler than the low-T$_{\rm eff}$ end of the G-dwarf Li-T$_{\rm eff}$ trend (which is based on detections).  Thus, while their stellar parameters have more uncertainty, they do not affect our conclusions.

\subsubsection{Supplemental Praesepe Lithium Abundances}

There have been several previous analyses of Li in the dwarfs of Praesepe, including BB88, Soderblom et~al.\ (1993a, hereafter S93a), Burkhart \& Coupry (1998, hereafter BC98), and Boesgaard, Roper, \& Lum (2013, hereafter B13).  Comparisons of our abundances to B88 find that for the one radial velocity member with a detection (KW 295) in both studies, we find a significantly (0.48 dex) stronger A(Li).  Correcting for the difference in T$_{\rm eff}$ of 82 K accounts for nearly half of the difference in A(Li), but the remaining is explained by an 18 m\AA\, stronger equivalent width for our line measurement.   In the small sample of B88, the only star we have not analyzed is the SB2 KW 142.  As with the analysis of the Hyades, we do not consider SB2s and the limited number of stars in common prevent us from properly placing KW 142 on a consistent scale.  

B13 focused on G dwarfs and performed Li synthesis, but they did not publish equivalent widths.  Therefore, we used our method of deriving synthetic equivalent widths described at the beginning of Section 6: we applied their abundances and T$_{\rm eff}$ to our COGs to give effective equivalent widths.  Adopting these equivalent widths and \textit{our} stellar parameters gave the rederived B13 A(Li).  Because no systematic differences were found between the seven stars in common with our sample, we adopted the rederived A(Li) of the four stars we did not observe directly (KW 23, KW 58, KW 181, KW 301).

For the 26 stars we have in common with S93a there is no overall significant systematic equivalent width or T$_{\rm eff}$ offset or trend with T$_{\rm eff}$..  Comparing to the corrected S93a A(Li) (see Soderblom et~al. 1995) similarly finds no systematic difference in their published A(Li) and ours.  Therefore, we reanalyzed the 27 Praesepe stars from S93a that are not in our sample or that of B13 and used these to supplement our Praesepe Li observations.  Again we directly applied their equivalent widths to our parameters and COGs.  Lastly, there are five additional Praesepe stars in S93a that we do not consider because Mermilliod et~al. (2009) find them to be either non-members (KW 34 and KW 258) or SB2s (KW 184, KW 496, and KW 533).

BC98 focus on hot Am stars of Praesepe, and while we focus on cooler stars, our samples share KW 286 and KW 350.  Their adopted stellar T$_{\rm eff}$ are distinct from ours and are typically several 100 K hotter.  For KW 286, their adopted T$_{\rm eff}$ is 399 K hotter than ours, which is a concerning difference, but it well explains their 0.45 dex richer A(Li).  Similarly, for KW 350 their adopted T$_{\rm eff}$ is 150 K hotter than ours, but this T$_{\rm eff}$ difference further exacerbates our 0.2 dex richer A(Li) for this star.  These T$_{\rm eff}$ differences may be related to BC98 determining T$_{\rm eff}$ from Stromgren photometry and be indicative of systematics introduced between different photometric systems, which are relatively minor in the T$_{\rm eff}$ range of interest in this paper but can become important at these higher temperatures.  While there are not enough stars to reliably test for systematics and place the BC98 abundances on a uniform scale with our own, as was done with the Hyades we reanalyzed KW 40, KW 279, and KW 538 with our parameters and assumed no systematic differences.

\subsubsection{Final Praesepe Lithium Sample}

For stars observed in more than one study, we give priority to the A(Li) from our own WIYN/Hydra data/analysis first, followed by our rederived abundances from B13, S93a, and lastly BC98.  The right panel of Figure 12 shows our final combined sample of Praesepe A(Li) with our spectral abundances in black, the reanalyzed B13 abundances in red, the S93a abundances in light blue, and the BC98 abundances in dark blue.  We have also marked stars with low $\sigma_{\rm Teff}$ that photometrically deviate from the Praesepe single star fiducial (see right panel of Figure 3) with overlaid yellow data points.  

Beginning with the A dwarfs, with these supplemental data they are predominantly rich in Li but show a very large scatter in abundances.   There do not appear to be any clear patterns other than a possibly increasing Li depletion with increasing T$_{\rm eff}$.

\begin{center}
\renewcommand{\baselinestretch}{1.1}
{\footnotesize \begin{longtable*}{l c c c c c c c c c c c c c c}
\multicolumn{15}{c}%
{{\bfseries \tablename\ \thetable{} - Stellar Lithium Data for Praesepe}} \\
\hline
ID & B-V & T$_{\rm eff}$ & $\sigma_{\rm Teff}$ & EqW & $\sigma$EqW & A(Li) & $\sigma$Li$_{S/N}$ & $\sigma$Li$_{\rm Teff}$ & S/N &  Exp. & \textit{v sin i} & \textit{v$_{rad}$} & Ref & Prime\\
\hline
\tiny KW &     \tiny(K)       &       \tiny(K)           &\tiny(m\AA)& \tiny(m\AA)     &       &                    &                         &     &       & \tiny(km s$^{-1}$)    &   \tiny(km s$^{-1}$)             &     &             \\
\endfirsthead
\multicolumn{15}{c}%
{{\bfseries \tablename\ \thetable{} -- continued from previous page}} \\
\hline
ID & B-V & T$_{\rm eff}$ & $\sigma_{\rm Teff}$ & EqW & $\sigma$EqW & A(Li) & $\sigma$Li$_{S/N}$ & $\sigma$Li$_{\rm Teff}$ & S/N &  Exp. & \textit{v sin i} & \textit{v$_{rad}$} & Ref & Prime\\
\hline
\tiny KW & &    \tiny(K)       &       \tiny(K)           &\tiny(m\AA)& \tiny(m\AA)     &       &                    &                         &     &       & \tiny(km s$^{-1}$)    &  \tiny(km s$^{-1}$)             &     &             \\
\hline
\hline
\endhead
\hline
\hline
   23  & 0.699 & 5627.9 & 53.8  & 32.24 &   -    & 1.773  &   -    &   0.080  &    -  &  -     &  5.1 & 35.21$\pm$0.23 & 2 & \checkmark\\
   27  & 0.731 & 5526.4 & 62.9  & 28.15 &  1.23  & 1.520  & 0.035  &   0.070  &   400 &  5h59m & $<$6 & 34.11$\pm$0.38 & 1 & \checkmark\\
   30  & 0.681 & 5686.8 &111.3  & 46.05 &  1.82  & 2.085  & 0.024  &   0.137  &   300 &  5h59m &    8 & 34.61$\pm$0.34 & 1 & \\
31$^a$ & 0.559 & 6111.2 & 39.7  & 46.17 &  -     & 2.529  &   -    &   0.038  &    -  &  -     &  10.5& 35.11$\pm$0.23 & 3 & \\ 
   32  & 0.796 & 5325.1 & 21.7  &   -   &  2.18  & 0.910  &   -    &   0.027  &   250 &  11h21m&    9 & 35.03$\pm$0.35 & 1 & \checkmark\\
   38  & 0.307 & 7115.8 &103.7  & 46.28 &  3.14  & 3.280  & 0.036  &   0.064  &   560 &  0h45m &  190 & 30.33$\pm$15.16 & 1 & \\
   49  & 0.580 & 6036.3 & 75.1  & 82.94 &  0.76  & 2.790  & 0.006  &   0.082  &   635 & 15h52m &   10 & 34.94$\pm$0.49 & 1 & \\
   58  & 0.683 & 5680.8 & 22.6  & 49.92 &  -     & 2.127  &  -     &   0.026  &    -  &  -     &  5.5 & 34.55$\pm$0.22 & 2 & \checkmark\\
90$^a$ & 0.709 & 5596.1 & 27.4  & 45.41 &  -     & 1.968  &   -    &   0.032  &    -  &  -     &  6.7 & 35.94$\pm$0.20 & 3 & \\ 
  100  & 0.583 & 6023.2 & 29.8  & 56.15 &  1.00  & 2.550  & 0.012  &   0.032  &   500 &  5h59m &   13 & 33.28$\pm$0.68 & 1 & \checkmark\\
  114  & 0.205 & 7569.6 & 31.1  & 12.58 &  1.68  & 2.920  & 0.080  &   0.040  &   715 &  0h45m &   90 & 34.24$\pm$12.84 & 1 & \checkmark\\
  124  & 0.336 & 6990.8 & 87.7  & 50.12 &  1.62  & 3.240  & 0.017  &   0.061  &   745 &  4h32m &   91 & 39.58$\pm$1.99 & 1 & \\
  146  & 0.407 & 6696.0 & 71.0  & 54.29 &  3.09  & 3.070  & 0.031  &   0.048  &   410 &  0h45m &  100 & 36.31$\pm$7.06 & 1 & \checkmark\\
  154  & 0.250 & 7364.6 & 83.4  & 20.20 &  2.83  & 3.040  & 0.068  &   0.058  &   580 &  5h17m &  166 & 28.09$\pm$20.55 & 1 & \\
  155  & 0.417 & 6654.3 & 19.9  &  9.81 &  1.45  & 2.160  & 0.083  &   0.015  &   490 &  5h17m &   30 & 35.06$\pm$1.69 & 1 & \checkmark\\
  162  & 0.571 & 6068.8 & 31.2  & 74.01 &  0.86  & 2.760  & 0.007  &   0.031  &   600 & 10h38m &   10 & 34.88$\pm$0.47 & 1 & \checkmark\\ 
  164  & 0.706 & 5605.3 & 14.2  & 42.37 &   -    & 1.934  &  -     &   0.018  &    -  &  -     &  3.9 & 34.57$\pm$0.29 & 3 & \checkmark\\
182$^a$& 0.660 & 5758.6 & 51.7  & 89.22 &   -    & 2.574  &  -     &   0.053  &    -  &  -     &  8.6 & 34.38$\pm$0.15 & 3 & \\
  196  & 0.598 & 5970.5 & 47.5  & 64.93 &   -    & 2.584  &  -     &   0.045  &    -  &  -     &  8.8 & 35.13$\pm$0.23 & 3 & \checkmark\\
  208  & 0.599 & 5978.6 & 50.3  & 87.79 &  0.69  & 2.770  & 0.005  &   0.047  &   715 & 19h48m &   12 & 33.78$\pm$0.57 & 1 & \checkmark\\
  213  & 0.796 & 5326.6 & 55.7  &    -  &  1.03  & 0.800  &   -    &   0.053  &   500 & 23h35m &    8 & 34.31$\pm$0.35 & 1 & \checkmark\\
  217  & 0.508 & 6300.9 & 30.2  & 47.02 &  1.06  & 2.690  & 0.012  &   0.024  &   530 &  6h27m &   18 & 32.28$\pm$0.94 & 1 & \checkmark\\
  222  & 0.490 & 6367.9 & 32.9  & 53.35 &  1.69  & 2.810  & 0.018  &   0.026  &   310 &  0h45m &   10 & 34.13$\pm$1.00 & 1 & \checkmark\\
  226  & 0.315 & 7081.6 & 57.0  & 42.93 &  2.22  & 3.220  & 0.024  &   0.036  &   600 &  0h45m &  115 & 34.22$\pm$16.50 & 1 & \checkmark\\
  238  & 0.506 & 6305.6 & 40.5  & 37.31 &  1.39  & 2.570  & 0.020  &   0.032  &   450 &  5h17m &   22 & 33.02$\pm$1.30 & 1 & \checkmark\\
  263  & 0.816 & 5268.2 & 36.0  & 32.21 &  1.98  & 1.080  & 0.070  &   0.050  &   277 &  5h00m &   13 & 34.51$\pm$0.34 & 1 & \checkmark\\
  271  & 0.307 & 7114.1 & 43.1  & 50.98 &  1.59  & 3.330  & 0.017  &   0.026  &   705 &  1h30m &   78 & 35.35$\pm$8.87 & 1 & \checkmark\\
  286  & 0.194 & 7623.0 & 28.9  &   -   &  1.00  & 2.600  &  -     &   0.020  &   675 &  5h17m &   35 & 32.18$\pm$3.10 & 1 & \checkmark\\
  288  & 0.602 & 5957.3 & 60.9  & 81.59 &  0.96  & 2.705  & 0.007  &   0.058  &   500 &  24h48m&    7 & 34.74$\pm$0.53 & 1 & \checkmark\\
  293  & 0.469 & 6447.7 & 86.8  & 30.53 &   -    & 2.584  &  -     &   0.081  &    -  &  -     &  31.9& 36.66$\pm$0.56 & 3 & \\
  295  & 0.409 & 6687.3 & 52.3  & 48.22 &  2.19  & 3.000  & 0.024  &   0.036  &   550 &  7h12m &   90 & 35.19$\pm$8.47 & 1 & \checkmark\\
  301  & 0.670 & 5723.6 & 51.6  & 64.57 &   -    & 2.331  &  -     &   0.056  &    -  &  -     &   7.2& 34.31$\pm$0.35 & 2 & \checkmark\\
  304  & 0.749 & 5468.5 & 19.7  & 28.80 &   -    & 1.439  &  -     &   0.030  &    -  &  -     &   -  & 33.78$\pm$0.26 & 3 & \checkmark\\
  318  & 0.279 & 7238.2 & 35.2  & 24.33 &  1.37  & 3.040  & 0.028  &   0.026  &   900 &  6h27m &   95 & 34.61$\pm$5.55 & 1 & \checkmark\\
  326  & 0.707 & 5601.4 & 57.4  & 46.11 &  0.68  & 1.980  & 0.009  &   0.070  &   700 & 24h03m &    9 & 34.05$\pm$0.42 & 1 & \checkmark\\
334$^a$& 0.739 & 5500.7 & 33.7  & 28.62 &   -    & 1.485  &   -    &   0.040  &    -  &  -     &   9.4& 36.58$\pm$0.32 & 3 & \\ 
  335  & 0.653 & 5781.4 & 29.9  & 77.45 &  1.00  & 2.500  & 0.008  &   0.030  &   500 & 10h17m &    9 & 34.30$\pm$0.40 & 1 & \checkmark\\
  336  & 0.731 & 5527.5 & 46.6  & 33.50 &   -    & 1.638  &  -     &   0.049  &    -  &  -     &  5.4 & 35.43$\pm$0.25 & 3 & \checkmark\\
  340  & 0.264 & 7305.5 & 74.2  & 30.46 &  2.37  & 3.200  & 0.039  &   0.053  &   760 &  0h45m &  200 & 37.01$\pm$10.34 & 1 & \checkmark\\
  370  & 0.351 & 6928.7 & 43.1  & 18.00 &  2.40  & 2.680  & 0.066  &   0.033  &   440 &  0h45m &   70 & 35.81$\pm$5.14 & 1 & \checkmark\\
  371  & 0.520 & 6253.8 & 88.2  & 24.78 &  1.37  & 2.310  & 0.030  &   0.074  &   450 &  3h25m &   24 & 35.82$\pm$0.70 & 1 & \\
  392  & 0.594 & 5984.1 & 86.9  & 16.01 &  1.37  & 1.760  & 0.050  &   0.070  &   365 &  8h00m &   10 & 35.01$\pm$0.37 & 1 & \\
  396  & 0.480 & 6405.3 & 88.3  & 27.56 &   -    & 2.491  &   -    &   0.088  &    -  &  -     &  15.8& 34.20$\pm$0.34 & 3 & \\
  403  & 0.786 & 5355.8 & 33.8  &   -   &  2.28  & 0.950  &   -    &   0.040  &   230 &  5h00m &   10 & 34.24$\pm$0.39 & 1 & \checkmark\\
  418  & 0.566 & 6084.0 & 16.7  & 96.59 &    -   & 2.944  &   -    &   0.016  &    -  &  -     &  8.1 & 33.99$\pm$0.27 & 3 & \checkmark\\ 
  421  & 0.529 & 6221.8 & 61.6  &110.94 &  2.44  & 3.150  & 0.016  &   0.043  &   210 &  1h55m &   12 & 33.45$\pm$0.93 & 1 & \checkmark\\
  429  & 0.292 & 7179.1 & 40.7  & 25.47 &  3.36  & 3.020  & 0.065  &   0.028  &   550 &  0h45m &  210 & 35.06$\pm$12.85 & 1 & \checkmark\\
  430  & 0.859 & 5143.5 &  6.3  & 37.30 &  1.87  & 1.010  & 0.035  &   0.010  &   280 & 5h59m  &   10 & 35.35$\pm$0.37 & 1 & \checkmark\\
  432  & 0.678 & 5698.3 & 60.0  & 69.59 &  1.27  & 2.350  & 0.011  &   0.066  &   380 & 12h15m &    8 & 33.14$\pm$0.51 & 1 & \checkmark\\
  454  & 0.477 & 6419.6 & 39.8  & 35.98 &  1.49  & 2.600  & 0.022  &   0.037  &   440 & 2h40m  &   21 & 33.98$\pm$1.24 & 1 & \checkmark\\
  459  & 0.399 & 6727.5 & 37.2  & 15.44 &  2.94  & 2.450  & 0.097  &   0.027  &   530 & 0h45m  &  150 & 36.90$\pm$12.27 & 1 & \checkmark\\
  466  & 0.657 & 5768.6 & 44.3  & 86.73 &  0.83  & 2.560  & 0.006  &   0.045  &   545 & 14h10m &   10 & 33.14$\pm$0.56 & 1 & \checkmark\\
  472  & 0.452 & 6514.1 & 40.0  & 19.11 &  3.57  & 2.100  & 0.097  &   0.035  &   250 & 2h40m  &   50 & 33.74$\pm$3.50 & 1 & \checkmark\\
  476  & 0.768 & 5410.5 & 26.9  & 36.60 &  2.18  & 1.540  & 0.025  &   0.040  &   240 & 5h00m  &   10 & 33.90$\pm$0.40 & 1 & \checkmark\\
  488  & 0.744 & 5484.8 & 23.9  & 35.63 &   -    & 1.650  &   -    &   0.039  &    -  &  -     &  4.5 & 34.93$\pm$0.24 & 3 & \checkmark\\ 
  541  & 0.590 & 5998.5 &117.4  & 79.35 &   -    & 2.729  &   -    &   0.112  &    -  &  -     &  4.3 & 34.93$\pm$0.18 & 3 & \\ 
\hline
\hline
\multicolumn{15}{l}{{Spectroscopic Binaries based on Mermilliod et~al.\ (2009), Patience et~al.\ (2002), and BC98}}\\
\hline
   40  & 0.206 & 7564.5 & 133.3 &  8.5  &   -    & 2.760  &   -    &   0.090  &    -  &  -     &   -  & -    & 4 & \\
  127  & 0.589 & 6004.5 & 62.1  &104.63 &   -    & 2.916  &   -    &   0.072  &    -  &  -     &  3.9 & 34.46$\pm$0.11 & 3 & \checkmark\\
  181  & 0.609 & 5932.8 & 89.4  & 59.89 &   -    & 2.498  &   -    &   0.089  &    -  &  -     & 11.9 & 34.64$\pm$0.10 & 2 & \\
275$^a$& 0.590 & 6000.4 & 37.6  & 70.44 &   -    & 2.658  &   -    &   0.042  &    -  &  -     &  4.1 & 34.58$\pm$0.17 & 3 & \\
  279  & 0.202 & 7585.6 &103.1  & 20.0  &   -    & 3.180  &   -    &   0.072  &    -  &  -     &   -  & -     & 4 & \\
  282  & 0.514 & 6275.5 & 34.2  & 39.43 &  2.01  & 2.575  & 0.027  &   0.027  &   325 & 0h45m  &   25 & 36.86$\pm$1.49 & 1 & \checkmark\\
  268  & 0.504 & 6314.3 & 92.2  & 33.01 &  0.78  & 2.515  & 0.012  &   0.073  &   700 & 6h27m  &   13 & 47.45$\pm$0.82 & 1 & \\
322$^a$& 0.697 & 5635.3 & 56.6  & 60.12 &   -    & 2.193  &  -     &   0.066  &    -  &  -     &  7.2 & 35.29$\pm$0.45 & 3 & \\
  325  & 0.630 & 5859.8 & 91.4  & 72.46 &   -    & 2.544  &  -     &   0.093  &    -  &  -     &  8.3 & 34.79$\pm$0.11 & 3 & \\ 
  341  & 0.527 & 6227.3 & 18.5  & 88.47 &  1.07  & 3.000  & 0.008  &   0.014  &   490 &  6h27m &   14 & 34.94$\pm$0.82 & 1 & \checkmark\\
  350  & 0.314 & 7083.2 &180.7  & 45.56 &  1.86  & 3.250  & 0.019  &   0.123  &   680 &  0h45m &   65 & 36.87$\pm$8.28 & 1 & \\
365$^a$& 0.672 & 5719.1 & 39.6  & 40.33 &   -    & 2.045  &  -     &   0.047  &    -  &  -     &  1.5 & 35.61$\pm$0.14 & 3 & \\
  368  & 0.729 & 5533.1 & 50.9  & 38.29 &   -    & 1.777  &  -     &   0.071  &    -  &  -     &  4.7 & 35.12$\pm$0.09 & 3 & \checkmark\\
  399  & 0.645 & 5809.5 & 57.5  & 85.71 &  0.81  & 2.590  & 0.006  &   0.058  &   590 & 14h55m &    8 & 33.07$\pm$0.52 & 1 & \checkmark\\
  416  & 0.444 & 6547.0 & 89.0  & 87.91 &  1.18  & 3.250  & 0.009  &   0.071  &   435 & 2h40m  &    8 & 17.8 $\pm$1.20 & 1 & \\
  434  & 0.752 & 5459.4 & 79.8  & 33.70 &  1.75  & 1.560  & 0.018  &   0.060  &   345 & 9h10m  &   15 & 27.45$\pm$0.33 & 1 & \\
  439  & 0.410 & 6682.6 & 17.9  & 70.79 &  1.63  & 3.210  & 0.013  &   0.012  &   290 & 1h55m  &    9 & 33.94$\pm$0.87 & 1 & \checkmark\\
  508  & 0.602 & 5955.9 & 38.0  & 82.41 &  1.00  & 2.710  & 0.010  &   0.036  &   500 & 8h00m  &    8 & 35.70$\pm$0.47 & 1 & \checkmark\\
  515  & 0.530 & 6218.9 & 91.4  & 76.23 &  1.32  & 2.910  & 0.012  &   0.067  &   390 & 0h45m  &    9 & 35.10$\pm$1.02 & 1 & \\
  538  & 0.301 & 7139.6 &252.2  & 22.00 &  -     & 2.910  &   -    &   0.153  &    -  &  -     &   -  & -      & 4 & \\
540$^a$& 0.696 & 5639.5 & 48.6  & 94.87 &   -    & 2.490  &   -    &   0.054  &    -  &  -     &  3.1 & 36.95$\pm$0.13 & 3 & \\
  549  & 0.490 & 6366.8 & 31.1  &117.17 &   -    & 3.300  &   -    &   0.013  &    -  &  -     &  10.6& 34.94$\pm$0.11 & 3 & \checkmark\\
\hline
\hline
\multicolumn{15}{l}{{3$\sigma$ Upper Limits}}  \\
\hline
   45  & 0.232 & 7446.4 & 35.1  &  5.62 &   -  &   2.47 &    -     &    -     &   900 & 4h32m  &  175 & 44.55$\pm$8.86 & 1 & \checkmark \\
   48  & 0.926 & 4962.3 &  8.1  &  5.85 &   -  &   0.41 &    -     &    -     &   245 & 12h06m &    8 & 34.24$\pm$0.54 & 1 & \checkmark\\
   70  & 0.794 & 5331.2 & 47.0  & 30.40 &   -  &   1.15 &    -     &    -     &    -  &  -     &  2.4 & 35.12$\pm$0.18 & 3 & \checkmark\\
   79  & 0.874 & 5103.2 &  7.7  &  4.75 &   -  &   0.56 &    -     &    -     &   310 & 5h59m  &    9 & 36.70$\pm$0.34 & 1 & \checkmark\\
  143  & 0.216 & 7518.4 & 67.1  &  4.21 &   -  &   2.37 &    -     &    -     &   870 & 3h47m  &   80 & 33.65$\pm$11.40 & 1 & \checkmark\\
  172  & 0.956 & 4885.4 & 11.8  &  6.82 &   -  &   0.41 &    -     &    -     &   210 & 12h06m & $<$6 & 34.40$\pm$0.50 & 1 & \checkmark\\
  183  & 1.001 & 4773.7 & 40.0  &  5.95 &   -  &   0.00 &    -     &    -     &   260 & 12h06m &    9 & 42.88$\pm$0.58 & 1 & \checkmark \\
  198  & 0.993 & 4791.8 & 10.7  &  7.29 &   -  &   0.21 &    -     &    -     &   210 & 11h06m &    7 & 34.37$\pm$0.51 & 1 & \checkmark\\
  209  & 0.999 & 4778.5 & 18.8  &  6.26 &   -  &   0.21 &    -     &    -     &   240 & 12h06m & $<$6 & 34.00$\pm$0.56 & 1 & \checkmark\\
  218  & 0.407 & 6694.2 &155.7  & 10.09 &   -  &   2.21 &    -     &    -     &    -  &  -     &   -  & 36.74$\pm$1.22 & 3 & \\
  227  & 0.423 & 6631.8 & 78.0  &  3.61 &   -  &   1.71 &    -     &    -     &   465 &  6h27m &   17 & 34.34$\pm$0.77 & 1 & \\
  237  & 1.005 & 4762.6 &   -   &  5.20 &   -  &   0.21 &    -     &    -     &   300 & 23h35m &    9 & 34.24$\pm$0.51 & 1 & \\
  239  & 0.453 & 6511.6 & 51.2  &  5.52 &   -  &   1.84 &    -     &    -     &   390 & 0h45m  &   34 & 35.90$\pm$1.89 & 1 & \checkmark\\
257$^a$& 0.786 & 5357.7 & 48.7  & 29.26 &   -  &   1.15 &    -     &    -     &    -  &  -     &  12.5& 35.40$\pm$0.44 & 3 & \\ 
  272  & 1.026 & 4713.4 &   -   & 12.35 &   -  &   0.21 &    -     &    -     &   125 & 5h59m  &    8 & 37.90$\pm$0.57 & 1 & \\
  299  & 1.099 & 4550.2 &   -   &  7.31 &   -  &   0.00 &    -     &    -     &   210 & 19h06m &    8 & 33.97$\pm$0.59 & 1 & \\
  313  & 0.892 & 5054.3 & 12.9  &  5.28 &   -  &   0.56 &    -     &    -     &   280 & 19h06m &   10 & 34.76$\pm$0.33 & 1 & \checkmark\\
  332  & 0.434 & 6585.2 & 16.2  &  4.35 &   -  &   1.77 &    -     &    -     &   525 &  6h27m &   37 & 35.20$\pm$0.83 & 1 & \checkmark\\
  344  & 0.868 & 5117.8 & 10.4  &  5.81 &   -  &   0.56 &    -     &    -     &   280 &  7h15m &   13 & 35.38$\pm$0.31 & 1 & \checkmark\\
  349  & 0.872 & 5108.8 &   -   &  5.41 &   -  &   0.56 &    -     &    -     &   270 & 19h06m & $<$6 & 35.04$\pm$0.32 & 1 & \\
  363  & 0.939 & 4928.1 & 20.3  &  6.13 &   -  &   0.41 &    -     &    -     &   250 & 18h21m & $<$6 & 34.62$\pm$0.53 & 1 & \checkmark\\
  375  & 0.212 & 7538.2 &212.6  &  7.08 &   -  &   2.65 &    -     &    -     &   625 & 0h45m  &  135 & 53.11$\pm$24.43 & 1 & \\ 
  448  & 0.886 & 5068.5 &  7.8  &  4.92 &   -  &   0.56 &    -     &    -     &   320 & 12h15m &   10 & 34.22$\pm$0.53 & 1 & \checkmark\\
  478  & 0.447 & 6533.5 & 69.8  &  8.28 &   -  &   2.04 &    -     &    -     &   400 & 2h40m  &   70 & 34.02$\pm$5.52 & 1 & \checkmark\\
  498  & 0.760 & 5438.3 & 81.2  & 26.59 &   -  &   1.25 &    -     &    -     &    -  &  -     &  4.7 & 34.10$\pm$0.23 & 3 & \\ 
 288W  & 1.225 & 4301.9 &   -   &  8.39 &   -  &   0.00 &    -     &    -     &   185 & 6h06m  &    9 & 35.24$\pm$0.59 & 1 & \\
 624W  & 1.057 & 4641.8 &   -   &  6.23 &   -  &   0.21 &    -     &    -     &   250 & 21h46m &    9 & 34.72$\pm$0.61 & 1 & \\
 792W  & 1.130 & 4484.4 &   -   &  6.99 &   -  &   0.0  &    -     &    -     &   220 & 0h45m  &    8 & 34.76$\pm$0.57 & 1 & \\
 899W  & 0.995 & 4787.5 &   -   &  8.76 &   -  &   0.21 &    -     &    -     &   180 & 8h00m  &   10 & 35.09$\pm$0.56 & 1 & \\
\hline
\hline
\multicolumn{15}{l}{{Spectroscopic Binary 3$\sigma$ Upper Limits}}  \\
\hline
   47  & 0.498 & 6338.0 & 56.3  &  8.67 &   -  & 1.75 &    -     &    -     &    -  &  -     &  12.4& 34.99$\pm$0.13 & 3 & \checkmark\\ 
  250  & 0.476 & 6423.1 & 35.6  &  4.59 &   -  & 1.70 &    -     &    -     &   490 &  4h32m &   36 & 34.10$\pm$0.96 & 1 & \checkmark\\
536$^a$& 0.490 & 6369.0 & 61.6  &  8.34 &   -  & 1.75 &    -     &    -     &    -  &  -     &  13.2& 36.90$\pm$0.19 & 3 & \\ 
\hline
\caption[Praesepe stellar Li data]{An (a) superscript marks stars that deviate from the single-star fiducial (see Figure 3).  Ref: (1) Our WIYN/Hydra sample, (2) Boesgaard, Roper, \& Lum (2013) (3) Soderblom et~al.\ (1993a), (4) Burkhart \& Coupry (1998).  The source for radial velocities and \textit{v sin i} of stars we did not observe is Mermilliod et~al.\ (2009).}
\end{longtable*}}
\end{center}

\vspace{-0.9cm}
On the cool side of the Li gap there are a handful of additional stars, including some very rich detections and very Li-poor upper limits.  However, none of these additional stars are single stars with low $\sigma_{\rm Teff}$, so their larger dispersion does not necessarily contrast with our own Praesepe Li-gap abundances.

Lastly, the reanalyzed sample from B13 and S93a primarily augment our Li plateau and G-dwarf A(Li) and they remarkably follow the general Li depletion trend observed in our own sample, further strengthening this trend.  We do, however, note the increased scatter found in stars that have low $\sigma_{\rm Teff}$ (relative to that observed in the Hyades).  A majority of this scatter appears to be generated by stars that photometrically deviate from the Praesepe single star fiducial.  Even though the two photometrically discrepant Hyades dwarfs do have A(Li) consistent with the trends, this suggests that these stars do not have reliable A(Li), similar to those with high $\sigma_{\rm Teff}$.  When not considering these stars, however, an apparent scatter in the G-dwarf A(Li) remains.  Could this remaining scatter simply be a result of errors?  For the two stars with the most discrepant A(Li) (KW 100 and KW 430), the total A(Li) uncertainty for KW 430 is only 0.040 dex, while it is at least 0.5 dex richer in Li than the nearly identical KW 433 and KW 314 that have Li upper limits.  Similarly, the total A(Li) uncertainty for KW 100 is only 0.035 dex, while it is $\sim$0.25 dex below the observed Li trend.  We reassuringly note, however, the strong consistency of the mean G-dwarf trends in our analysis and that of T13 and S93a. 

\section{Hyades \& Praesepe Lithium Comparison and Discussion}

The Hyades and Praesepe have indistinguishable age and [Fe/H].  Do they also have the same Li morphology, or are there differences that would require additional factors to explain?  We now examine this question by comparing their A(Li) from A dwarfs to K dwarfs.

\begin{figure*}[htp]
\begin{center}
\includegraphics[scale=0.66]{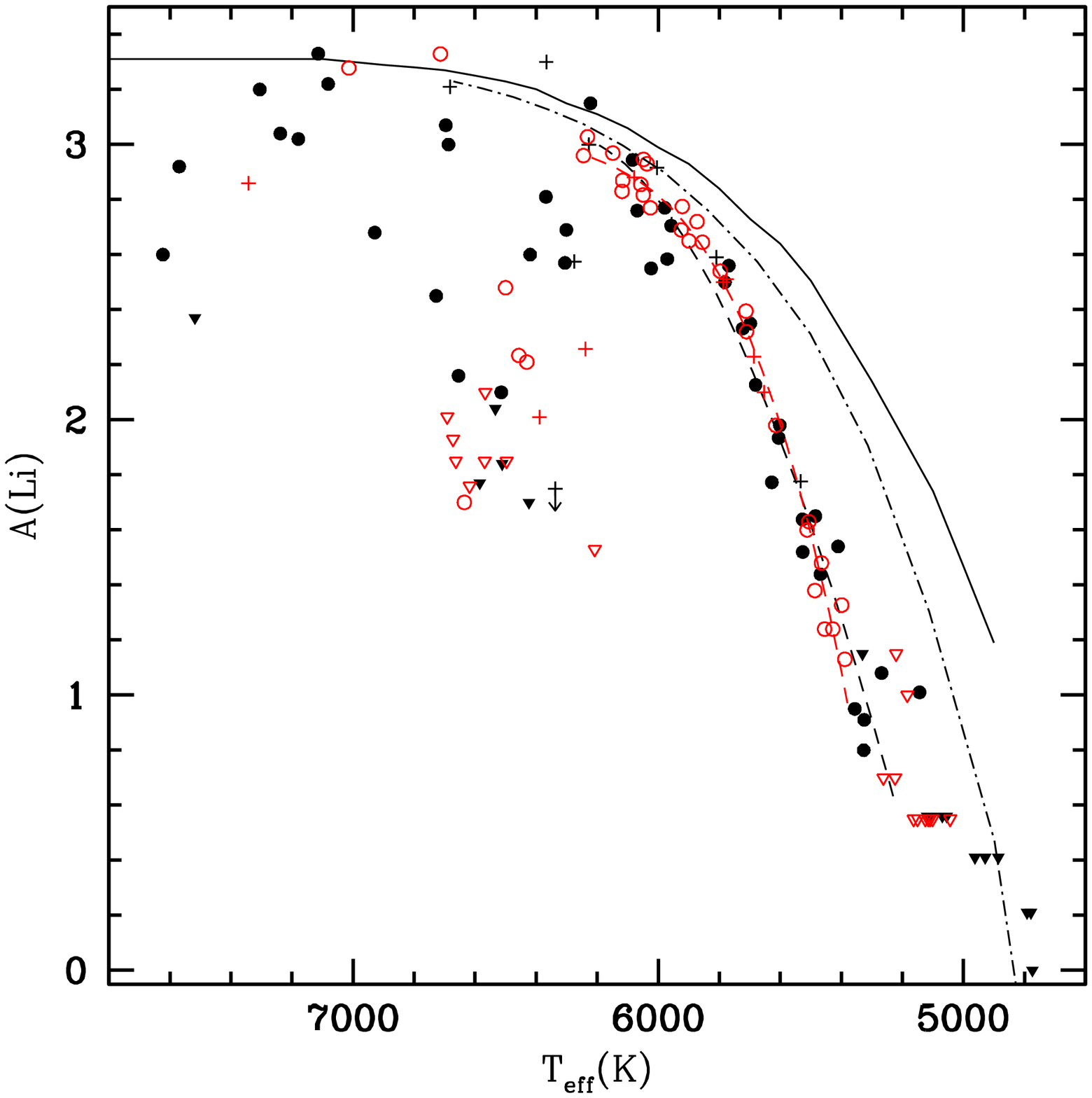}
\end{center}
\vspace{-0.45cm}
\caption{A direct comparison of the Hyades (open red) and Praesepe (solid black) A(Li) presented in the right panels of Figures 10 and 12.  As discussed in the text, the prime samples show tighter Li-T$_{\rm eff}$ relations compared to when stars with larger errors (e.g., large $\sigma_{\rm Teff}$ due to binarity or other reasons) are included, and thus fundamentally are more likely to represent the true cluster Li-T$_{\rm eff}$ relations.  We thus show only each cluster's prime samples; that is, only single-star members (circles) and binaries (pluses) with low T$_{\rm eff}$ error are presented and the different sources are not distinguished.  The dramatic consistency of the two clusters is made clear.  The depletion trends (shown by the red and black dashed curves for the Hyades and Praesepe, respectively) in the coolest G dwarfs do appear to diverge in the two clusters, but it is not significant in the temperature range (T$>$5350 K) where there are Li detections in both clusters and can be explained by the increased errors in these faint and heavily-depleted dwarfs.  Standard Li depletion models for the appropriate age and metallicity are also shown, from P97 (solid) and SP14 (dashed-dot).}
\end{figure*}
 
Figure 13 shows the prime set of A(Li) for each cluster plotted together against T$_{\rm eff}$.  All stars are highly probable PM and photometric members.  Recalling the discussion of the previous subsections, these prime sets include stars that have small $\sigma_{\rm Teff}$ ($<$75 K, from the T$_{\rm eff}$ derived using the 10 UBVRI color combinations), that have shown no evidence of binarity in their spectra, and that are photometrically consistent with each cluster's single star fiducials.  To be clear, stars are excluded if they have large $\sigma_{\rm Teff}$ ($>$75 K), if they have insufficient photometry to measure a $\sigma_{\rm Teff}$, if they have exhibited any evidence of binarity in fxcor, deviate from the cluster fiducial, or if they are SPTLBs; such stars were shown to have larger scatter in the Li-T$_{\rm eff}$ relation compared to the prime set.  Radial velocity discrepant stars and those shown to have variable radial velocities (SB1s) in other multi-epoch studies are not excluded if they have small $\sigma_{\rm Teff}$.  Recall that these Hyades and Praesepe PM members lying outside of the cluster radial velocity peak showed the same Li-T$_{\rm eff}$ behavior as radial velocity members.

In general we find that the two clusters' Li morphologies are similar and/or complement each other: in T$_{\rm eff}$ ranges where both clusters have significant numbers of stars, the two cluster Li-T$_{\rm eff}$ patterns are indistinguishable, whereas in T$_{\rm eff}$ ranges where only one cluster has significant numbers of stars, that cluster's Li-T$_{\rm eff}$ pattern fits in very accurately with the Li-T$_{\rm eff}$ pattern(s) defined by one or both clusters in the neighboring T$_{\rm eff}$ ranges.  These remarkable concordances might suggest combining the two clusters to derive a more refined Li-T$_{\rm eff}$ relation for this age and metallicity.  In more detail:

\subsection {A Dwarfs}

Both clusters show a similar Li morphology for T$_{\rm eff}>$7000 K, with most stars congregating near A(Li) = 3.0 to 3.3, though one Hyad and two stars in Praesepe show significant Li depletions with A(Li) in the range 2.4 to 2.9 or lower.  As is widely known, surface abundances of A stars can be affected by microscopic diffusion processes, mass loss, magnetic fields, and rotation-induced circulation.  Such relative uniformity is thus remarkable, and this results from our focusing only on stars with low $\sigma_{\rm Teff}$.  

Looking for trends, both samples combined do begin to suggest that in A dwarfs the depletion gradually increases with increasing T$_{\rm eff}$. Does this represent a true depletion trend or is it indicative of systematics in our color-temperature relation at high T$_{\rm eff}$?  We remind the reader that this is the T$_{\rm eff}$ where the systematic differences between the Deliyannis et~al.\ (2002) relation and the Ram{\'i}rez \& Mel{\'e}ndez relation become significant (see Figure 4).  Adopting their relation would bring the hot F (cool A) stars at 7100 K to 7400 K and accordingly increase their A(Li) by $\sim$0.3 dex to $\sim$3.6 dex, which is very Li rich.  For the stars at higher T$_{\rm eff}$ the more rapid increase of T$_{\rm eff}$ with B-V would not produce a A(Li) plateau but would significantly mitigate this apparent A-dwarf trend.  Either way, there remains an apparent trend of Li depletion with increasing T$_{\rm eff}$, although this is not a robust conclusion due to the sample size.

\subsection{Li Gap}

In the Li gap we noted in each cluster several outliers above and below the Li-gap trends.  However, combining both clusters allows us to draw stronger conclusions than from either cluster alone.  

On the cool side of the Li gap, single stars form a very tight Li-T$_{\rm eff}$ relation, whereas (the few) binaries with small $\sigma_{\rm Teff}$ mostly fall below this tight relation.  (One Praesepe low $\sigma_{\rm Teff}$ binary, KW 549, lies significantly above the trend at A(Li) = 3.30, T$_{\rm eff}$ = 6367 K.)  This is inconsistent with our finding in G dwarfs, where binaries and single stars with small $\sigma_{\rm Teff}$ remain consistent.  Is this the result of some key difference in the Li depletion processes in these different T$_{\rm eff}$ ranges?  Does binarity not affect G-dwarf depletion processes but does affect Li-gap F-dwarf depletion processes?  This is an interesting possibility, but because of this we will exclude all binaries, regardless of their $\sigma_{\rm Teff}$, in further analysis of the Li gap in the Hyades and Praesepe. 

\begin{figure}[htp]
\begin{center}
\includegraphics[scale=0.42]{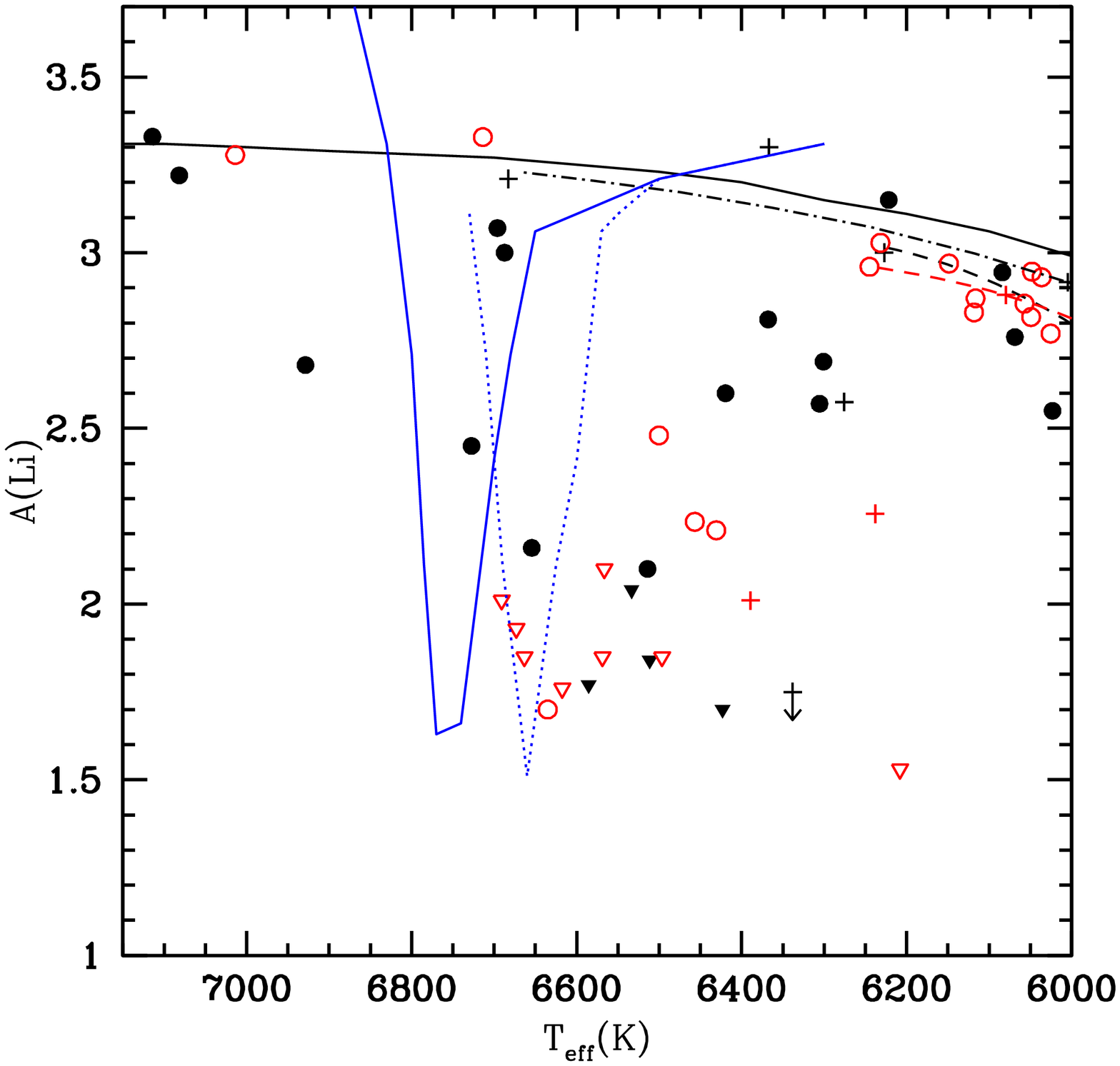}
\end{center}
\vspace{-0.45cm}
\caption{Here we focus on the Li-gap data from our combined Hyades and Praesepe prime samples, using the same data point types and plotting the same models and trends shown in Figure 13.  The apparent wall on the hot side of the Li gap is shown near 6700 K.  In blue we illustrate two models of Li diffusion from Richer \& Michaud (1993) with [Fe/H]=0.0 (solid blue) and [Fe/H]=+0.18 (dotted blue) with an A(Li)$_{\rm init}$=3.31.  The steep depletion trend in the [Fe/H]=+0.18 is remarkably consistent with the observations.  We also note at higher T$_{\rm eff}$ (e.g., the solar model near 6900 K) the diffusion model predicts A(Li) above the A(Li)$_{\rm init}$, which may help explain the super-Li-rich star discussed in the Introduction in NGC 6633 at T$_{\rm eff}$=7086 K and A(Li)=4.29.}
\end{figure}

The hot side of the Li gap is similar in both clusters in that both have substantial 
or rising A(Li) in the T$_{\rm eff}$ range of 6700 to 7100 K, and both clusters 
show severely depleted A(Li) in the middle of the gap, between T$_{\rm eff}$ = 
6500 and 6700 K.  Whereas
previous studies defined the Li-T$_{\rm eff}$ morphology of the hot side
of the Li gap using stars that have turned out to have high
$\sigma_{\rm Teff}$, our restricting the sample to only stars with low
$\sigma_{\rm Teff}$ and combining both clusters should in principle yield a
clearer picture of the Li-T$_{\rm eff}$ morphology.  Six single star detections, one
low $\sigma_{\rm Teff}$ binary, and three upper limits seem to define a nearly
vertically rising trend between 6635 and 6715 K, from A(Li) = 1.70 (in vB
90) to 3.33 (in vB 20).  It is tempting to believe that this vertical wall
redefines the hot edge of the Li gap, and that slightly hotter stars
remain at high levels of A(Li); the six stars between 7000 and 7305 K have
A(Li) between 3.0 and 3.35.

Figure 14 shows that this wall is consistent with the diffusion models of 
Richer \& Michaud (1993), adopting an A(Li)$_{\rm init}$ of 3.31.  Although, 
as these authors acknowledge, the diffusion gap is too narrow, there is 
impressive agreement between the hot edge of the models and our observations. 
Two sets of models are shown, with [Fe/H] = 0.0 and +0.18.  The ones with 
+0.18 (close to the Hyades/Praesepe metallicity) agree better with the 
data.  Assuming a slightly higher A(Li)$_{\rm init}$, such as 3.5 to 3.6, 
consistent with the possibility of Galactic Li production, would make the 
models agree even better with the wall.  A caveat might be the uncertainty 
of the predicted T$_{\rm eff}$ due to uncertainties in the model physics (Richer 
\& Michaud 1993) and to other factors like the color-T$_{\rm eff}$ relation.  Also, we have 
discussed a preponderance of evidence that favors rotational mixing as the 
dominant Li depletion mechanism at cooler T$_{\rm eff}$, and indeed 
rotationally-induced mixing can inhibit diffusion.  However, we 
cannot rule out diffusion as the cause of the wall.

\begin{figure*}[htp]
\begin{center}
\includegraphics[scale=0.69]{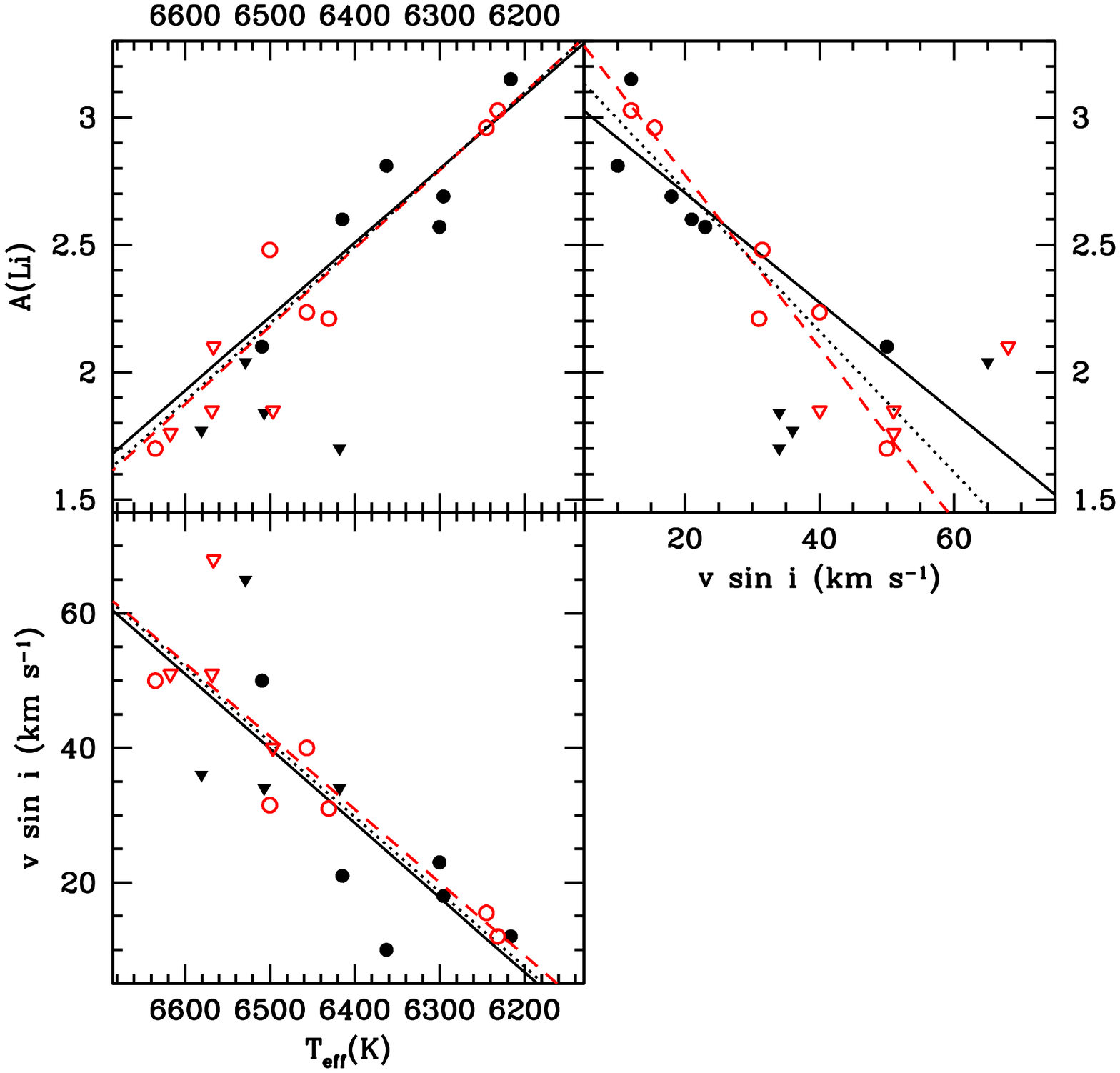}
\end{center}
\vspace{-0.45cm}
\caption{The linear correlations found in the cool sides of the Li gap in both the Hyades (open red) and Praesepe (solid black).  The dashed red lines represent the linear correlations found for the Hyades and the solid black lines represent the linear correlations found for Praesepe.  The upper limits are also shown for reference, but they are not used in determining the correlations involving A(Li).  The similarity of the correlations found in the Hyades and Praesepe show that there is no significant difference in the characteristics of either cluster's Li-gap depletion.  The correlations based on both cluster's data are shown as black dotted lines.}
\end{figure*}

All that being said,
the lower A(Li) = 2.68 of KW 370 suggests another possibility for the Li
morphology of the hot side of the Li gap: that the rise of A(Li) is more gradual,
going from 1.70 at T$_{\rm eff}$ = 6635 K (vB 90) to 3.3 at T$_{\rm eff}$ =
7000 K.  In that context, the four stars (3 single stars and 1 binary with
low $\sigma_{\rm Teff}$) near 6700 K that have A(Li) = 3.00 to 3.33 would be
remarkable in their Li-preserving properties.  In this scenario, the
middle of the gap would be wider (6500 to 6800 K instead of 6500 to 6700 K) and
would exhibit a huge variety of A(Li), exceeding a range $>$1.7 dex.  The
implications for constraining models are quite different: a vertical wall
is different from a more gradual rise, and a huge range of A(Li) in the
middle of the Li gap is different from the high A(Li) merely reflecting
the top of a vertical wall.  Unfortunately, our data do not allow us to
distinguish between these two possibilities.

The rising A(Li) trends on the cool side of the Li gaps are consistent with each other and complement each other, with the Hyades showing detections having lower A(Li) in the T$_{\rm eff}$ range 6400 to 6500 K, and Praesepe showing detections with larger A(Li) in the T$_{\rm eff}$ range 6300 to 6400 K.  Such complementarity is rather remarkable. In fact, as mentioned just above, the combination of both clusters' prime samples almost suggests a somewhat tight Li-T$_{\rm eff}$ relation for the cool side of the gap.  However, some scatter that does exceed the formal errors is apparent, even when focusing only on single stars.  This scatter is greater than any present in the G dwarfs.  As pointed out in Section 6.1.3, this is especially true at the hot edge of the Li plateau near 6250 K, where the A(Li) show a very striking range of nearly 2 dex at virtually the same T$_{\rm eff}$. 

Cummings et~al.\ (2012) found that the cool edge of the Li gap, where the significant depletion begins, had very similar T$_{\rm eff}$ in our preliminary Hyades and Praesepe results in comparison to the gap edges in the much older and much more metal-rich cluster NGC 6253 (age=3.0 Gyr, [Fe/H]=+0.43) and the moderately older ($\sim$1.5 Gyr) clusters NGC 3680, NGC 752, and IC 4651.  We note that this is when their stars' T$_{\rm eff}$ are converted back to the T$_{\rm eff}$ that they had at the Hyades age, using Y$^2$ isochrones.  These clusters span a large range in both age and metallicity (-0.08$<$[Fe/H]$<$+0.43) and demonstrate that after 700 Myr the stars at the cool edge of the Li gap do not subsequently deplete significant Li, which would broaden the gap.  This may also be true in gaps younger than the Hyades and Praesepe, but there is limited analysis of the gap in such clusters.  Because the T$_{\rm eff}$ of a star of a given mass depends on metallicity, however, Cummings et~al.\ (2012) concluded that the mass of the cool edge of the Li gap does depend on metallicity and results in an approximately linear correlation.  Fran{\c c}ois et~al.\ (2013) expanded upon this with the Li analysis of the metal-poor NGC 2243 ([Fe/H]=-0.54$\pm$0.10) and found that even at this extreme metallicity there remains a linear correlation between the mass at the cool edge of the gap and metallicity all of the way up to the very metal rich NGC 6253.  The agreement between the Li gaps in the Hyades and Praesepe strengthens these conclusions.

As summarized in Cummings et~al.\ (2012), a variety of evidence favors rotationally-induced mixing over other proposed mechanisms as the main physical cause of the Li depletions in the gap.  Therefore, it is of interest to see whether any correlations exist between \textit{v sin i} and A(Li), in particular for the cool side of the Li gap, which we define to be 6200 to 6635 K.  We acknowledge the challenge in defining the precise transition from the cool side to the hot edge of the gap -- and cannot know if such a precise edge even exists -- so the low Li detection in vB 90 (T$_{\rm eff}$=6635 K, A(Li)=1.70) is included on the cool side but the just slightly hotter stars (KW 155, KW 459, and so on) with sharply rising A(Li) are not.  

Figure 15 shows A(Li) and \textit{v sin i} versus T$_{\rm eff}$, and A(Li) versus \textit{v sin i}.  The red dashed line is a linear fit for the Hyades using only the Li detections (open red).  Stars with Li upper limits (red inverted triangles) are shown here only for reference; any information about Li depletion contained in these stars (such as departure from linearity) remains largely unknown, at least for now.  Another caveat is that the \textit{v sin i} can significantly underestimate the true rotation rates.  Fortunately, the Li detections cover a broad range of \textit{v sin i}, T$_{\rm eff}$, and A(Li).  Hotter stars are more depleted in Li (upper-left panel, linear correlation coefficient r=-0.95).  Hotter stars also rotate much faster (lower-left panel, r=0.89; this comparison does include the Li upper limits); thus, faster rotators are more depleted in Li (upper-right panel, r=-0.97).  Higher order, non-linear fits were not found to be appropriate in any of the correlations, but the preponderance of Li upper limits at the rapid rotation (and hot) end in the \textit{v sin i} versus T$_{\rm eff}$ relation suggests a nonlinear drop at that end.  We stress that these fits are correlations only and that no direct causal relationship has been established.  Furthermore, in the top two panels, the A(Li) show more scatter around the fits than can be fully accounted for by the errors.  Does this suggest that both T$_{\rm eff}$ and rotation must be considered simultaneously? 

For Praesepe, the linear fits shown as a solid black line give r=-0.87 for T$_{\rm eff}$ versus A(Li), r=-0.91 for \textit{v sin i} versus A(Li), and r=0.76 for T$_{\rm eff}$ versus \textit{v sin i} (including Li upper limits).  Interestingly, the three upper limits in the \textit{v sin i} versus A(Li) plot suggest a downturn (or a steeper slope in better agreement with that observed in the Hyades).

All correlations are very similar between the two clusters, as is evident from Figure 15.  Thus, not only are the Li-T$_{\rm eff}$ and Li-rotation trends qualitatively similar in the two clusters, they are also quantitatively similar.  Therefore, we can improve the statistics by combining all 12 Li-gap star detections (dotted lines in Figure 15).  These final correlations using all of the data are A(Li)=21.88--T$_{\rm eff}$$\times$0.00303; A(Li)=3.270--\textit{v sin i}$\times$0.0277; and \textit{v sin i}=--683.1+T$_{\rm eff}$$\times$0.1114.  With this more significant sample, we can also consider all variables simultaneously with a 3-dimensional linear regression and find that A(Li)=12.69--0.001526$\times$T$_{\rm eff}$--0.01557$\times$\textit{v sin i}.  This simple model improves on the residuals of the 2-dimensional fits as demonstrated by its coefficient of determination (r$^2$) of 0.92.  The remaining residuals ($\sim$0.1 dex) can be explained by the abundance errors and the limitations of using \textit{v sin i} as a measurement of true rotation rate.  This illustrates that when the T$_{\rm eff}$ and \textit{v sin i} are considered together they can act as a reliable predictor of A(Li) on the cool side of the Li gap.  Consistent analysis of Li gaps in clusters of differing metallicity and age will further help to understand these relationships.

What might be the physical cause of this Li-T$_{\rm eff}$-\textit{v sin i} correlation?
We suggest it is at least qualitatively consistent with the basic tenets
of the Yale rotational models, which tie mixing and the resulting
Li-depletion to redistribution and loss of stellar angular momentum, for
the following reasons.  On the cool side of the Li gap, angular momentum
on the ZAMS is a steeply increasing function of T$_{\rm eff}$ (Barnes
2007; Reinhold \& Gizon 2015).  Periods of stars near the ZAMS increase 
by a factor of 3 to 4 from the red edge to the blue edge of the cool side of 
the Li gap, and
since mass and radius also increase, angular momentum increases by an even
greater factor.  Note that this increase in angular momentum is quite sizable compared
to, for example, stellar mass, which increases only 15\%\, from the red edge
to blue edge.  By the age of the Hyades, stars in this T$_{\rm eff}$
range have spun down by factors of 2 to 3, as
measured by their periods.  Thus, the amount of angular momentum lost between the ZAMS and
650 Myr is a steeply increasing function of T$_{\rm eff}$.  To
rough approximation, if this angular momentum loss is tied to mixing and Li depletion,
the Li depletion should increase with T$_{\rm eff}$.

We want to point out some additional factors that may be relevant when considering
the complexities of rotational mixing and 
Li depletion in the gap.  These include a) the depth of the SCZ (and
thus how much mixing is done by convection versus by rotational mixing,
although the SCZ is already relatively shallow at the red edge and becomes
shallower toward the blue edge, so rotational mixing has to do most
of the work associated with mixing throughout the whole T$_{\rm eff}$ range),
b) the depth of the Li preservation region (which becomes shallower toward 
the blue edge), and c) issues such as whether it takes longer for faster rotators 
to spin down, even if they lose more angular momentum than slower rotators do as 
they age and spin down.

These various subtleties notwithstanding, it is worth attempting just one
example of a quantitative comparison with theory, in particular, with the models of 
Pinsonneault et~al.\ (1990).  If the rotational models of Pinsonneault et~al.\ (1990) are to reproduce the cool
side of the Li gap, then their Figure 15 suggests that the initial angular
momentum must increase by a factor of $\sim$8 from the cool (red) edge to the hot (blue)
edge of the Li gap (6200 to 6635 K, as defined above).  We test this
prediction using the following assumptions and (rough) approximations.  We
use the (B-V)-rotational period (P) relation of Barnes (2007), which was derived from
open cluster data, and choose 100 Myr because stars of the relevant masses
have settled onto the MS by that age.  This much gives P$_{\rm red}$/P$_{\rm blue}$ = 3.3.
To convert to angular momentum, we assume the structure of a star at one
edge is roughly homologous with that at the other edge, so that the
moments of inertia depend on MR$^2$ in the same way, and that any internal
differential rotation with depth is similar in the two stars.  Y$^2$
isochrones with [Fe/H] = +0.15 are used to identify appropriate masses and
radii at both edges.  (Although the time dependence in the Barnes (2007)
relation is separable, using an age where the stars have just settled
onto their stable MS T$_{\rm eff}$ helps ensure we are using appropriate masses.)
These assumptions allow a simple (approximate)
analytical expression for the ratio of angular momenta (J).  Using the values
identified above, the result is J$_{\rm blue}$/J$_{\rm red}$ = 5.3 .  In view of all the
simplifying assumptions, this is remarkably close to predicted value of 8,
and thus most encouraging for these types of models.  In
view of our high-precision Li-T$_{\rm eff}$-\textit{v sin i} relation, we call for new
detailed theoretical exploration of the formation and evolution of the
cool side of the Li gap.

\subsection{The Li Plateau}

Both clusters have A(Li) of $\sim$3.0 between 6000 and 6200 K.  The Hyades is more richly populated in this region, but it retains a very tight and consistent A(Li) trend with depletion moderately increasing with decreasing T$_{\rm eff}$.  The Li plateau for Praesepe has fewer stars and an apparent increased scatter that includes a number of moderate outliers.  These stars do not have large $\sigma_{\rm Teff}$ and are predominantly single stars that are consistent with the Praesepe single star fiducial.  These remaining outliers cannot be explained by errors, with the errors of the Praesepe plateau stars being consistent with the Hyades plateau stars.  Does this suggest a true increase in Li scatter for these plateau stars in Praesepe versus the Hyades?  It is an interesting possibility, but again we find that the mean Li trends are in remarkable agreement between the two clusters.
 
\subsection {G Dwarfs (and K Dwarfs)}

Each cluster shows a very tight G-dwarf Li-T$_{\rm eff}$ trend, but with the Hyades trend appearing even tighter.  This moderate increase in the Praesepe G-dwarf Li scatter cannot be explained by an increase in its abundance errors because they are not larger than those of the Hyades.  Is this scatter significant and does it represent a distinction between these very similar clusters?  For example, Somers \& Pinsonneault (2016) briefly considered the effects on Li depletion that differing rotational histories would have on two clusters of identical age and composition, and they find that differing cluster distributions of initial angular momenta can lead to differing observed Li spreads in clusters as old as M67.  Distinct distributions of initial angular momenta have now been observed with Coker et~al. (2016) finding that h Persei (13 Myr) is rich with rapid rotators in comparison to the younger Orion Nebula Cluster (1 Myr).

To test the statistical significance of this observed scatter in Praesepe versus the Hyades, we combined the low $\sigma_{\rm Teff}$ G-dwarf data sets of both clusters and fit a third-order trend to the total sample.  Analyzing each cluster's residual distribution relative to this fit with a two-sample Kolmogorov-Smirnov test finds a p-value of 0.499.  Even when we include all photometrically discrepant G dwarfs with low $\sigma_{\rm Teff}$, which are preferential Li outliers in Praesepe, this decreases the p-value to only 0.395.  This means that we cannot rule out that these A(Li) are drawn from the same sample.  Therefore, the Li morphologies of the two clusters are indistinguishable.  These Praesepe outliers are still of interest, but the apparent increase in scatter of Praesepe remains insignificant and the slight departures of the corresponding trends (dashed lines) suggested in both the hottest and coolest G dwarfs are not significant either.  Lastly, for K dwarfs both clusters show significant depletion with mostly low Li upper limits.

The identical [Fe/H] and G-dwarf morphologies, in addition to identical trends in the cool side of the Li gap, suggest that the two clusters must have had similar A(Li)$_{\rm init}$.  For this consistency of A(Li) to remain across such a large range of T$_{\rm eff}$ and Li depletions, it is unlikely that there were differences in their A(Li)$_{\rm init}$ that have been perfectly balanced out across this significant parameter space.  However, this is difficult to test directly because we cannot assume that \textit{any} of the stars are undepleted.  Even if some are, we cannot identify them confidently.  For example, we know that while SPTLBs on the Li Plateau are minimally depleted, the depletion is nonetheless greater than zero.  This is suggested by the P97 model predictions, the complexities of component interactions in the SPTLB, and the empirical fact that some cluster stars near 6700 to 7200 K seem to have even higher A(Li) than do the SPTLBs.  (A cautionary remark: in stars near 7000 K, diffusion may levitate Li to the surface from layers below, thereby enriching the surface A(Li) to values \textit{above} the initial value, so that these stars are also unreliable indicators of the cluster A(Li)$_{\rm init}$, Deliyannis, Steinhauer, \& Jeffries 2002.)  In spite of these challenges, circumstantial evidence for similar A(Li)$_{\rm init}$ for the two clusters comes from Galactic Li production models.  Li production processes such as spallation, SNe II, AGB stars, and novae have been thought to correlate with the production of Fe (e.g., Romano 1999; Ryan et~al.\ 2001; Travaglio et~al.\ 2001), so we might expect that two clusters with identical [Fe/H] will have formed with a similar A(Li)$_{\rm init}$. 

Standard theory predicts that G-dwarf Li depletion depends on [Fe/H], but differences in overall metallicity will also cause differences in Li depletion even in stars with the same [Fe/H].  Therefore, in two stars with the same [Fe/H] but different [alpha/Fe], the more alpha-rich star will deplete more Li (e.g., Swenson et~al.\ 1994).  Because the two clusters have identical Li morphologies and [Fe/H], we might then also expect that the two clusters will have similar [alpha/Fe]. Indeed, Carrera \& Pancino (2011) find that both clusters have a solar [$\alpha$/Fe].  However, this result is based on only 3 red giants in each cluster.  

Focusing on the degree of the G-dwarf Li depletion, and as discussed in Section 6.1, it is clear that the Hyades and Praesepe G dwarfs have depleted far more of their Li than predicted by the P97/SP14 standard models.  If Galactic Li production means that our assumed A(Li)$_{\rm init}$ for the standard Li isochrones should be even higher than the meteoritic value of 3.31 that was assumed, then the discrepancy between the models and observations is even larger.  rotationally-induced mixing is likely to be important, as evidenced by the higher A(Li) seen in the SPTLBs.  However, gravity waves may also play an important role, especially at later times when the effects of rotational mixing diminish.  Regardless, the tightness of the G-dwarf Li morphology suggests that the correct mechanism(s) produce very little, if any, scatter at this age.  Therefore, if rotational mixing creates both the cool side of the Li gap and the G-dwarf Li depletion, it must be able to create less Li scatter in the G dwarfs as compared to the Li-gap dwarfs.  This might be possible if, for example, the higher-massed Li gap stars formed with a larger range in initial angular momentum than the G dwarfs did, as suggested by Barnes (2007) and Reinhold \& Gizon (2015).  Another possibility is the rotational models described in Somers \& Pinsonneault (2015), where rotation correlates with radial inflation during the PMS and can strongly affect Li depletion.  This creates the Li scatter of order 1 dex observed in the well-studied young Pleiades, with faster rotating G dwarfs undergoing more inflation and hence significantly less depletion during the PMS.  In comparison, the predicted spread in F-dwarf PMS depletion is relatively minor.  Is it by chance that at the age of the Hyades the G-dwarf A(Li) reach a brief equilibrium?  Specifically, does the subsequent rotationally-induced mixing that occurs during the MS and that causes faster rotators to deplete relatively more Li balance the effects of inflated radii that dominate during PMS Li depletion?

Looking at the broader picture, in comparison to both the Hyades and Praesepe the young Pleiades-aged clusters show much larger Li scatter in their G dwarfs (e.g., Soderblom et~al.\ 1993b), and very old clusters like M67 also show greater scatter (Pasquini et~al.\ 2008; Jones et~al.\ 1999), which based on error analysis is very likely to be real.  The responsible Li depletion mechanisms must be able to explain this evolving scatter.  One of our primary goals in this series of studies is to test the predicted [Fe/H] dependence of Li depletion in the standard model.  This can also provide insight on the metallicity dependence, if any, of these additional depletion mechanisms.  



\section{Summary}

Applying Y$^2$ isochrones to Johnson photometry of the Hyades and
Praesepe, we find that the turnoff ages of these clusters, at 635$\pm$25
Myr and 670$\pm$25 Myr, respectively, are indistinguishable.

We present WIYN/Hydra spectra with high signal-to-noise for 37 Hyades and 78
Praesepe dwarfs that are highly probable members as inferred from
previously published PMs and radial velocities.  We measured
radial velocities for all of our sample stars, finding 33 Hyades and 67
Praesepe stars that are consistent with single-star membership (``radial
velocity members").  Using only these subsamples, we derive cluster means
of 39.1$\pm$0.2($\sigma_\mu$) and 34.7$\pm$0.2($\sigma_\mu$) for the Hyades and
Praesepe, respectively, in outstanding agreement with previous studies.

To strengthen further our abundance results we also performed a thorough
analysis of each star's photometry.  Instead of basing the spectral
parameters on only one color, we examined all 10 colors derivable from
UBVRI photometry.  We created empirical color-color relations between each
color and B-V using each cluster's single-star fiducial.  This allowed us
to combine B-V with nine additional effective B-V colors transformed from the
other colors.  This creates a robust multicolor based B-V and T$_{\rm eff}$. 
Across these multiple colors, some of the Hyades and Praesepe stars have high 
$\sigma_{\rm Teff}$ ($>$75 K), which could be
caused by binarity or atmospheric peculiarities.  Indeed, a large fraction
of known spectroscopic binaries in both clusters are found to have high $\sigma_{\rm Teff}$.

Considering only Praesepe radial velocity members with low $\sigma_{\rm Teff}$ 
that are rotating slowly (\textit{v sin i} $<$ 25 km s$^{-1}$), and using 16 isolated 
Fe I lines we find (from 415 Fe I measurements in 39 stars) a cluster mean
[Fe/H]=0.156$\pm$0.004 ($\sigma_\mu$), which is quantitatively consistent with our
result for 37 Hyades proper motion members of [Fe/H]=0.146$\pm$0.004($\sigma_\mu$).
Daytime solar spectra taken with the same instrumentation were used to
define solar gf values, thus placing our results on a purely
self-consistent scale relative to the Sun.  Overall, the majority of the
lines are remarkably well-behaved across a large range in
T$_{\rm eff}$.  We discarded those lines that were not well-behaved, stronger
than 150 m\AA, or too noisy ($<$ 3$\sigma$ according to the $\sigma$ relation of
D93).  The stellar [Fe/H] show no trend with T$_{\rm eff}$ over a range of T$_{\rm eff}$
that exceeds 1700 K.

Spectral Li synthesis of our stars finds the well-established Li-T$_{\rm eff}$
morphologies in these two clusters, with Li-rich A dwarfs, heavily
depleted Li-gap F dwarfs, a richer Li plateau in late F/early G dwarfs,
increasing Li depletion with decreasing T$_{\rm eff}$ in G dwarfs, and K dwarfs
having undetectable Li.  We supplement our sample of A(Li) with analysis
of spectra from T93 of stars that we ourselves did not
observe, having applied the same spectral synthesis techniques.
Additionally, we reanalyzed published Li equivalent widths and synthetic
A(Li) from other Li studies of these clusters using our own stellar
parameters and COGs, and placed all A(Li) on a consistent scale with our
own observations.  The resulting total samples constitute the largest
self-consistently analyzed samples of A(Li) in both clusters.

While T93 found that some binaries have larger Li scatter, we find
(more broadly) that stars with large $\sigma_{\rm Teff}$
have larger Li scatter about the mean Li-T$_{\rm eff}$ trends than do stars with low
$\sigma_{\rm Teff}$, and so we exclude such stars from further consideration (except
for some SPTLBs, which typically lie above the mean trends, as expected
from models with rotational mixing combined with tidal circularization
theory).  For each cluster we define a ``prime" subsample consisting of
probable members with low $\sigma_{\rm Teff}$ and no measurable spectroscopic
evidence of contamination from a secondary (from \textit{fxcor}), even if they are known
binaries from other studies.  The resulting tight Li-T$_{\rm eff}$ relations are most
encouraging for our future WIYN/Hydra studies of clusters that have little
or no proper motion and binarity information available, where we can 
impose the restriction to consider only stars that have radial velocities
consistent with single star membership and a low $\sigma_{\rm Teff}$.

Using only the prime samples leads to the conclusion that where both clusters have a
significant number of members their Li-T$_{\rm eff}$ morphologies are indistinguishable 
over the entire T$_{\rm eff}$ range under consideration, which spans 3000 K (4800 to 7800 
K).  Each cluster's Li morphology is also highly complementary to that of the other
cluster, in the sense that where one cluster has a paucity of stars, the
other cluster's Li-T$_{\rm eff}$ trend merges remarkably with the Li-T$_{\rm eff}$ trends on
either side of the T$_{\rm eff}$ range in question.  We thus propose that the
combined prime samples for both clusters offers stronger constraints for
models than does the sample from either cluster alone.  Armed with the
combined prime samples, we re-evaluate and refine knowledge of the Li-T$_{\rm eff}$
morphologies from A dwarfs to K dwarfs.

A dwarfs (T$_{\rm eff}$ $>$ 7000 K) suggest decreasing A(Li) with increasing T$_{\rm eff}$, but
there are too few stars to be sure.  We are aware of no models that predict such a trend.

Previously, the hot side of the Li gap was defined mostly by Hyads that we
now know have large $\sigma_{\rm Teff}$.  If instead we use the combined prime
samples, two possible morphologies emerge.  One is that a nearly vertical
``wall" of A(Li) rises in a tiny T$_{\rm eff}$ range from A(Li) = 1.70 at 6635 K to
3.33 at 6715 K.  Such a wall could be consistent with diffusion.  
Although, as we discuss, there is a large variety of evidence suggesting
the rotational mixing is the dominant Li depletion mechanism for cooler
gap dwarfs, diffusion cannot be ruled out as the cause of the wall.  The
second possibility is that the rise in A(Li) is more gradual from A(Li) =
1.70 at 6635 K to 3.30 at 7000 K, in which case, remarkably, four stars near
6700 K are minimally depleted (A(Li) $\geq$ 3.0) and lie well above their
neighbors.  Unfortunately, we cannot distinguish between these two
possibilities.

The middle of the Li gap (6500 to 6700 K for possibility 1, above, or
6500 to 6800 K for possibility 2) is at least as deep as A(Li) = 1.7, and is
thus at least 1.6 dex below the A(Li)$_{\rm meteoritic}$ = 3.31, or at least
1.7 to 1.8 dex below A(Li)$_{\rm init}$ if the initial cluster value is at 3.4 to 3.5.

The cool side of the Li gap (6635 to 6200 K) shows rising Li with
decreasing T$_{\rm eff}$.  Most binaries deviate from the much tighter Li
morphology defined by the single stars; perhaps binarity is somehow
related to Li depletion in this T$_{\rm eff}$ range.  Considering only single stars
from the prime samples, we find that Hyades and Preasepe each have
linear anti-correlations between A(Li) and \textit{v sin i}, as was found for the
Hyades by BT86 and Boesgaard (1987), that are very similar for the two
clusters.  We also find similar anti-correlations between A(Li) and T$_{\rm eff}$
for the two clusters, as well as similar correlations between \textit{v sin i} and
T$_{\rm eff}$.  There is moderate scatter in these correlations that cannot be fully explained by the
errors.  However, consideration of the full combined sample from both
clusters in 3-dimensions results in a tighter relation when considering Li, \textit{v sin i}, and T$_{\rm eff}$
simultaneously.  Considering both the errors and the
limitations of using \textit{v sin i} to represent rotation rates, this 3-dimensional
relation may be able to explain most of observed scatter.  These patterns
are consistent with the basic tenets of the Yale-style rotational models.
In this T$_{\rm eff}$ range, angular momentum on the ZAMS is a steeply increasing
function of T$_{\rm eff}$.  By the age of the Hyades, stars in this T$_{\rm eff}$ range have
spun down by a factor of 2 to 3, in terms of their periods.  Thus, the amount
of angular momentum lost between 100 and 650 Myr is also a steeply
increasing function of T$_{\rm eff}$.  To first approximation, the rotational
models we have discussed tie mixing and Li depletion to loss of angular
momentum, providing a natural explanation for the cool side of the Li gap.
(We have also mentioned some additional factors that may be important.)  
A semi-quantitative comparison with the rotational models of P90 lends very
encouraging support to these types of models.  We
call for a new detailed theoretical exploration of the formation and
evolution of the cool side of the Li gap, in view of the refinements we
have provided of the Li-T$_{\rm eff}$-\textit{v sin i} relation.

The combined sample of stars cooler than the gap shows a Li plateau near 
A(Li) $\sim$ 3.0 between 6200 and 6000 K.

G dwarfs in both clusters show decreasing A(Li) with decreasing T$_{\rm eff}$, with
very little scatter.  Unlike on the cool side of the Li gap, G dwarf
binaries (with low $\sigma_{\rm Teff}$) fall right on the trends defined by single
stars.  (Recall that, by contrast, stars with large $\sigma_{\rm Teff}$ show
significant scatter around these trends.)  The two cluster trends are
indistinguishable.  It might appear to the eye that Praesepe has more
scatter but a KS test cannot rule out that both clusters are drawn from the
same sample.  If additional parameters besides metallicity and age exist
that govern cluster G-dwarf depletion trends, such as the distribution of
initial angular momenta, then we have not detected evidence for the
importance of such parameters.  Or it would seem that any additional parameters that are
important are also consistent in these two clusters.

The strong consistency in the Li-T$_{\rm eff}$ trends in both clusters
merits further discussion and perhaps also some speculation.  While the
strong consistency in Hyades and Praesepe A(Li) suggests that clusters
with similar [Fe/H] likely undergo very similar Li depletion patterns at
similar age across a broad range of T$_{\rm eff}$, this also suggests that they
have consistent A(Li)$_{\rm init}$.  Of course, more complex scenarios are
conceivable.  For example, one cluster might have formed with a higher
A(Li)$_{\rm init}$, which would require Galactic Li production that somehow differs
for the same [Fe/H].  This would also require that all of its stars depleted more 
Li during their lifetime (maybe because of higher typical initial
stellar angular moments) in
just the right way so as to match the other cluster's Li morphology at the
current age.  For now we accept the far simpler scenario until
new information comes along that demands greater complexity.  This simpler
scenario purports both that Galactic Li production (that is, the cluster
A(Li)$_{\rm init}$) correlates with metallicity and that stellar Li depletion
(both standard and non-standard) is identical in clusters of
identical age and metallicity.  We fully acknowledge the dangers of drawing
general conclusions from a sample of only two clusters, and we encourage
the study of larger numbers of clusters that have identical age and
metallicity.

This study provides a strong foundation for further analysis of
Hyades-aged clusters.  Specifically, studying Hyades-aged clusters of
differing metallicity can test the potential effects of differing [Fe/H]
on Li depletion, testing model predictions and constraining future models,
and can explore whether there might be evidence for differences in the
cluster A(Li)$_{\rm init}$.

\vspace{0.7cm}
We thank the referee (David James) for his thorough comments on this paper.  J.D.C. acknowledges support from the 
National Science Foundation (NSF) through grant AST-1211719.  C.P.D. acknowledges support from the NSF from grants 
AST-0607567 and AST-1211699.  This 
research has made use of the WEBDA database, operated at the Department of Theoretical Physics and Astrophysics 
of the Masaryk University.  We would also like to thank the WIYN observatory staff, and
especially Di Harmer, whose dedication and skillful work helped us obtain
these excellent spectra.

\end{document}